\documentclass[12pt]{article}
\usepackage{graphicx}
\usepackage[dvipsnames]{xcolor}
\usepackage[section]{placeins}
\usepackage[top=1in, left=1in, right=1in, bottom=1in]{geometry}
\usepackage[parfill]{parskip}	% Activate to begin paragraphs with an empty line rather than an indent
\usepackage{caption}
\usepackage{subcaption}
\usepackage[export]{adjustbox}
\usepackage{amssymb}
\usepackage{bbm}
\usepackage{bm}
\usepackage{dsfont}
\usepackage{amssymb}
\usepackage{amsmath}
\usepackage{amsfonts}

\usepackage{enumitem}
\usepackage{setspace}
\usepackage{booktabs}
\usepackage[compact]{titlesec}

\usepackage{lscape}

\usepackage{accents}

\usepackage[nocitation]{apacite}
\usepackage{natbib}
\bibpunct{(}{)}{;}{a}{}{,}
\usepackage{hyperref}
\usepackage{titling}
\usepackage{pifont}
\usepackage[capposition=top]{floatrow}
\newcommand{\subtitle}[1]{%
  \posttitle{%
    \par\end{center}
    \begin{center}\large#1\end{center}
    \vskip0.5em}%
}
\usepackage{amsthm}
\usepackage{amsmath} %maths
\usepackage{multirow}
\usepackage{multicol}
\usepackage{mathrsfs}

\usepackage{tikz}
\usetikzlibrary{shapes.geometric, arrows.meta, positioning, fit}

  {\end{smallmatrix}\right)}

\usepackage{algorithmicx}
\usepackage{algorithm}
\usepackage[noend]{algpseudocode}
\usepackage{algpseudocode} 
\usepackage{outlines}

\hypersetup{hidelinks}

\newtheorem{theorem}{Theorem}

\newtheorem{algorithm}[theorem]{Algorithm}

\def\ci{\perp\!\!\!\perp}

\newtheorem{claim}{Claim}
\newtheorem{remark}{Remark}
\newtheorem{subtheorem}{Proposition}

\usepackage{mathrsfs}

\usepackage{color}

\usepackage{array}
\newcolumntype{C}[1]{>{\centering\arraybackslash}m{#1}}

\begin{document}
\pagestyle{plain}

\newtheoremstyle{mystyle}% name
{\topsep}% Space above
{\topsep}% Space below
{\it}% Body font
{}% Indent amount
{\bf}% Theorem head font
{.}%Punctuation after theorem head
{.5em}%Space after theorem head
{}% theorem head spec
\theoremstyle{mystyle}
\newtheorem{assumptionex}{Assumption}
\newenvironment{assumption}
  {\pushQED{\qed}\renewcommand{\qedsymbol}{}\assumptionex}
  {\popQED\endassumptionex}
\newtheorem{assumptionexp}{Assumption}
\newenvironment{assumptionp}
  {\pushQED{\qed}\renewcommand{\qedsymbol}{}\assumptionexp}
  {\popQED\endassumptionexp}
\renewcommand{\theassumptionexp}{\arabic{assumptionexp}$'$}

\newtheorem{assumptionexpp}{Assumption}
\newenvironment{assumptionpp}
  {\pushQED{\qed}\renewcommand{\qedsymbol}{}\assumptionexpp}
  {\popQED\endassumptionexpp}
\renewcommand{\theassumptionexpp}{\arabic{assumptionexpp}$''$}

\newtheorem{assumptionexppp}{Assumption}
\newenvironment{assumptionppp}
  {\pushQED{\qed}\renewcommand{\qedsymbol}{}\assumptionexppp}
  {\popQED\endassumptionexppp}
\renewcommand{\theassumptionexppp}{\arabic{assumptionexppp}$'''$}

\renewcommand{\arraystretch}{1.3}

\newcommand{\argmin}{\mathop{\mathrm{argmin}}}
\makeatletter
\newcommand{\grande}{\bBigg@{2.25}}
\newcommand{\enorme}{\bBigg@{5}}

\newcommand{\blind}{0}

\newcommand{\tit}{Adaptive Orthogonalization for Stable Estimation of the
Effects of Time-Varying Treatments}

\if0\blind

{\title{\tit
\thanks{The authors thank Zach Branson, Ambarish Chattopadhyay, Eric Cohn, Larry Han, Jie Hu, Kwangho Kim, Kosuke Imai, Marshall Joffe, Bijan Niknam, Marc Ratkovic, Jamie Robins, Zhu Shen, Davide Viviano,
and Yi Zhang for helpful comments and suggestions. This work was partially supported by the Alfred P. Sloan Foundation (G-2020-13946) and the Patient-Centered Outcomes Research
Initiative (PCORI, ME-2022C1-25648).
}\vspace*{.3in}}
\author{
Yige Li 
\thanks{Department of Health Care Policy, Harvard University.}	   
\and
Mar\'ia de los Angeles Resa 
\thanks{Department of Statistics, Columbia University.}
\and 
Jos\'e R. Zubizarreta$^{*}$  
\thanks{Department of Health Care Policy, Biostatistics, and Statistics, and CAUSALab, Harvard University.
Email: \url{zubizarreta@hcp.med.harvard.edu}.} 
}

\date{} 

\maketitle
}\fi

\if1\blind
\title{\tit}
\date{} 
\maketitle
\fi

\begin{abstract}
Inferring the causal effects of time-varying treatments is often hindered by highly variable inverse propensity weights, particularly in settings with limited covariate overlap. Building on the key framework of \cite{imai2015robust}, we establish sufficient balancing conditions for identification in longitudinal studies of treatment effects and propose a novel estimator that directly targets features of counterfactual or potential covariates. Instead of balancing observed covariates, our method balances the components of covariates that are orthogonal to their history, thereby isolating the new information at each time point. This strategy directly targets the joint distribution of potential covariates and prioritizes features that are most relevant to the outcome. We prove that the resulting estimator for the mean potential outcome is consistent and asymptotically normal, even in settings where standard inverse propensity weighting fails. Extensive simulations show that our estimator attains efficiency comparable to that of g-computation while providing superior robustness to model misspecification. We apply our method to a longitudinal study of private versus public schooling in Chile, demonstrating its stability and interpretability in estimating their effects on university admission scores.
\end{abstract}

%\vspace*{.3in}

\begin{center}
\end{center}
\clearpage
% \doublespacing

\singlespacing
\pagebreak
\thispagestyle{empty} 
\tableofcontents
\pagebreak
\doublespacing

\setcounter{page}{2}

\section{Introduction}
\label{s:intro}
\subsection{Estimating time-varying treatment effects}

Many pressing questions in biomedicine and the social sciences relate to treatments that are administered repeatedly over time rather than at a single point \citep{robins1986new}.
A canonical example is the estimation of the effect of treatment strategies for HIV-positive patients, where decisions to initiate zidovudine or prophylaxis vary over time depending on a patient's evolving clinical history, such as CD4 lymphocyte counts \citep{hernan2001marginal}.
When estimating the effects of time-varying treatments, one-step covariate adjustment methods, including matching, regression or weighting, may yield biased effect estimators if they incorrectly adjust for post-treatment covariates \citep{rosenbaum1984consequences}. 
Generally speaking, unbiased estimation of the effects of treatments that vary in time requires covariate adjustment that respects the temporal structure of the data.

In essence, under exchangeability and positivity assumptions that all the time-varying confounders are measured and that each individual has a positive probability of receiving the treatment at each time point given their treatment and covariate histories, the distributions of potential outcomes can be nonparametrically identified. Marginal structural models (MSMs) are commonly used to model these potential outcomes \citep{robins1998marginal}. If MSMs are correctly specified, their parameters are identifiable \citep{hernan2001marginal}. 
Identification can be achieved by the g-formula \citep{robins2000robust} or inverse probability weighting (IPW) \citep{cole2008constructing}, each giving rise to a class of estimators.

On the one hand, estimators based on the g-formula require correct specification of the outcome-generating mechanism to be unbiased \citep{bang2005doubly}. For example, the iterated conditional expectation (ICE) method is a recursive g-computation algorithm that requires fitting a sequence of models for the outcome conditional on past treatment and covariate histories \citep{robins2000robust}. The choice of functional form for these models introduces a fundamental bias-variance trade-off. 
Parametric approaches are susceptible to bias from model misspecification, whereas nonparametric approaches, while more flexible, are limited by the curse of dimensionality and slow convergence rates \citep{robins1997toward}. 
% While parametric g-formula approaches are sensitive to model misspecification, nonparametric implementations may suffer from the curse of dimensionality and increased variance \citep{robins1997toward}.

On the other hand, IPW estimators require modeling the conditional probabilities of treatment assignment given past covariates and treatments.
However, even with a correctly specified treatment model, the resulting inverse probability weights can inadequately adjust for confounding in settings with limited covariate overlap or extreme treatment assignment probabilities. 
When the estimated probabilities are close to zero or one, the resulting weights can become highly unstable, producing heavy-tailed distributions and non-convergent asymptotic variances \citep{ma2020robust}. 
Investigators can process weights to improve their stability; however, tactics such as trimming are highly dependent on cutoff choices \citep{li2018addressing} and need a debiasing step using outcome models to achieve asymptotic normality \citep{ma2020robust}.

These challenges have motivated the development of advanced methods for time-varying treatments with improved efficiency and robustness. 
The principle behind these methods is to use doubly robust estimators, such as augmented inverse probability weighting (AIPW), which are consistent as long as either the treatment or the outcome model is correctly specified \citep{bang2005doubly}.
Machine learning algorithms fit well into this framework, providing model flexibility while preserving theoretical performance guarantees \citep{chernozhukov2018double}. 
Building upon the AIPW framework, targeted maximum likelihood estimation (TMLE; \citealt{vanderlaan2011targeted}) combines machine learning with influence function theory to achieve both double robustness and asymptotic efficiency. 
Together, these advances provide researchers with principled statistical methods for studying the causal effects of time-varying treatments. 
However, these estimators can still be unstable when the positivity assumption is nearly violated, and their use of black-box algorithms can make diagnostics difficult and obscure how the data are aggregated to produce causal effect estimates.

To address these challenges, a distinct line of research focuses on balancing covariates directly through weighting.
Notably, \citet{imai2015robust} devise a set of balance conditions for longitudinal studies based on counterfactual or potential covariates, enabling direct estimation of inverse propensity scores without computing reciprocals. 
Following these principles, several methods have been proposed \citep{zhou2020residual,kallus2021optimal,avagyan2021stable,yiu2022joint}; however, since the potential covariates are not fully observed, in practice, most implementations balance the means of observed covariates across treatment paths.
This limitation can compromise performance, particularly under limited covariate overlap, because these balancing weights fail to correctly adjust for time-varying covariates in the outcome model.  
Moreover, these estimators generally exhibit lower efficiency than correctly specified g-computation formula approaches. 
 
These considerations have motivated developments, such as dynamic covariate balancing, a method proposed by \citet{viviano2024dynamic}. Their approach estimates treatment effects by recursively projecting expected potential outcomes onto past histories and mitigating bias by sequentially balancing time-varying covariates. Our method focuses on the covariate process directly. We simultaneously capture the shifts of time-varying covariates by projecting post-treatment covariates onto their ancestors and balancing the resulting residuals. This approach targets population joint covariate distribution by isolating outcome-relevant orthogonal components. To enhance interpretability, we use a simple Hájek estimator that does not rely on estimated regression coefficients and provides straightforward diagnostics.

\subsection{Contribution and outline}\label{sec::contribution}

This paper articulates sufficient balance conditions for identification of time-varying treatment effects under a semi-parametric outcome model. It also proposes a method that implements these conditions using potential covariates by balancing the components of observed covariates that are orthogonal to their history, thereby isolating the new information at each time point.
Specifically, building on \citet{imai2015robust}, our method balances time-varying, outcome-relevant covariates towards their estimated mean potential covariates.
This is accomplished by balancing the residuals obtained from projecting each covariate onto its ancestors towards explicit targets to control chance error. 
By imposing balance on these residuals, this estimator achieves asymptotic normality for mean potential outcomes.

To enhance estimator stability, the proposed approach finds weights of minimal variance subject to the aforementioned balance requirements. This directly controls the mean squared error (MSE) and ensures the estimator's asymptotic normality.
Under regularity conditions, the approach is robust to outcome model misspecification, as the dual formulation of the optimization problem corresponds to modeling the inverse propensity score with a penalized loss function. 
Moreover, the proposed approach avoids extrapolation by restricting weights to be nonnegative. 
Empirical studies demonstrate that it matches or outperforms standard approaches, including the g-computation formula, IPW, doubly robust estimators, and some balancing methods, in terms of bias and variance, across a range of simulation scenarios.

The paper proceeds as follows. Section \ref{Framework} outlines the setup, notation, estimands, and assumptions.
Section \ref{Methods} presents identification strategies and our proposed estimation procedure.
Section \ref{Practical} discusses practical considerations during implementation. 
Section \ref{Empirical} evaluates the approach in three simulation studies. Section \ref{CaseStudy} applies the approach to a longitudinal study of the effect of attending a private subsidized school on students' university admission test scores in Santiago, Chile. 
Finally, Section \ref{Discuss} concludes, discussing limitations and providing future directions.

%\section{Overview of marginal structural models}
\section{Framework}\label{Framework}

%%%%%%%%%%%%%%%%%%%%
%%%%%%%%%%%%%%%%%%%%
\subsection{Setting and notation}

We consider a random sample $\mathcal S$ of $n$ independent and identically distributed individuals from a population $\mathcal P$ followed over $T+1$ time points. 
At each time point $t \in \mathcal{T} := \{1,2,...,T\}$, individual $i \in \mathcal{I} := \{1,2,...,n\}$ receives a time-dependent binary treatment $Z_{it}$. 
The treatment $Z_{t}$ may depend on its past treatments $\bar{Z}_{t-1} := (Z_{1}, ..., Z_{t-1})$ and past covariates $\bar{X}_t := (X_{1}, ..., X_{t})$, and may affect its future treatment $\underaccent{\bar}{Z}_{t+1} := (Z_{t+1}, \cdots, Z_{T})$, future covariates $\underaccent{\bar}{X}_{t+1} := (X_{t+1}, \cdots, X_{T})$ and continuous outcome $Y$ observed at the end of follow up at time $T+1$.
The treatment $Z_t$ and the covariates $X_t$ respectively take values from the sets $\mathcal{{Z}}_t = \{0, 1\}$ and $\mathcal{{X}}_t = \mathbb R^{P_t}$, where $P_t$ is the dimension of $X_t$;
$\bar{Z}_{t}$ takes values from $\mathcal{\bar{Z}}_t=\mathcal{{Z}}_1\times\dots\times\mathcal{{Z}}_t$ and $\bar{X}_{t}$ from $\mathcal{\bar{X}}_t=\mathcal{{X}}_1\times\dots\times\mathcal{{X}}_t$. 
Let $1\{\cdot\}$ be the indicator function, so that $1\{\bar{Z}_T = \bar{z}_T\}$ equals $1$ if the observed treatment path $\bar{Z}_T$ is equal to $\bar{z}_T$, and $0$ otherwise.

Building upon the potential outcomes framework formalized by \cite{rubin1974estimating} and extended to longitudinal contexts by \cite{robins1986new}, we denote  $Y(\bar{z}_T)$ as the potential outcome observed under the treatment path $\bar{z}_T$, where $\bar{z}_T \in \mathcal{\bar{Z}}_T$. Similarly, $X_{t}(\bar{z}_{t-1})$ represents the potential value of covariates at time $t$ given the treatment history $\bar{z}_{t-1} \in \mathcal{\bar{Z}}_{t-1}$. We define $\bar{X}_{t}(\bar{z}_{t-1})$ as the history of potential covariates $(X_{1}, X_2(z_1), ...X_t(\bar{z}_{t-1}))$ and $\underaccent{\bar}{X}_{t+1}(\bar{z}_{t})$ as the future potential covariates $(X_{t+1}(\bar{z}_{t}), ..., X_T(\bar{z}_{T-1}))$. 

%%%%%%%%%%%%%%%%%%%%
%%%%%%%%%%%%%%%%%%%%
\subsection{Estimands and assumptions}

For nonparametric identification of $\operatorname{E}_{\mathcal P}[Y(\bar{z}_T)]$, we make Assumptions \ref{ignorability}-\ref{consistency} \citep{robins1998marginal}.

\begin{assumption} \label{ignorability}Sequential ignorability: potential outcomes and covariates are conditionally independent of current treatment assignment, given past treatments and covariates.

a. $Y(\bar{z}_T)\ci Z_{1}\mid X_{1}, ~ Y(\bar{z}_T)\ci Z_{t}\mid(\bar{Z}_{t-1}, \bar{X}_{t}), \;\forall \bar{z}_T\in \mathcal{\bar{Z}}_{T}, \; \forall t \in \{2,...,T\}.$ 

b. ${X}_{2}({z}_{1})\ci Z_{1}\mid {X}_{1}, ~ \underaccent{\bar}{X}_{t+1}(\bar{z}_{t})\ci Z_{t}\mid(\bar{Z}_{t-1},\bar{X}_{t}), \;\forall \bar{z}_T\in \mathcal{\bar{Z}}_{T}, \; \forall t \in \{2,...,T-1\}.$ 
\end{assumption}

\begin{assumption} \label{positivity}
Positivity: if the joint density of past treatments and past covariates is positive, then the conditional probability of current treatment given the past is positive.

$f_{\mathcal P}(\bar{z}_{t-1},\bar{x}_t)>0 \Rightarrow \mathrm{P}(z_{t}\mid\bar{z}_{t-1},\bar{x}_t)>0, \; \forall (\bar{z}_{t},\bar{x}_t) \in \bar{\mathcal Z}_t\times \bar{\mathcal X}_t, \; \forall t \in \{2,...,T\}.$
\end{assumption}

\begin{assumption} \label{consistency}
Consistency/Stable unit treatment value assumption (SUTVA).

a. $Y = \sum_{\bar{z}_T \in \bar{\mathcal{Z}}_T}1\{\bar{Z}_{T} = \bar{z}_T\} Y(\bar{z}_T).$ 
\; b. $X_{t} = \sum_{\bar{z}_{t-1} \in \bar{\mathcal{Z}}_{t-1}}1\{\bar{Z}_{t-1} = \bar{z}_{t-1}\} X_{t}(\bar{z}_{t-1}).$
\end{assumption}

\section{Identification, estimation, and implementation}\label{Methods}

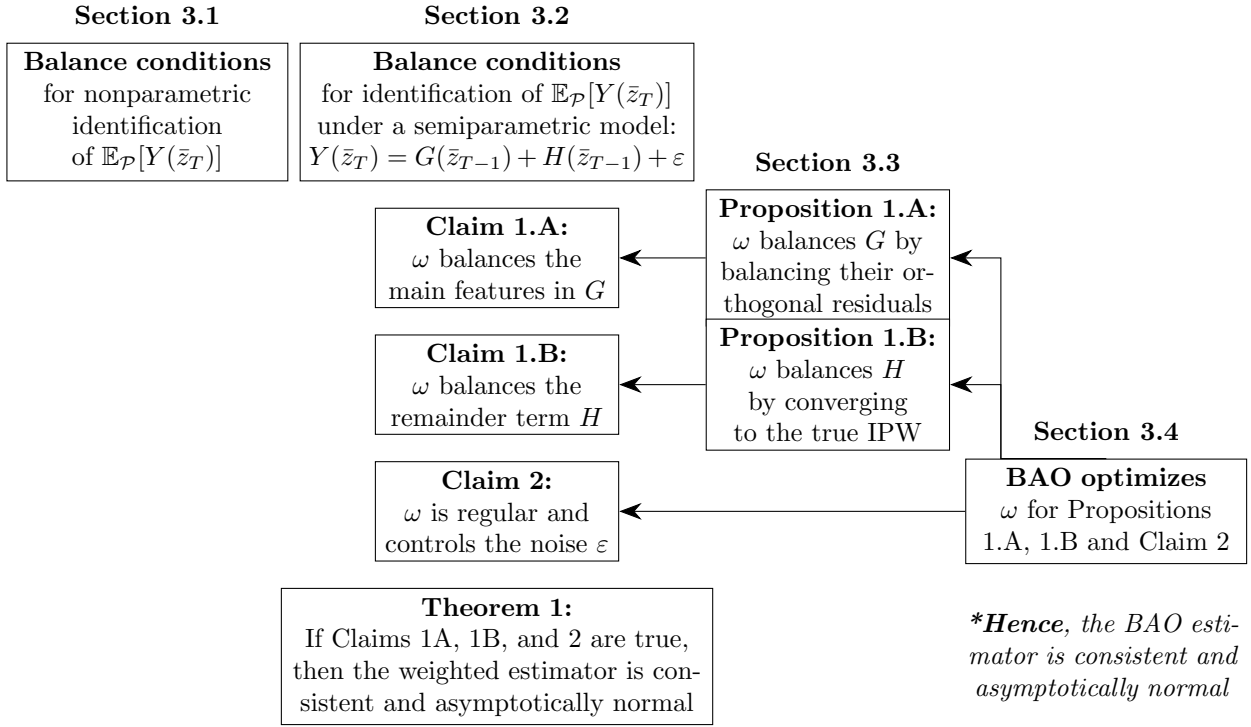
\begin{figure}[!htbp]
\centering
\caption{Overview of Section \ref{Methods}.}
\label{fig::roadmap}
\footnotesize
\begin{tikzpicture}[
    node distance=1.5cm and 1cm,
    startstop/.style={rectangle, minimum width=0.01cm, minimum height=0.01cm, text centered, draw=black, fill=white, text width=3.5cm, align=center},
    block/.style={rectangle, minimum width=3cm, minimum height=1.2cm, text centered, draw=black, fill=white, text width=3.cm, align=center},
    decision/.style={rectangle, minimum width=0.1cm, minimum height=0.1cm, text centered, draw=black, fill=white, text width=2.8cm, align=center},
    line/.style={draw, -{Stealth[length=3mm]}}
]

% --- PART 1: The Theoretical Contract (Section 3.1) ---
\node (goal) [startstop, text width=3.5cm] {\textbf{Balance conditions} \\for nonparametric \\identification of $\mathbb{E}_{\mathcal P}[Y(\bar{z}_T)]$};
\node (model) [block, right=0.15cm of goal, text width=5cm] {\textbf{Balance conditions} \\ for identification of $\mathbb{E}_{\mathcal P}[Y(\bar{z}_T)]$ \\under a semiparametric model: \\$Y(\bar{z}_T) = G(\bar{z}_{T-1}) + H(\bar{z}_{T-1})+ \varepsilon$};

% The two main conditions
\node (cond1a) [decision, below =0.35cm of model, text width=3cm] {\textbf{Claim \ref{con::bal}.A:} \\$\omega$ balances the main features in $G$};
\node (cond1b) [decision, below =0.35cm of cond1a, text width=3cm] {\textbf{Claim \ref{con::bal}.B:} \\$\omega$ balances the remainder term $H$};
\node (cond2) [decision, below =0.35cm of cond1b, text width=3cm] {\textbf{Claim \ref{con::wei}:} \\$\omega$ is regular and controls the noise $\varepsilon$};

% The main theorem that relies on the conditions
\node (thm1) [block, below=0.35cm of cond2, fill=white, text width=5.5cm] {\textbf{Theorem \ref{thm::1}:} \\
If Claims \ref{con::bal}A, \ref{con::bal}B, and \ref{con::wei} are true, then the weighted estimator is consistent and asymptotically normal};

% --- PART 2: Fulfilling the Contract with BAO (Section 3.2) ---
% The specific propositions for BAO
\node (prop1a) [block, right=1.15cm of cond1a, text width=3cm] {\textbf{Proposition \ref{prop1}:} $\omega$ balances $G$ by balancing their orthogonal residuals};
\node (prop1b) [block, right=1.15cm of cond1b, text width=3cm] {\textbf{Proposition \ref{prop2}:} \\ $\omega$ balances $H$ by converging to the true IPW};
\node (bao_opt) [block, right=4.6cm of cond2, text width=3.5cm] {\textbf{BAO optimizes} $\omega$ for Propositions \ref{prop1}, \ref{prop2} and Claim \ref{con::wei}};

% The final conclusion
\node (conclusion) [startstop, right=2.8cm of thm1, fill=white, draw=white, text width=4.5cm] {\textit{\textbf{*Hence}, the BAO estimator is consistent and asymptotically normal}};

% --- Draw Arrows ---
% \path [line] (model) -- node[pos=0.3, left] {} (thm1);
% \path [line] (cond1a) -- node[pos=0.3, right] {} (thm1);

% Arrows from conditions to the main theorem
% \draw[line] (model.south) -- ++(0,-0.15) -| (thm1.west);
% \draw[line] (cond1a.south) -- ++(0,-0.15) -| (thm1.west);
% \draw[line] (cond1b.south) -- ++(0,-0.15) -| (thm1.west);
% \draw[line] (cond2.south) -- ++(0,-0.15) -| (thm1.west);

% Arrows showing how BAO fulfills the conditions
% \draw[line, dashed] (prop1b.west) -- ++(-0.5,0) |- (cond1a.south) node[pos=0.25, below, sloped] {for $h_T$};
% \draw[line, dashed] (prop1a.east) -- ++(0.5,0) |- (cond1a.south) node[pos=0.25, below, sloped] {for $g_t$};

% Final implication
% \path [line, very thick] (thm1) -- (conclusion);

% --- Draw Boxes to Group Sections ---
\node[draw=white, fit=(goal), label={[yshift=0.cm]above:\textbf{Section \ref{sec::sufficient}}}] {};
\node[draw=white, fit=(model), label={[yshift=0.cm]above:\textbf{Section \ref{sec::estimation}}}] {};
\node[draw=white, fit=(prop1a)(prop1b), label={[yshift=0.cm]above:\textbf{Section \ref{sec::adjustments}}}] {};
\node[draw=white, fit=(bao_opt), label={[yshift=0.cm]above:\textbf{Section \ref{sec::implementation}}}] {};

\draw[line] (bao_opt.north) -- ++(-1.4,0) |- (prop1a.east);
\draw[line] (bao_opt.north) -- ++(-1.4,0) |- (prop1b.east);
\path [line] (prop1a) -- node[pos=0.5, left] {} (cond1a);
\path [line] (prop1b) -- node[pos=0.5, right] {} (cond1b);
\path [line] (bao_opt) -- node[pos=0.5, right, align=center] {} (cond2);

\end{tikzpicture}
\end{figure}

This section is organized as follows. Drawing on \citet{imai2015robust}, Section \ref{sec::sufficient} introduces balance conditions that enable the nonparametric identification of mean potential outcomes. Section \ref{sec::estimation} presents sufficient conditions for a more restrictive but broad class of semiparametric models and, through Theorem \ref{thm::1}, establishes the consistency and asymptotic normality of the resulting weighting estimator. Section \ref{sec::adjustments} addresses the practical challenge of satisfying the required balance conditions by proposing a covariate orthogonalization procedure. Section \ref{sec::implementation} then outlines a convex optimization approach for finding the weights. 
The results in this Section are summarized in Figure \ref{fig::roadmap}.

%%%%%%%%%%%%%%%%%%
%%%%%%%%%%%%%%%%%%

\subsection{Identification of potential outcomes and balance conditions}\label{sec::sufficient}

We wish to learn the mean potential outcome in population $\mathcal P$ under treatment path $\bar{z}_T\in\bar{\mathcal{Z}}_T$, $\mathrm{E}_{\mathcal P}[Y(\bar{z}_T)]$.
Under Assumptions \ref{ignorability}a, \ref{positivity}, and \ref{consistency}a, we can identify this estimand using IPW \citep{robins2000marginal} as follows,
\begin{eqnarray}
\mathrm{E}_{\mathcal P}[Y(\bar{z}_T)] = \int Y \frac{1\{\bar{Z}_T = \bar{z}_T\}\mathrm{P}(\bar{z}_T)}{\pi(\bar{x}_T;\bar{z}_T)}f_{\mathcal P}(Y| \bar{X}_T=\bar{x}_T,  \bar{z}_T) f_{\mathcal P}(\bar{X}_T=\bar{x}_T| \bar{Z}_{T}= \bar{z}_T)dYd\bar{x}_T, \label{eqn::IdByPS_3.1}
\end{eqnarray}
where $\pi(\bar{x}_T;\bar{z}_T)$ denotes the time-varying propensity score; that is, the product of the treatment assignment probabilities given the past covariates and treatments, $\pi(\bar{x}_T;\bar{z}_T):=\prod_{t = 1}^T \mathrm{P}(Z_t = z_t|\bar{X}_t = \bar{x}_t, \bar{Z}_{t-1}=\bar{z}_{t-1})$.
While Equation \eqref{eqn::IdByPS_3.1} provides identification through the weights $\frac{1\{\bar{Z}_T = \bar{z}_T\}\mathrm{P}(\bar{z}_T)}{\pi(\bar{X}_T;\bar{z}_T)}$, the resulting estimators are often unstable due to poor covariate overlap across possible treatment paths. 
To develop a more stable weighting procedure, we supplement the identification representation provided by Equation \eqref{eqn::IdByPS_3.1} with the following representation derived from the g-formula \citep{robins2000robust}.

Under Assumptions \ref{ignorability}a and \ref{consistency}a, the g-formula identifies the causal estimand as
\begin{eqnarray}
\mathrm{E}_{\mathcal P}[Y(\bar{z}_T)] = \int \mathrm{E}_{\mathcal P}[Y | \bar{X}_T = \bar{x}_T,  \bar{z}_T] \prod_{t = 1}^T f_{\mathcal P}(X_t = x_t| \bar{X}_{t-1} = \bar{x}_{t-1}, \bar{Z}_{t-1} = \bar{z}_{t-1})d\bar{x}_T. \label{eqn::IdByG_3.1} 
\end{eqnarray}
Further, under additional Assumptions \ref{ignorability}b and \ref{consistency}b, as shown in Section S1.1 in the Supplementary Materials, this estimand is given by:
\begin{eqnarray}
\mathrm{E}_{\mathcal P}[Y(\bar{z}_T)] = \int \mathrm{E}_{\mathcal P}[Y| \bar{X}_T(\bar{z}_{T-1}) = \bar{x}_T, \bar{Z}_T = \bar{z}_T]f_{\mathcal P}(\bar{X}_T(\bar{z}_{T-1}) = \bar{x}_T)d\bar{x}_T.\label{eqn::IdByBC_3.1} 
\end{eqnarray}

% A key property of the propensity score we would like our procedure to inherit is that the resulting inverse product weights $\frac{1\{\bar{Z}_T = \bar{z}_T\}\mathrm{P}(\bar{z}_T)}{\pi(\bar{X}_T;\bar{z}_T)}$ balance any function $\psi\{\bar{x}_T;\bar{z}_T\}$ in the subpopulation under treatment path $\bar{z}_T$ relative to its mean potential value under $\bar{z}_T$ (see \cite{imai2015robust} and Section S1.1 in the Supplementary Materials for details),

A key property of inverse propensity weighting, which our procedure aims to inherit, is its ability to balance any function $\psi(\bar{x}_T; \bar{z}_T)$ toward its expectation over the corresponding potential covariate distribution, as explained in \cite{imai2015robust} and extended in Section S1.1 of our Supplementary Materials:
\begin{eqnarray}\label{eqn::balAll}
\int \psi\{\bar{x}_T;\bar{z}_T\}\frac{1\{\bar{Z}_T = \bar{z}_T\}\mathrm{P}(\bar{z}_T)}{\pi(\bar{x}_T;\bar{z}_T)} f_{\mathcal P}( \bar{x}_T|\bar{z}_T)d\bar{x}_T= \int \psi\{\bar{x}_T;\bar{z}_T\}f_{\mathcal P}(\bar{X}_T(\bar{z}_{T-1}) =\bar{x}_T) d\bar{x}_T.
\end{eqnarray}

Use $\mathrm{E}_{\mathcal{P}_\omega}[\cdot]$ to denote the expectation over population $\mathcal P$ using generic weights $\omega$.
Now write $\omega(\bar{x}_T;\bar{z}_T)$ instead of  $\frac{1\{\bar{Z}_T = \bar{z}_T\}\mathrm{P}(\bar{z}_T)}{\pi(\bar{x}_T;\bar{z}_T)}$ in the right hand side of Equation \eqref{eqn::IdByPS_3.1} and define the weighted mean observed outcome under treatment path $\bar{z}_T$ as
\begin{eqnarray}
\mathrm{E}_{\mathcal{P}_{\omega}}[Y|\bar{Z}_T = \bar{z}_T] :=  \int Y\omega(\bar{x}_T;\bar{z}_T) f_{\mathcal P}(Y| \bar{X}_T=\bar{x}_T, \bar{z}_T)f_{\mathcal P}(\bar{X}_T = \bar{x}_T|\bar{Z}_T = \bar{z}_T) dYd\bar{x}_T. \label{eqn::IdByW}
\end{eqnarray}
Integrating the outcome in \eqref{eqn::IdByW}, 
\begin{eqnarray}
\mathrm{E}_{\mathcal{P}_{\omega}}[Y|\bar{Z}_T = \bar{z}_T]=\int \mathrm{E}_{\mathcal P}[Y| \bar{X}_T = \bar{x}_T, \bar{Z}_T = \bar{z}_T] \omega(\bar{x}_T;\bar{z}_T) f_{\mathcal P}(\bar{X}_T = \bar{x}_T|\bar{Z}_T = \bar{z}_T) d\bar{x}_T. \label{eqn::IdByW2}
\end{eqnarray}

From the expressions in Equations \eqref{eqn::IdByBC_3.1} and \eqref{eqn::IdByW2}, it follows that if we find weights such that  $\mathrm{E}_{\mathcal{P}_{\omega}}[Y|\bar{Z}_T=\bar{z}_T] = \mathrm{E}_{\mathcal P}[Y(\bar{z}_T)]$ then 
we can nonparametrically identify the estimand as a particular case of \eqref{eqn::balAll}.
That is, if $\mathrm{E}_{\mathcal P}[Y| \bar{X}_T = \bar{x}_T, \bar{Z}_T = \bar{z}_T]$ is a function of the covariates and treatment paths $\bar{x}_T$ and $\bar{z}_T$, say $ \mathrm{E}_{\mathcal P}[Y| \bar{X}_T = \bar{x}_T, \bar{Z}_T = \bar{z}_T] = \varphi \{\bar{x}_T;\bar{z}_T\}$, a necessary and sufficient balance condition for identifying $\mathrm{E}_{\mathcal P}[Y(\bar{z}_T)]$ using $\mathrm{E}_{\mathcal{P}_{\omega}}[Y|\bar{Z}_T=\bar{z}_T]$ is that
$\varphi \{\bar{x}_T;\bar{z}_T\}$ is balanced to its mean potential value under $\bar{z}_T$,
\begin{eqnarray}\label{eqn::balY}
\int \varphi\{\bar{x}_T;\bar{z}_T\}\omega(\bar{x}_T;\bar{z}_T) f_{\mathcal P}(\bar{X}_T = \bar{x}_T| \bar{Z}_T = \bar{z}_T)d\bar{x}_T = \int \varphi\{\bar{x}_T;\bar{z}_T\}f_{\mathcal P}(\bar{X}_T(\bar{z}_{T-1}) =\bar{x}_T) d\bar{x}_T.
\end{eqnarray}
Condition \eqref{eqn::balY} highlights the 
essential role of balancing outcome-relevant covariate features. The task now is to specify the model for $\mathrm{E}_{\mathcal P}[Y| \bar{X}_T = \bar{x}_T, \bar{Z}_T = \bar{z}_T]$ and then solve for the corresponding weights $\omega(\bar{x}_T;\bar{z}_T)$ that achieve this balance.

%%%%%%%%%%%%%%%%%%
%%%%%%%%%%%%%%%%%%

\subsection{Sufficient balancing requirements under a model class}\label{sec::estimation}

To reconcile generalizability and feasibility, we assume an additive structure that accommodates effect heterogeneity, allowing covariates at any time $t = 1, \ldots, T$ to act as effect modifiers,
\begin{eqnarray}\label{eqn::model_3.2}
Y(\bar{z}_T) =\alpha_{0,\bar{z}_T} + \sum_{t = 1}^T \alpha_{t,\bar{z}_{T}}^\top g_t\{X_t(\bar{z}_{t-1})\} +h_T\{\bar{X}_T(\bar{z}_{T-1});\bar{z}_T\}+\varepsilon, 
\end{eqnarray}
where $\varepsilon$ is a random error with mean zero and finite variance, $g_t\{\cdot\}$ is an unknown functional form for $X_t(\bar{z}_{t-1})$, and $h_T\{\cdot\}$ is the true remainder term. 
The coefficients $\alpha_{t,\bar{z}_{T}}$ may vary across treatment paths $\bar{z}_{T} \in \bar{\mathcal Z}_{T}$, accommodating flexible effect modification. 
This model can be viewed as a linear approximation to a more general model $\mathrm{E}_{\mathcal P}[Y(\bar{z}_T)| \bar{X}_T(\bar{z}_{T-1}), \bar{Z}_T = \bar{z}_T] = \alpha_{0,\bar{z}_T} +  G\{\bar{X}_T(\bar{z}_{T-1});\bar{z}_T\} + h_T\{\bar{X}_T(\bar{z}_{T-1});\bar{z}_T\}$, where $G\{\bar{X}_T(\bar{z}_{T-1});\bar{z}_T\}$ admits an additive representation $\sum_{t = 1}^T\alpha_{t,\bar{z}_T}^\top g_t\{X_t(\bar{z}_{t-1})\}$. 

In the weighted outcome $\mathrm{E}_{\mathcal{P}_{\omega}}[Y|\bar{Z}_T = \bar{z}_T]$, the weights $\{\omega(\bar{x}_T;\bar{z}_T): \bar{x}_T \in \bar{\mathcal X}_T\}$ correspond to the overall population. 
Let $\mathrm {E}_{\mathcal{S}_{\omega}}[Y|\bar{Z}_T = \bar{z}_T] := \sum_{i  = 1}^n 1\{\bar{Z}_{iT} = \bar{z}_T\}\omega_i Y_i$ denote a weighted estimator on the sample $\mathcal{S}$ of size $n$.
Our goal is to find weights $\{\omega_i:\bar{Z}_{iT} = \bar{z}_T\}$ for each $i = 1, ..., n$ such that $\mathrm {E}_{\mathcal{S}_{\omega}}[Y|\bar{Z}_T = \bar{z}_T]$ is consistent and asymptotically normal for $\mathrm{E}_{\mathcal P}[Y(\bar{z}_T)]$.
Theorem \ref{thm::1} establishes this result under Claims \ref{con::bal}-\ref{con::wei}, which we substantiate in the propositions in Section \ref{sec::adjustments} and solve for with the optimization problem in Section \ref{sec::implementation} (see Figure \ref{fig::roadmap}).  

\begin{claim}[Covariate Balance] \label{con::bal}
The weighted sample means of the covariate functions $g$ and $h$, $\mathrm{E}_{\mathcal{S}_{\omega}}[g_t\{X_t\} \mid \bar{z}_T]$ and $\mathrm{E}_{\mathcal{S}_{\omega}}[h_T\{\bar{X}_T;\bar{Z}_T\} \mid \bar{z}_T]$, are $\sqrt{n}$-consistent and asymptotically normal estimators of their respective population potential covariate targets, $\mathrm{E}_{\mathcal{P}}[g_t\{X_t(\bar{z}_{t-1})\}]$ and $\mathrm{E}_{\mathcal{P}}[h_T\{\bar{X}_T(\bar{z}_{T-1});\bar{z}_T\}]$:
\begin{eqnarray*}
    \text{Claim \ref{prop1}:} && \hspace{-0.6cm} \sqrt{n}(\mathrm {E}_{\mathcal{S}_{\omega}}[g_t\{X_t\} | \bar{Z}_T = \bar{z}_T] - \mathrm{E}_{\mathcal P}[g_t\{X_t(\bar{z}_{t-1})\}]) \overset{d}{\rightarrow} \text{Normal}(0,V_t), ~t = 1,...,T,\\
    \text{Claim \ref{prop2}:}  && \hspace{-0.6cm} \sqrt{n}(\mathrm {E}_{\mathcal{S}_{\omega}}[h_T\{\bar{X}_T;\bar{Z}_T\}|\bar{Z}_T =\bar{z}_T] - \mathrm{E}_{\mathcal P}[h_T\{\bar{X}_T(\bar{z}_{T-1});\bar{z}_T\}]) \overset{d}{\rightarrow} \text{Normal}(0,V_0),
\end{eqnarray*} 
where the asymptotic variance matrices $V_t$, $t = 0,...,T$, are finite and positive definite.
\end{claim}

\begin{claim}[Weight Regularity] \label{con::wei}
The weights $\{\omega_i: \bar{Z}_{iT} = \bar{z}_T\}_{i = 1}^n$ are non-negative and sum to one. The maximum value is $o(\frac{1}{\sqrt n})$ and the effective sample size $\frac{(\sum_{i = 1}^n 1\{\bar{Z}_{iT} = \bar{z}_T\}\omega_i)^2}{\sum_{i = 1}^n 1\{\bar{Z}_{iT} = \bar{z}_T\}\omega_i^2}$ is $ O(n)$.
\end{claim}

Claim \ref{con::bal} is the central requirement: the weights must balance the features of the covariate distribution that are relevant to the outcome. 
Claim \ref{con::wei} ensures the weights are well-behaved, preventing any single observation from unduly influencing the estimate and ensuring the validity of a weighted central limit theorem.
In Propositions \ref{prop1} and \ref{prop2} of Section \ref{sec::adjustments}, we verify that our proposed weights satisfy Claims \ref{con::bal} and \ref{con::wei}, respectively.
We then directly solve for these conditions and those in Claim \ref{con::wei} with the optimization program presented in Section \ref{sec::implementation}.
We now state our main theoretical result.

\begin{theorem}\label{thm::1}
Suppose the potential outcome model described in Equation \eqref{eqn::model_3.2} holds, and assume the weights satisfy Claims \ref{con::bal} and \ref{con::wei}. 
Then, the weighted estimator $\mathrm{E}_{\mathcal{S}_{\omega}}[Y \mid \bar{Z}_T = \bar{z}_T]$ is consistent and asymptotically normal for $\mathrm{E}_{\mathcal{P}}[Y(\bar{z}_T)]$. 

\begin{proof}
Under outcome model \eqref{eqn::model_3.2}, $\mathrm{E}_{\mathcal P}[Y(\bar{z}_T)]$ and $\mathrm {E}_{\mathcal{S}_{\omega}}[Y|\bar{z}_T]$ can be written as

$\mathrm{E}_{\mathcal P}[Y(\bar{z}_T)] = \alpha_{0,\bar{z}_T}  + \sum_{t = 1}^T \alpha_{t,\bar{z}_{T}}^\top \mathrm{E}_{\mathcal P}[g_t\{X_t(\bar{z}_{t-1})\}] +  \mathrm{E}_{\mathcal P}[h_T\{\bar{X}_T(\bar{z}_{T-1});\bar{z}_T\}]+\mathrm{E}_{\mathcal P}[\varepsilon]$ and

$\mathrm {E}_{\mathcal{S}_{\omega}}[Y|\bar{z}_T]=  \alpha_{0,\bar{z}_T} \mathrm {E}_{\mathcal{S}_{\omega}}[1|\bar{z}_T]+ \sum_{t = 1}^T \alpha_{t,\bar{z}_{T}}^\top\mathrm {E}_{\mathcal{S}_{\omega}}[g_t\{X_t\} | \bar{z}_T] +\mathrm {E}_{\mathcal{S}_{\omega}}[h_T\{\bar{X}_T;\bar{Z}_T\}|\bar{z}_T]+\mathrm {E}_{\mathcal{S}_{\omega}}[\varepsilon|\bar{z}_T]$.

For the first term in $\mathrm {E}_{\mathcal{S}_{\omega}}[Y|\bar{z}_T]$, since $\sum_{i=1}^n 1\{\bar{Z}_{iT} = \bar{z}_T\} \omega_i = 1$, $\mathrm {E}_{\mathcal{S}_{\omega}}[1|\bar{z}_T] - 1=0$. 

For the fourth term, since $\frac{(\sum_{i = 1}^n1\{\bar{Z}_{iT} = \bar{z}_T\}\omega_i)^2}{\sum_{i=1}^n 1\{\bar{Z}_{iT} = \bar{z}_T\}\omega_i^2} = O(n)$ and $\max(\{1\{\bar{Z}_{iT} = \bar{z}_T\}\omega_i\}_{i = 1}^n) = o(\frac{1}{\sqrt n})$ (Claim \ref{con::wei}), 
by the Lindeberg's condition for the Central Limit Theorem, $\mathrm {E}_{\mathcal{S}_{\omega}}[\varepsilon|\bar{z}_T] = \sum_{i = 1}^n 1\{\bar{Z}_{iT} = \bar{z}_T\}\omega_i(\varepsilon_i -\mathrm{E}_{\mathcal P}[\varepsilon])+\mathrm{E}_{\mathcal P}[\varepsilon]$ is asymptotically Gaussian. 

For the second and third terms, by Claims \ref{prop1} and \ref{prop2}, $\mathrm {E}_{\mathcal{S}_{\omega}}[g_t\{X_t\}|\bar{z}_T] -\mathrm{E}_{\mathcal P}[g_t\{X_t(\bar{z}_{t-1})\}] = O_p(n^{-\frac 1 2})$, $\mathrm {E}_{\mathcal{S}_{\omega}}[h_T\{\bar{X}_T;\bar{Z}_T\}|\bar{z}_T] -\mathrm{E}_{\mathcal P}[h_T\{\bar{X}_T(\bar{z}_{t-1});\bar{z}_T\}] = O_p(n^{-\frac 1 2})$, and $\sum_{t = 1}^T \alpha_{t,\bar{z}_T}^\top \mathrm {E}_{\mathcal{S}_{\omega}}[g_t\{X_t\}|\bar{z}_T]+\mathrm {E}_{\mathcal{S}_{\omega}}[h_T\{\bar{X}_T;\bar{Z}_T\}|\bar{z}_T]$ is asymptotically Gaussian. 

Thus, $\mathrm {E}_{\mathcal{S}_{\omega}}[Y|\bar{z}_T] -\mathrm{E}_{\mathcal P}[Y(\bar{z}_T)]=O_p(n^{-\frac 1 2})$ and $\mathrm {E}_{\mathcal{S}_{\omega}}[Y|\bar{z}_T] $ is asymptotically Gaussian. 
\end{proof}
\end{theorem}

%%%%%%%%%%%%%%%%%%
%%%%%%%%%%%%%%%%%%

\subsection{Balancing procedure when covariates depend on ancestors}\label{sec::adjustments}

In order to balance $g_1\{X_1\},...,g_T\{X_T\}$, we need the following conditions to hold,\begin{equation}\label{eqn::balcon_3.2}
   \mathrm{E}_{\mathcal{P}_{\omega}}[g_t\{X_t\} | \bar{Z}_T = \bar{z}_T] = \mathrm{E}_{\mathcal P}[g_t\{X_t(\bar{z}_{t-1})\}], ~t = 1,...,T.
\end{equation}
These conditions have mean potential covariates as targets, which are unknown.
To our knowledge, existing balancing approaches implement more restrictive, sequential mean independence conditions, without reference to targets on potential covariates, 
\begin{eqnarray}
\mathrm{E}_{\mathcal{P}_{\omega}}[g_1\{X_1\} | \bar{z}_T] = \mathrm{E}_{\mathcal P}[g_1\{X_1\}]; ~ \mathrm{E}_{\mathcal{P}_{\omega}}[g_t\{X_t\} | \bar{z}_T] = \mathrm{E}_{\mathcal{P}_{\omega}}[g_t\{X_t\} | \bar{Z}_{t-1} = \bar{z}_{t-1}], ~t = 2,...,T. \label{eqn::balconred_3.2}
\end{eqnarray}
The conditions in \eqref{eqn::balconred_3.2} requires that, after weighting, covariates are mean independent of future treatments. This condition, however, is not sufficient for \eqref{eqn::balcon_3.2}. The failure arises because it does not correctly account for the internal dependence structure of the covariates. To see this, consider an example of $T = 2$ and suppose, without loss of generality, that $Z_1$ does not affect $X_2$, i.e., $\mathrm{E}_{\mathcal P}[g_2\{X_2(z_1)\}] = \mathrm{E}_{\mathcal P}[g_2\{X_2(z_1')\}]$ for $z_1\neq z_1'$. This example shows that \eqref{eqn::balconred_3.2} alone does not guarantee $\mathrm{E}_{\mathcal{P}_{\omega}}[g_2\{X_2\} | z_1]=\mathrm{E}_{\mathcal{P}_{\omega}}[g_2\{X_2\} | z_1']$.

By definition, $\mathrm{E}_{\mathcal{P}_{\omega}}[g_2\{X_2\} |z_1]  =\int 1\{Z_1 = z_1\}\omega g_2\{X_2\}f_{\mathcal P}(X_2|X_1,z_1)f_{\mathcal P}(X_1|z_1)d\bar{X}_2$. Let $\varphi_2\{x_1;z_1\} := \mathrm{E}_{\mathcal P}[g_2\{X_2\}|x_1,z_1] = \int g_2\{X_2\}f_{\mathcal P}(X_2|x_1,z_1) dX_2$ and let $R_2$ denote the residual $R_2 :=g_2\{X_2\}-\varphi_2\{X_1;Z_1\}$. Although \eqref{eqn::balconred_3.2} balances $g_1\{X_1\}$ and $g_2\{X_2\}$ such that $\mathrm{E}_{\mathcal{P}_{\omega}}[g_1\{X_1\} | \bar{z}_2]=\mathrm{E}_{\mathcal P}[g_1\{X_1\}]$ and $\mathrm{E}_{\mathcal{P}_{\omega}}[g_2\{X_2\} | \bar{z}_2]=\mathrm{E}_{\mathcal{P}_{\omega}}[g_2\{X_2\}|z_1]$, the two components $\varphi_2\{X_1;Z_1\}$ and $R_2$ are not guaranteed to be balanced between the two groups $\{i: Z_{i1} = z_1\}$ and $\{i: Z_{i1} = z_1'\}$. Consequently, 
there exist cases where $\mathrm{E}_{\mathcal{P}_{\omega}}[g_2\{X_2\} | z_1]\neq \mathrm{E}_{\mathcal{P}_{\omega}}[g_2\{X_2\} | z_1']$ and thus \eqref{eqn::balconred_3.2} does not imply \eqref{eqn::balcon_3.2}. 

Now, consider under what conditions \eqref{eqn::balconred_3.2} implies \eqref{eqn::balcon_3.2}. If we additionally assume $X_2\ci X_1|Z_1$, together with the ignorability assumption $X_2(z_1)\ci Z_1|X_1$, then we have $X_2(z_1)\ci (Z_1,X_1)$. In this special case, the balance target of $g_2\{X_2\}$  is identifiable from the sample, since the target becomes the conditional mean $\mathrm{E}_{\mathcal P}[g_2\{X_2(z_1)\}] = \mathrm{E}_{\mathcal P}[g_2\{X_2\}|z_1]$. 

In our balancing procedure to achieve \eqref{eqn::balcon_3.2}, we do not assume $X_2\ci X_1|Z_1$ but instead assume the mean independence of $R_2$ and $X_1$ given $Z_1$. In Proposition \ref{prop1}, we show that our balance procedure uses the structures in the time-varying covariates to decompose $g_t\{X_t\}$ into $\varphi_t\{\bar{X}_{t-1};\bar{Z}_{t-1}\}:=\mathrm{E}_{\mathcal P}[g_t\{X_t\}|\bar{X}_{t-1},\bar{Z}_{t-1}]$ and $R_t:= g_t\{X_t\}-\mathrm{E}_{\mathcal P}[g_t\{X_t\}|\bar{X}_{t-1},\bar{Z}_{t-1}]$. 

If $g_t\{X_t\}$ is continuous and $\varphi_t\{\bar{X}_{t-1};\bar{Z}_{t-1}\}$ is linear such that for some coefficients $\beta_{X_t\sim \bar{X}_{t-1}|z_{t-1}}$, $\varphi_t\{\bar{X}_{t-1};\bar{Z}_{t-1}\} = \sum_{\bar{z}_{t-1}\in \bar{\mathcal Z}_{t-1}} 1\{\bar{Z}_{t-1} = \bar{z}_{t-1}\} \beta_{X_t\sim \bar{X}_{t-1}|z_{t-1}}\bar{g}_{t-1}\{X_{t-1}(\bar{z}_{t-2})\}$, where the history $\bar{g}_{t-1}\{X_{t-1}(\bar{z}_{t-2})\} := (g_1\{X_1\},,...,{g}_{t-1}\{X_{t-1}(\bar{z}_{t-2})\})$, then balancing $\varphi_t\{\bar{X}_{t-1};\bar{Z}_{t-1}\}$ only depends on balancing $\bar{g}_{t-1}\{X_{t-1}\}$ without additional balance requirements. Then $R_t(\bar{z}_{t-1}) = g_t\{X_t(\bar{z}_{t-1})\} - \beta_{X_t\sim \bar{X}_{t-1}|z_{t-1}}\bar{g}_{t-1}\{X_{t-1}(\bar{z}_{t-2})\}$ can be interpreted as the residual component of projecting $g_t\{X_t(\bar{z}_{t-1})\}$ onto its history $\bar{g}_{t-1}\{X_{t-1}(\bar{z}_{t-2})\}$, within the strata $Z_{t-1} = z_{t-1}$.
If $g_t\{X_t\}$ is binary, then $\varphi_t\{\bar{X}_{t-1};\bar{Z}_{t-1}\}=\sum_{\bar{z}_{t-1}\in \bar{\mathcal Z}_{t-1}}\mathrm{P}(g_t\{X_t\} = 1|\bar{X}_{t-1},\bar{z}_{t-1};\beta_{X_t\sim \bar{X}_{t-1}|z_{t-1}})$.
The following theorem states that if we can find weights that balance the baseline covariates $g_1\{X_1\}$ and the residuals $\{R_t\}_{t = 2}^T$ to their targets, then 
Claim \ref{prop1} holds. 

\setcounter{subtheorem}{0}
\begin{subtheorem}~\label{prop1}
\begin{enumerate}[label=(\Roman*)]
    \item Population balance (identification). 
    Suppose the following mean independence condition holds 
\begin{eqnarray}\label{eqn::v}
\mathrm{E}_{\mathcal P}[\mathrm{E}_{\mathcal P}[\cdots\mathrm{E}_{\mathcal P}[\mathrm{E}_{\mathcal P}[R_t|\bar{X}_{t-1}, \bar{z}_{t-1}]|\bar{X}_{t-2}, \bar{z}_{t-2}]\cdots|X_1,z_1]] = \mathrm{E}_{\mathcal P}[R_t|\bar{z}_{t-1}], ~t = 2,..., T, ~~
\end{eqnarray}
where  $R_t(\bar{z}_{t-1}) := g_t\{X_t(\bar{z}_{t-1})\} - {\beta}_{X_t\sim \bar{X}_{t-1}|z_{t-1}}\bar{g}_{t-1}\{X_{t-1}(\bar{z}_{t-2})\}$ are the residuals at time $t$. If the weights $\omega$ satisfy the following population balance condition
\begin{eqnarray}
\Big\{\mathrm{E}_{\mathcal{P}_{\omega}}[g_1\{X_1\}|\bar{z}_T] = \mathrm{E}_{\mathcal P}[g_1\{X_1\}]; ~\mathrm{E}_{\mathcal{P}_{\omega}}[R_t|\bar{z}_T] =  \mathrm{E}_{\mathcal P}[R_t|\bar{z}_{t-1}], ~ t = 2,...,T\Big\}, \label{eqn::balconimp}
\end{eqnarray}
then the balance condition in Equation \eqref{eqn::balcon_3.2} holds.  

\item Sample balance (estimation). Suppose the variance-covariance matrix of $g_t\{X_t\}$ is finite and non-singular for each $t = 1,...,T$, and the estimator $\widehat{\beta}_{X_t\sim \bar{X}_{t-1}|z_{t-1}}$ is $\sqrt{n}$-consistent for ${\beta}_{X_t\sim \bar{X}_{t-1}|z_{t-1}}$, $t = 2,...,T$; then the sample balance condition 
\begin{eqnarray}
&& \hspace{-1cm}\Big\{\mathrm{E}_{\mathcal{S}_{\omega}}[g_1\{X_1\}|\bar{z}_T] -  \mathrm{E}_{\mathcal S}[g_1\{X_1\}] = o(n^{-\frac{1}{2}}); \mathrm{E}_{\mathcal{S}_{\omega}}[\widehat{R}_t| \bar{z}_T] -  \mathrm{E}_{\mathcal S}[\widehat{R}_t|\bar{z}_{t-1}]= o(n^{-\frac{1}{2}}), t = 2,...,T\Big\}\qquad
\label{eqn::balconsam}
\end{eqnarray}
implies Claim \ref{prop1}, where $\widehat{R}_t(\bar{z}_{t-1}) := g_t\{X_t(\bar{z}_{t-1})\} - \widehat{\beta}_{X_t\sim \bar{X}_{t-1}|z_{t-1}}\bar{g}_{t-1}\{X_{t-1}(\bar{z}_{t-2})\}$. 
\end{enumerate}
\begin{proof}
The proof is in Section S1.2 in the Supplementary Materials.
\end{proof}
\end{subtheorem}

\begin{remark}\label{rem::condition}
In Proposition \ref{prop1}, we achieve  balance for $\bar{g}_T\{X_T\}$ by balancing the residuals of their projections. As shown in Supplementary Materials Section S1.2, this procedure is equivalent to balancing the observed covariates to the imputed estimates of their potential values. 
This procedure ensures the asymptotic normality stated in Claim \ref{prop1}, provided that $g_t\{X_t(\bar{z}_{t-1})\}$ satisfies condition \eqref{eqn::v}.  
If, however, condition \eqref{eqn::v} is not met and the balance target $\mathrm{E}_{\mathcal P}[\bar{g}_T\{X_T(\bar{z}_{T-1})\}]$ is achieved only approximately (that is, with some bias) the conditions in \eqref{eqn::balconsam} nevertheless guarantee that each covariate is mean independent of future treatments, in the sense that $\mathrm{E}_{\mathcal{S}_{\omega}}[g_t\{X_t\} |\bar{z}_T] - \mathrm{E}_{\mathcal{S}_{\omega}}[g_t\{X_t\} | \bar{z}_{t-1}] = o(n^{-1/2})$.  
Moreover, the flexibility afforded by approximate balance, as discussed in Proposition \ref{prop1}(II), allows for effective management of the bias-variance trade-off. 
\end{remark}

In particular, we solve the following convex optimization problem and call this approach balancing by adaptive orthogonalization (BAO), 
%Let $\{\omega^{\text{bao}}_i: \bar{Z}_{iT} = \bar{z}_T\}_{i = 1}^n$ denote the solution to,
\begin{eqnarray}
\Big\{\hspace{-0.5cm} && \hspace{-0.25cm} \underset{\omega}{\text{minimize}}\sum_{i: \bar{Z}_{iT} =\bar{z}_{T}} \omega_i^2 : \big|\mathrm{E}_{\mathcal{S}_{\omega}}[g_1\{X_1\}|\bar{z}_T] -  \mathrm{E}_{\mathcal{S}}[g_1\{X_1\}]\big|\preceq \delta_{1}, ~\big|\mathrm{E}_{\mathcal{S}_{\omega}}[\widehat{R}_t|\bar{z}_T] -  \mathrm{E}_{\mathcal{S}}[\widehat{R}_t|\bar{z}_{t-1}]\big| \preceq \delta_{t},  \nonumber \\
&& \hspace{-0.25cm} t=2,...,T, \text{ where the tolerances } \delta_1, ..., \delta_T = o(n^{-\frac{1}{2}}) \text{ are vectors of non-negative values}\Big\}. \qquad  
\label{eqn::opt0}
\end{eqnarray}

In Proposition \ref{prop2}, we will show that if the true propensity score has a well-behaved tail near 0, and the weights from \eqref{eqn::opt0} converge in probability to the true propensity scores, then $h_T\{\bar{X}_T;\bar{Z}_T\}$ is also balanced relative to its target $\mathrm{E}_{\mathcal P}[h_T\{\bar{X}_T(\bar{z}_{T-1});\bar{z}_T\}]$. 
The form of the objective function in \eqref{eqn::opt0} will be discussed in Section \ref{sec::implementation}.

\begin{assumption}[Tail behavior of $\pi(\bar{X}_{T};\bar{z}_T)$ at 0; \citealt{ma2020robust}]\label{thin} 
% ~\\
% a. $\pi(\bar{X}_{T};\bar{z}_T)$ has a regular varying tail with index $\gamma_0-1 >0$ at 0: $\lim_{q\downarrow 0}\frac{\mathrm{P}({\pi}(\bar{X}_{T};\bar{z}_T)\leq qs)}{\mathrm{P}({\pi}(\bar{X}_{T};\bar{z}_T)\leq q)} =s^{\gamma_0-1}$ for all $s>0$.

% b. The regular varying tail in a is relatively thin, $\gamma_0\geq 2$.

The cumulative distribution function of $\pi(\bar{X}_{T};\bar{z}_T)$ is regularly varying at 0. Specifically, there exists a constant $\gamma_0\geq 2$ such that for all $s>0$: $\lim_{q\downarrow 0}\frac{\mathrm{P}({\pi}(\bar{X}_{T};\bar{z}_T)\leq qs)}{\mathrm{P}({\pi}(\bar{X}_{T};\bar{z}_T)\leq q)} =s^{\gamma_0-1}$. 
\end{assumption}

\begin{assumption}[Distribution of $h_T\{\bar{X}_T;\bar{Z}_T\}$; Assumption 2, \citealt{ma2020robust}]\label{bounded}
% ~\\
For some $\epsilon>0$, $\mathrm{E}_{\mathcal P}\big[|h_T\{\bar{X}_T;\bar{Z}_T\}|^{\gamma_0 \vee 2+\epsilon}\big|{\pi}(\bar{X}_{T};\bar{z}_T) = q, \bar{Z}_T = \bar{z}_T\big]$ is uniformly bounded, where the conditional distribution of $h_T\{\bar{X}_T;\bar{Z}_T\}$ given ${\pi}(\bar{X}_{T};\bar{z}_T) = q$ and $\bar{Z}_T = \bar{z}_T$ converges to an existing distribution as $q\downarrow 0$.
\end{assumption}

\begin{assumption}\label{linear}
There exists a vector $\beta_{h\sim g}$ such that $\|{h}_T\{\bar{\boldsymbol X}_T;\bar{z}_T\}-\bar{g}_T\{\boldsymbol{X}_T(\bar{z}_{T-1})\}\beta_{h\sim g}\|_2 = O_p(1)$, where $\boldsymbol{X}_t, {h}_T\{\bar{\boldsymbol X}_T;\bar{z}_T\}, \bar{g}_T\{\boldsymbol{X}_T\}$ are data matrices with $n$ rows.
\end{assumption}

\begin{subtheorem} \label{prop2}
Under Assumptions \ref{thin}, \ref{bounded}, and \ref{linear}, as well as the regularity conditions in Assumption S1 (Supplementary Materials), the weights derived from \eqref{eqn::opt0} satisfy Claim \ref{prop2}.
\begin{proof}
The proof consists of the following parts, with detailed arguments provided in Section S1.3 in the Supplementary Materials.
\begin{enumerate}[label=(\Roman*)]
    \item Under Assumptions \ref{thin} and \ref{bounded}, $\mathrm{E}_{\mathcal{S}_{\pi^{-1}}}[h_T\{\bar{X}_T;\bar{Z}_T\}|\bar{z}_T]$ is $\sqrt{n}$-consistent for $\mathrm{E}_{\mathcal P}[h_T\{\bar{X}_T(\bar{z}_{T-1});\bar{z}_T\}]$ and is asymptotically Gaussian.
    
    \item Under the regularity conditions extended from Assumption 1 in \citet{wang2020minimal} and Assumptions 2 and 4 in \citet{chattopadhyay2023onestep} and listed in the Supplementary Materials, the weights converge to the vector of true inverse propensity scores, $n^{-1/2}\|{1}\{\bar{\boldsymbol Z}_{T} = \bar{z}_T\} n\omega^{\text{bao}} - {1}\{\bar{\boldsymbol Z}_{T} = \bar{z}_T\}{\pi}^{-1}(\bar{\boldsymbol X}_{T};\bar{z}_T)\|_2 = o_p(1)$, where $\bar{\boldsymbol Z}_T$ and $\bar{\boldsymbol X}_T$ are matrices with $n$ rows, ${1}\{\bar{\boldsymbol Z}_{T} = \bar{z}_T\}\omega^{\text{bao}}$ is a vector of length $n$ solved from \eqref{eqn::opt0}.

    \item With (I) and (II), $\mathrm{E}_{\mathcal{S}_{\omega^{\text{bao}}}}[h_T\{\bar{X}_T;\bar{z}_T\}|\bar{z}_T]$ is consistent for $\mathrm{E}_{\mathcal P}[h_T\{\bar{X}_T(\bar{z}_{T-1});\bar{z}_T\}]$. 
    Moreover, under Assumption \ref{linear}, $\mathrm{E}_{\mathcal{S}_{\omega^{\text{bao}}}}[h_T\{\bar{X}_T;\bar{z}_T\}|\bar{z}_T]$ is asymptotically Gaussian.    
\end{enumerate}
\vspace{-1.05cm}
\end{proof}
\end{subtheorem}

\subsection{Implementation through constrained convex optimizations}\label{sec::implementation}

Thus far, we have established balance conditions for the weighted estimator $\mathrm{E}_{\mathcal{S}_{\omega}}[Y|\bar{z}_T]$ and highlighted the importance of constraining the weights to be non-negative and normalized. Beyond ensuring consistency and asymptotic normality, we also seek to obtain optimal weights that minimize specific loss functions in finite samples. This consideration motivates the objective function in \eqref{eqn::opt0}.

Under outcome model \eqref{eqn::model_3.2}, the estimation error of the weighted estimator is given by
\begin{eqnarray*}
\Delta  &:=& \mathrm{E}_{\mathcal{S}_{\omega}}[Y| \bar{z}_T] - \mathrm{E}_{\mathcal P}[Y(\bar{z}_T)]\\
&=& \underbrace{\sum_{t = 1}^T \alpha_{t,\bar{z}_T}^\top \Big\{\mathrm{E}_{\mathcal{S}_\omega}[g_t\{X_{t}\}|\bar{z}_T] - \mathrm{E}_{\mathcal P}[g_t\{X_{t}(\bar{z}_{t-1})\}]\Big\}}_{\Delta_g}+\\
&& \underbrace{\mathrm{E}_{\mathcal{S}_\omega}[h_T\{\bar{X}_{T};\bar{z}_T\}|\bar{z}_T] - \mathrm{E}_{\mathcal P}[h_T\{\bar{X}_{T}(\bar{z}_{T-1});\bar{z}_T\}]}_{\Delta_h} + \underbrace{\mathrm{E}_{\mathcal{S}_\omega}[\varepsilon|\bar{z}_T]-\mathrm{E}_{\mathcal P}[\varepsilon]}_{\Delta_{\varepsilon}},
\end{eqnarray*}
where the random error $\varepsilon$ has mean 0 and finite variance. The mean squared error loss can be decomposed as
$\mathrm{E}_{\mathcal P}[\Delta^2] = \mathrm{E}_{\mathcal P}[(\Delta_g+\Delta_h)^2] + \mathrm{E}_{\mathcal P}[\Delta_{\varepsilon}^2]+\mathrm{E}_{\mathcal P}[2(\Delta_g+\Delta_h)\Delta_{\varepsilon}] $.
If the random error $\varepsilon_i$ is i.i.d. with variance $\sigma^2$, then
$\mathrm{E}_{\mathcal P}[\Delta_{\varepsilon}^2] = \sum_{i = 1}^n 1\{\bar{Z}_{iT} = \bar{z}_T\}\omega^2_i\mathrm{E}_{\mathcal P}[\varepsilon^2_i]= \sum_{i = 1}^n 1\{\bar{Z}_{iT} = \bar{z}_T\}\omega^2_i\sigma^2$.
This motivates us to minimize the $\ell_2$-norm (or the variance) of the weights subject to imposing constraints on the covariate imbalances. 

Hence, as previewed in \eqref{eqn::opt0}, the optimal weights for treatment path $\bar{z}_T$ are given by 
\begin{eqnarray}
\bigg\{\underset{\omega}{\text{minimize}} \sum_{i \in \mathcal{I}_{\bar{z}_{T}}} \omega_i^2 :  \bigg|\sum\limits_{i\in \mathcal{I}_ {\bar{z}_{T}}} \omega_{i}\widehat{R}_{it}-
\sum\limits_{i\in \mathcal{I}_{\bar{z}_{t-1}} }\frac{1}{\left\vert{ \mathcal{I}_{\bar{z}_{t-1}}}\right\vert}\widehat{R}_{it}
\bigg| \preceq \delta_{t}, ~ t=1,...,T; \sum\limits_{i \in \mathcal{I}_{\bar{z}_{T}}} \omega_{i}  =  1, ~ \omega_i \geq  0\bigg\}, \label{eqn::opt1}
\end{eqnarray}  
where $\widehat{R}_{i1} := g_1\{X_{i1}\}$, $\widehat{R}_{it} := g_t\{X_{it}\} - \widehat{\beta}_{X_{t}\sim \bar{X}_{t-1}|z_{t-1}}\bar{g}_t\{{X}_{i,t-1}\}$, $\mathcal{I}_{\bar{z}_0} := \mathcal{I}$, $\mathcal{I}_{\bar{z}_t} := \{i \in \mathcal{I} : \bar{Z}_{it}=\bar{z}_t\}$ and $\delta_{t}	\succeq \textbf{0}$ is a vector of imbalance tolerances determined by the investigators.

To obtain weights for all treatment paths, we can either solve \eqref{eqn::opt1} separately for each path, or collectively solve \eqref{eqn::opt} for all paths together by weighting them according to their sample proportions $\mathrm{P}_{\mathcal S}$,
\begin{eqnarray}
\bigg\{\underset{\omega}{\text{minimize}} \sum_{\bar{z}_T\in \mathcal{\bar{Z}}_{T}}\sum_{i \in \mathcal{I}_{\bar{z}_T}} [\mathrm{P}_{\mathcal S}(\bar{z}_{T})\omega_i]^2 :  &&\bigg|\sum\limits_{i\in \mathcal{I}_{\bar{z}_{T}}} \omega_{i}\widehat{R}_{it} 
-
\sum\limits_{i\in \mathcal{I}_{\bar{z}_{t-1}} }\frac{1}{\left\vert{ \mathcal{I}_{\bar{z}_{t-1}}}\right\vert}\widehat{R}_{it}
\bigg| \preceq\delta_{t,\bar{z}_{T}}, ~t=1,...,T, \nonumber \\
&&\sum\limits_{i \in \mathcal{I}_{\bar{z}_{T}}} \omega_{i}  =  1, ~\forall \bar{z}_{T} \in \mathcal{\bar{Z}}_{T}; ~ \omega_i \geq  0, ~ i \in \mathcal{I} \bigg\}. \label{eqn::opt}
\end{eqnarray}  

In programs \eqref{eqn::opt1} and \eqref{eqn::opt}, we minimize the $\ell_2$-norm to obtain the most stable set of weights for a specified level of covariate balance.  
Non-negativity constraints ensure that these adjustments are made by interpolating within the observed data, rather than by extrapolating beyond the range of support.  
When the optimization problem is infeasible, adjusting the tolerance parameter $\delta$ yields valuable diagnostic information—namely, whether adequate overlap exists or if some degree of extrapolation is necessary to achieve balance.  
A similar perspective is discussed by \citet{mattei2015discussion}. See also \citet{platt2012positivity} and \citet{fogarty2016discrete} for related considerations.

%%%%%%%%%%%%%%%%%%%%
%%%%%%%%%%%%%%%%%%%%
%%%%%%%%%%%%%%%%%%%%
%\vspace{-.25cm}

% \newpage
\section{Implementation algorithm and practical guidelines}\label{Practical}

In Section \ref{Methods}, we discussed the weights optimization programs \eqref{eqn::opt1}-\eqref{eqn::opt}. Here, we present the implementation procedure in Algorithm \ref{algorithm1} and highlight key practical considerations.

\begin{algorithm}
\caption{Balancing by Adaptive Orthogonalization (BAO)}
\label{algorithm1}
\begin{spacing}{1.}
\begin{algorithmic}[0]
\State \textbf{Step 1.} Specify the covariate adjustment functions $g_t\{X_t\}$ for each $t = 1, \ldots, T$.
\State \textbf{Step 2.} Determine the optimal tolerance $\delta^*$ by tuning over a candidate set $\mathcal{D}$. The tuning procedure is adapted from Algorithm 1 of \cite{wang2020minimal}.
\State \textbf{Step 3.} Solve the optimization programs \eqref{eqn::opt1} or \eqref{eqn::opt} using $\delta^*$ to obtain the weights $\omega^{\text{BAO}}$.
\State \textbf{Step 4.} Assess covariate balance with $\omega^{\text{BAO}}$. If additional adjustments are needed, refine $g_t\{X_t\}$ for $t = 1, \ldots, T$ and return to Step 2; otherwise, proceed to Step 5.
\State \textbf{Step 5.} Estimate the mean potential outcomes or the parameters in marginal structural models (MSMs) using $\omega^{\text{BAO}}$.
\State \textbf{Step 6.} Compute standard errors of the point estimates by bootstrapping the sample and resolving the optimization problem for each resampled dataset with the same tolerance $\delta^*$.
\end{algorithmic}
\end{spacing}
\end{algorithm}

\subsection{Functional forms for adjustment}

The number of balancing constraints, $K = 2^T P_{\bar{g}_T}$, increases exponentially with the number of treatment periods ($T$), which can produce infeasible or unstable weights. Consequently, a parsimonious specification of covariate functions is essential. A practical approach is to start with linear forms for $g_t(X_t)$ and use the diagnostic step (Step 4) to assess whether higher-order imbalances remain. This iterative process is generally adequate as it does not require the outcomes and the estimator is robust to modest misspecification of these functions.

\subsection{Tuning algorithm for approximate balance}

The tolerance parameters $\{\delta_t\}$ control the bias-variance trade-off. We adapt the bootstrap-based tuning algorithm proposed by \citet{wang2020minimal}, which selects the value of $\delta$ that minimizes average covariate imbalance across bootstrap samples. This data-driven procedure automates the selection of the degree of approximate balance.

%%%%%%%%%%%%%%%%%%%%
%%%%%%%%%%%%%%%%%%%%
\subsection{Extension for censored data}
In settings with right-censoring, let $C_t = 1$ indicate that an individual is lost to follow-up at time $t$. We assume that censoring is sequentially ignorable given the history of past treatments and covariates. To address censoring, we modify the balancing procedure so that it is applied only to the subset of individuals who remain uncensored at each relevant time point. Specifically, the residual balancing conditions from
\eqref{eqn::balconsam} are extended to:
\begin{eqnarray}
&& \hspace{-0.75cm} \Big\{\mathrm{E}_{\mathcal{S}_{\omega}}[g_1\{X_1\}|\bar{z}_T,C_T = 0] =  \mathrm{E}_{\mathcal{S}}[g_1\{X_1\}]; ~ \mathrm{E}_{\mathcal{S}_{\omega}}[\widehat{R}_t(\bar{z}_{t-1})|\bar{z}_T,C_T = 0] =  \mathrm{E}_{\mathcal{S}}[\widehat{R}_t(\bar{z}_{t-1})|\bar{z}_{t-1},C_{t-1} = 0]; \nonumber\\
&& \hspace{-0.75cm}~\text{ where } \widehat{R}_t(\bar{z}_{t-1}) := g_t\{X_t(\bar{z}_{t-1})\} - \widehat{\beta}_{X_t\sim \bar{X}_{t-1}|z_{t-1},c_{t-1} = 0}\bar{g}_{t-1}\{X_{t-1}(\bar{z}_{t-2})\}, t = 2,...,T\Big\}, \label{eqn::balconsam_cen}
\end{eqnarray}
where the residuals $\widehat{R}_t$ are estimated from regressions fitted exclusively on the uncensored population at time $t-1$. This approach ensures that the final weights are constructed for the complete-case sample and, under the sequential ignorability assumption, yield results that generalize to the full population (proof is in the Supplementary Materials, Section S2).

%%%%%%%%%%%%%%%%%%%%
%%%%%%%%%%%%%%%%%%%%
\subsection{Specification and fitting of MSMs}

After estimating the mean potential outcomes $\mathrm{E}_{\mathcal{P}}[Y(\bar{z}_T)]$ for all treatment paths $\bar{z}_T \in \bar{\mathcal{Z}}_T$, it is often useful to summarize the treatment effects using a parsimonious MSM, denoted $\text{msm}(\bar{z}_T; \boldsymbol{\tau})$, where $\boldsymbol{\tau}$ represents a small set of interpretable parameters. Common specifications include the additive model, $\text{msm}(\bar{z}_T; \boldsymbol{\tau}) = \tau_0 + \sum_{t=1}^T \tau_t z_t$, and the cumulative model, $\text{msm}(\bar{z}_T; \boldsymbol{\tau}) = \tau_0 + \tau_1 \sum_{t=1}^T z_t$. 

When fitting an MSM, estimates for treatment paths with few observations may be imprecise and disproportionately influential. To address this, we recommend using weighted least squares, assigning each estimated mean potential outcome a weight proportional to its sample prevalence along the corresponding path, or alternatively adopting robust regression techniques to mitigate the impact of unstable estimates.

%%%%%%%%%%%%%%%%%%%%
%%%%%%%%%%%%%%%%%%%%
%%%%%%%%%%%%%%%%%%%%
\section{Simulation studies}\label{Empirical}

In three simulation studies, we assess the performance of the proposed method alongside a variety of widely used approaches. Each method is evaluated in terms of bias, root mean squared error (RMSE), and the coverage and length of the 95\% confidence intervals.

%%%%%%%%%%%%%%%%%%%%
%%%%%%%%%%%%%%%%%%%%
\subsection{Methods}\label{sec::methodsCompare}
\begin{itemize}
    \item Unadjusted: IPW without covariate adjustment. 
	\item IPW via logistic regression (LR):
		We consider three weighting schemes: standard LR weights, $\{\prod_{t = 1}^T\frac{1}{\mathrm{P}(Z_{it} | Z_{i,t-1}, X_{it})}\}_{i\in \mathcal I}$;
		stabilized LR weights, $\{\prod_{t = 1}^T\frac{\mathrm{P}(Z_{it} | \bar{Z}_{i,t-1})}{\mathrm{P}(Z_{it} | Z_{i,t-1}, X_{it})}\}_{i\in \mathcal I}$;
		and truncated LR weights, where the stabilized weights are truncated at the 95\textsuperscript{th} percentile, $\omega_i = \min\{\omega_i, q_{0.95}(\{\omega_i\}_{i\in \mathcal I})\}$. All propensity score models are correctly specified. 
	\item IPW via super learning (SL) and ensemble learning (EL): Implemented with the  \texttt{SuperLearner} package (v2.0-28) in   \texttt{R}, using base learners (\texttt{glm}, \texttt{knn}, \texttt{nnet}, \texttt{gbm}) as in \cite{gruber2015ensemble}. 
		We consider standard, stabilized, and truncated weighting schemes, analogous to LR. The correct propensity score model (logit linear) is included among the base learners. Variances are computed via the robust sandwich estimator.
	\item Covariate balancing propensity score (CBPS): Estimated using the \texttt{CBPS} package (v0.4), which outperformed latter versions in our replications. Two specifications are considered: exact (CBPSe) and overidentified (CBPSo), with all models correctly specified. Variances are estimated by the robust sandwich estimator.
    \item Residual balancing weighting (RBW): Computed using the \texttt{rbw} package (v0.3.2). Residuals of covariates are estimated through OLS regression. Variances are assessed via 100 bootstrap samples.
	\item Iterative conditional expectation (g-computation formula): Uses OLS regression in each iteration, stratifying by treatment paths (g stra.) or pooling over all paths (g pool.). Variances are estimated from 1,000 bootstrap samples.
	\item Longitudinal targeted maximum likelihood estimation (LTMLE): As implemented in the \texttt{ltmle} package \citep{ltmle2017}. Propensity score models are correctly specified; outcome models are linear. Variances are estimated by bootstrapping (100 samples). 
    \item Balancing by adaptive orthogonalization (BAO): Implemented with the \texttt{sbw} package. The standardized tolerance parameter \(\delta\) is selected from $\{0.001, 0.01, 0.05\}$, as described in Section \ref{Practical}. For all covariates, \(\beta_{X_t \sim \bar{X}_{t-1}|z_{t-1}}\) is estimated by OLS. Variances are computed from 100 bootstrap samples.
\end{itemize}

%%%%%%%%%%%%%%%%%%%%
%%%%%%%%%%%%%%%%%%%%
\subsection{Studies}

We set $T = 2$ in Studies 1 and 3, and $T = 3$ in Study 2. Figure S1 in the Supplementary Materials illustrates the relationships among variables in all three studies. For IPW and CBPS, we fit the true MSMs using the corresponding weights. For the g-computation formula, LTMLE, and BAO, we first estimate mean potential outcomes by stratification and then fit the MSMs to these estimates.

%%%%%%%%%%%%%%%%%%%%
\subsubsection{Study 1: correctly specified models}

Study 1 adapts a design from \citet{imai2015robust}, considering correct specification of both the treatment and outcome mechanisms.  
The study has a four-dimensional continuous, time-varying covariate ($X_t$), a binary treatment ($Z_t$), and a continuous outcome ($Y$), measured over $T=2$ time points.  
Specifically, for each individual $i = 1, \ldots, n$ and time $t = 1, 2$, covariates are generated as $X_{it} = (U_{it} X^0_{it1},\ U_{it} X^0_{it2},\ |U_{it} X^0_{it3}|,\ |U_{it} X^0_{it4}|)$, where $X^0_{itp} \overset{\text{i.i.d}}{\sim} \text{Normal}(0, 1)$ for $p = 1, \ldots, 4$, $U_{i1} = 1$, and $U_{i2} = (2 Z_{i1} + 5)/3$.  
Treatments are assigned as $Z_{it} \sim \text{Bernoulli}(\text{expit}\{-Z_{i, t-1} + X_{it1} - 0.5 X_{it2} + 0.25 X_{it3} + 0.1 X_{it4} + (-0.5)^t\})$, with $Z_{i0} = 0$.  
The outcome is generated as  
$
Y_i = 250 - 10 \sum_{t=1}^{2} Z_{it} + \sum_{t=1}^{2} (27.4 X_{it1} + 13.7 X_{it2} + 13.7 X_{it3} + 13.7 X_{it4}) + \varepsilon_i,
$
where $\varepsilon_i \overset{\text{i.i.d.}}{\sim} \text{Normal}(0, 25)$. 
The true MSM is given by $\mathrm{E}_{\mathcal{P}}[Y(\bar{z}_2)] = \tau_0 + \tau_1 z_1 + \tau_2 z_2$, with $(\tau_0, \tau_1, \tau_2) = (308.30, 4.57, -10.00)$.

%%%%%%%%%%%%%%%%%%%%
\subsubsection{Study 2: extended time horizon}

Study 2 extends Setting 1 to $T=3$ time points. For each $i = 1, \ldots, n$, we draw latent variables $U_i \overset{\text{i.i.d}}{\sim} \textrm{Uniform} (1, 5)$ and $X^0_{itp} \overset{\text{i.i.d}}{\sim} \text{Normal}(0,1)$, $t = 1,2,3$, $p = 1,\ldots,4$. For baseline covariates at $t = 1$, $X_{i1} = (\frac{1}{U_i}X^0_{i11}, \frac{1}{U_i}X^0_{i12}, \frac{1}{U_i}|X^0_{i13}|, \frac{1}{U_i}|X^0_{i14}|)$. For covariates at $t = 2,3$, $X_{it}$ consists of $X_{i,t-1,1}+\frac{1}{U_i}X^0_{it1}+Z_{i,t-1}$, $X_{i,t-1,2}+\frac{1}{U_i}X^0_{it2}+Z_{i,t-1}$, $X_{i,t-1,3}+\min\{X_{i,t-1,3},\max(-X_{i,t-1,3},\frac{1}{U_i}X^0_{it3})\}+Z_{i,t-1}$, and $X_{i,t-1,4}+\min\{X_{i,t-1,4},\max(-X_{i,t-1,4},\frac{1}{U_i}X^0_{it4})\}+Z_{i,t-1}$. Treatments are generated as $Z_{it} \sim \text{Bernoulli}(\text{expit}\{-Z_{i,t-1}+X_{it1}-0.5X_{it2}+0.25X_{it3}+0.1X_{it4}+(-0.5)^t\})$ with $Z_{i0}=0$, and the outcome is $Y_i=250-10\sum_{t=1}^{2}Z_{it} + 58.5Z_{i3} + (27.4X_{i31} + 13.7X_{i32} + 13.7X_{i33} + 13.7X_{i34}) + U_i + \varepsilon_i$ where $\varepsilon_{i}\overset{\text{i.i.d}}{\sim} \text{Normal}(0,5^2)$. The true MSM is $\mathrm{E}_{\mathcal P}[Y(\bar{z}_3)] = \widetilde{\tau}_0+\widetilde{\tau}_{1}\sum_{t = 1}^3 z_t$ with $(\widetilde{\tau}_0, \widetilde{\tau}_{1})= (261.80, 58.50)$.

%%%%%%%%%%%%%%%%%%%%
\subsubsection{Study 3: misspecified outcome model}

Study 3 is adapted from the design of \citet{hainmueller2012entropy} and introduces misspecification through a non-linear outcome with $T=2$. Specifically, for each $i = 1, \ldots, n$ at $t = 1$, covariates are generated as $X_{i1[1:2]} \overset{\text{i.i.d}}{\sim} \text{Normal}([0, 0]^\top, [(2, 1)^\top, (1, 1)^\top])$, $X_{i13} \overset{\text{i.i.d}}{\sim} \chi^2_1$, and $X_{i14} \overset{\text{i.i.d}}{\sim} \mathrm{Bernoulli}(0.5)$. At $t = 2$, covariates are $X_{i2[1:2]} \sim \text{Normal}(X_{i1[1:2]} + 0.1, 1)$, $X_{i23} \sim \chi^2_1(X_{i13})$, and $X_{i24} \sim \mathrm{Bernoulli}(\text{expit}\{X_{i14}\})$. Treatments are assigned as $Z_{i1} \sim \text{Bernoulli}(\text{expit}\{X_{i11}+2X_{i12}-0.5X_{i13}+X_{i14}\})$ and $Z_{i2} \sim \text{Bernoulli}(\text{expit}\{2Z_{i1}+X_{i21}+2X_{i22}-0.5X_{i23}+X_{i24}\})$. The outcome is defined as $Y_i = \sum_{t=1}^{2}(X_{it1}+X_{it2}+X_{it3})^2 + \varepsilon_i$, where $\varepsilon_i \overset{\text{i.i.d}}{\sim} \text{Normal}(0,1)$. The true MSM is $\mathrm{E}_{\mathcal P}[Y(\bar{z}_2)] = \tau_0 + \tau_1 z_1 + \tau_2 z_2$ with $(\tau_0, \tau_1, \tau_2) = (27.82, 0.00, 0.00)$.

For all studies, we simulate datasets with sample sizes $n \in \{500, 1000, 2000\}$. Each simulation setting is replicated 1,000 times to evaluate estimator performance.

%%%%%%%
%%%%%%%
%%%%%%%
\subsection{Results} \label{sim_results}
 
For Study 1, Table \ref{tab:first} reports the estimated parameters $(\widehat{\tau}_0, \widehat{\tau}_1, \widehat{\tau}_2)$. As sample size increases, all methods show improved performance, with reductions in both bias and RMSE. Among the IPW methods, SL consistently achieves the lowest bias and RMSE. Weight stabilization provides minimal benefit, while truncation significantly increases bias. CBPSo outperforms CBPSe, and both perform similarly to the stabilized SL method. In this correctly specified setting, the g-computation formula and LTMLE are highly efficient, yielding small RMSEs. Our proposed method, BAO, performs comparably to LTMLE.

Results for the intercept $\widetilde{\tau}_0$ and cumulative effect $\widetilde{\tau}_1$ in Study 2 are presented in Table \ref{tab:secondcum}. While IPW using logistic regression (LR) is nearly unbiased, its RMSE remains relatively high. In contrast, IPW approaches based on SL or EL achieve lower RMSEs compared to LR. The application of weight truncation increases both bias and RMSE. Consistent with Study 1, CBPSo performs similarly to the IPW methods, while the g-computation formula (pooled and stratified), LTMLE, and BAO all exhibit relatively low bias and RMSE.

The results for Study 3 appear in Table \ref{tab:third}. This study examines robustness to outcome model misspecification, with all models deliberately misspecified as linear. For $\tau_1$, CBPSe and RBW yield smaller RMSEs than the IPW methods, while the pooled g-computation formula and BAO exhibit relatively low bias and RMSE. For $\tau_2$, the g-computation formula remains biased due to misspecification. In contrast, all IPW and CBPS methods display negligible bias and achieve coverage rates near 95\%. Both LTMLE and BAO also produce small bias and RMSE, with BAO maintaining coverage close to 95\%. We note, however, that bias in BAO's intercept estimates does not decrease with increasing sample size.

\begin{figure}[!htbp]
% \vspace{-0.25cm}
\begin{center}
\begin{subfigure}{.32\textwidth}
\includegraphics[scale = 0.535]{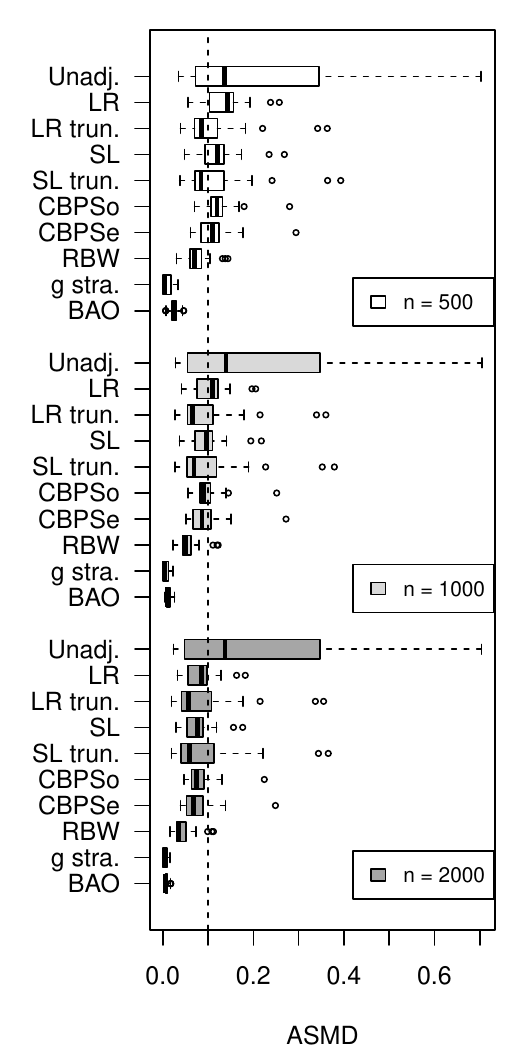}
\end{subfigure}
\begin{subfigure}{.32\textwidth}
\includegraphics[scale = 0.535]{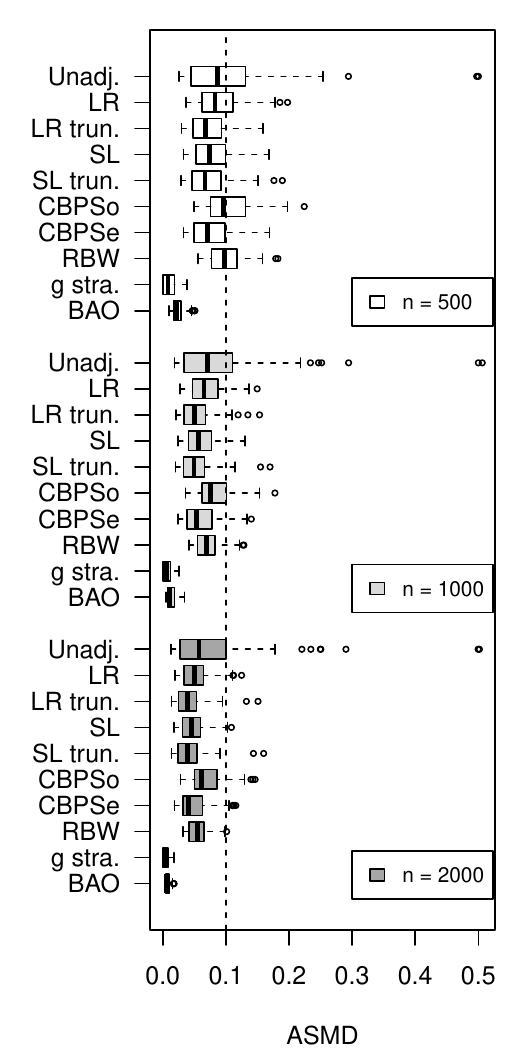}
\end{subfigure}
\begin{subfigure}{.32\textwidth}
\includegraphics[scale = 0.535]{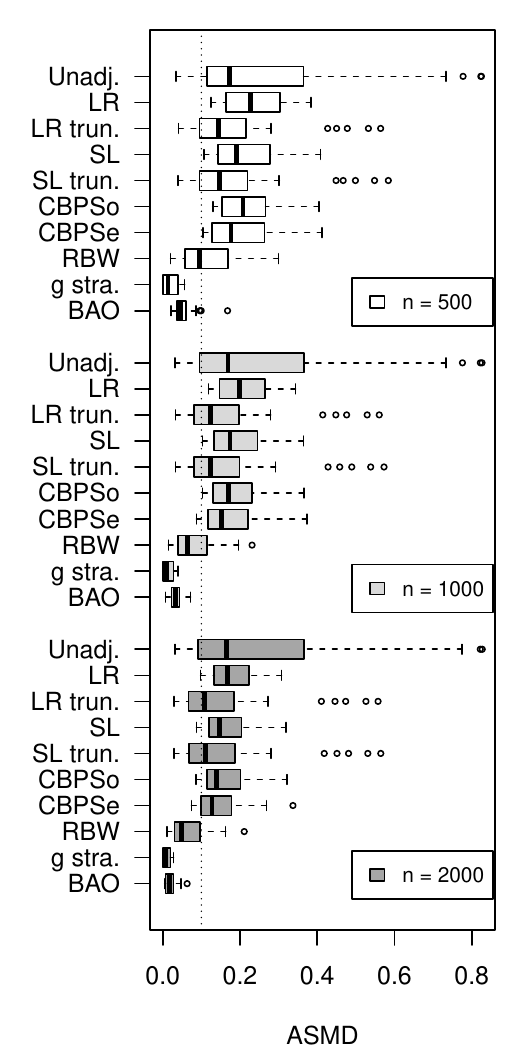}
\end{subfigure}
\begin{subfigure}{.32\textwidth}
\includegraphics[scale = 0.535]{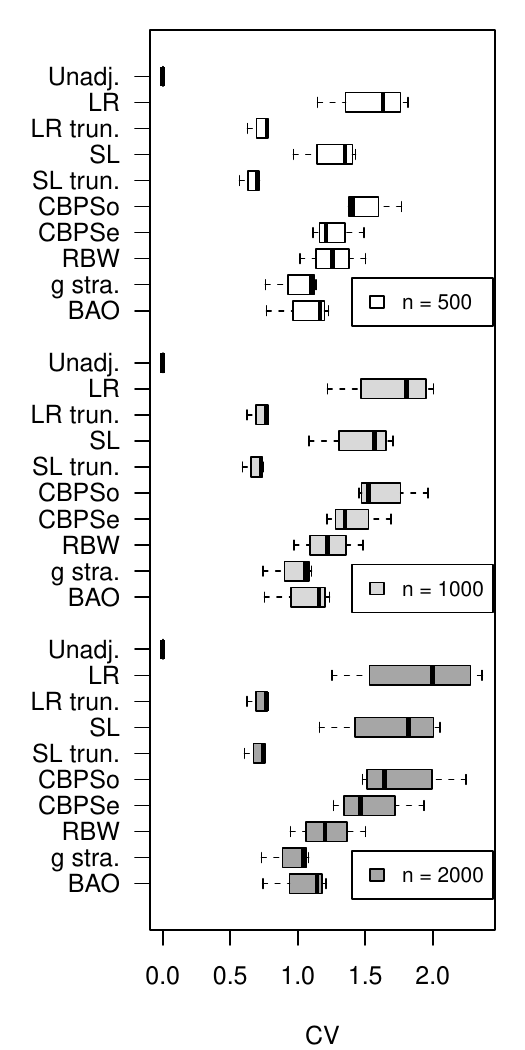}
\caption{First setting}
\end{subfigure}
\begin{subfigure}{.32\textwidth}
\includegraphics[scale = 0.535]{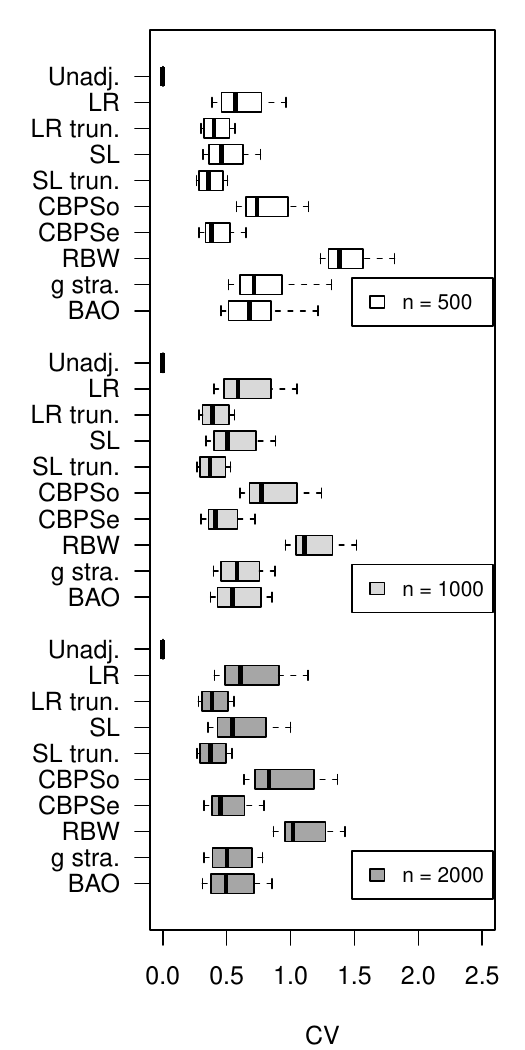}
\caption{Second setting}
\end{subfigure}
\begin{subfigure}{.32\textwidth}
\includegraphics[scale = 0.535]{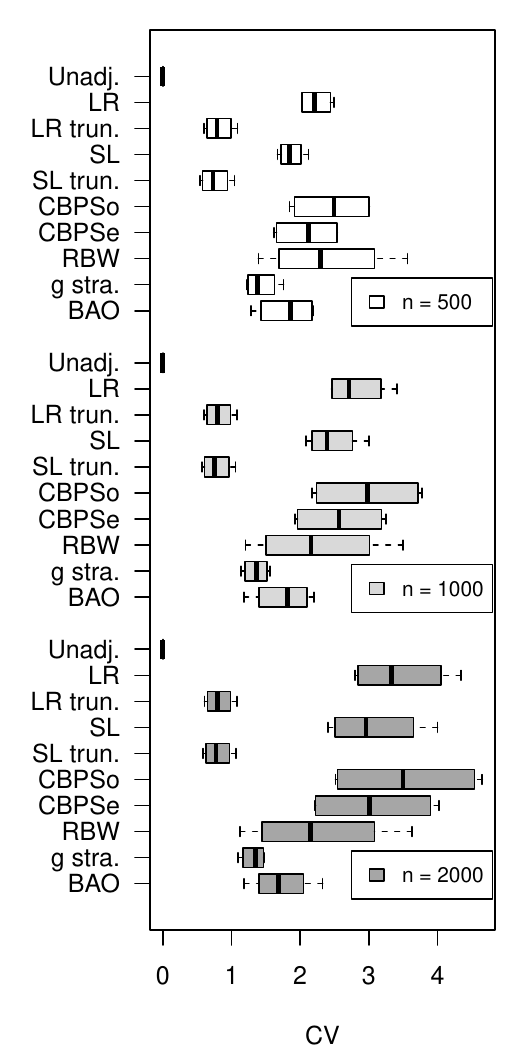}
\caption{Third setting}
\end{subfigure}
\caption{Covariate imbalance of $\bar{X}_T$ (measured by absolute standardized mean difference, ASMD) and weights variation (measured by coefficient of variation, CV) across all $2^T$ treatment paths in three studies. Values are averaged over 1000 simulated replicates.}\label{fig::asmd}
\end{center}
\vspace{-1cm}
\end{figure}

To further evaluate the performance of the different weighting approaches, we examine both the variability of the weights and the covariate imbalances after weighting, as measured by $\{|\mathrm{E}_{\mathcal{S}_{\omega}}[X_{tp} | \bar{z}_T]  - \mathrm{E}_{\mathcal{S}_{\omega}}[X_{tp}|\bar{z}_{t-1}]|: p = 1,\ldots,4;~ t = 1,\ldots,T;~ \bar{z}_T \in \bar{\mathcal Z}_T\}$, presented in Figure \ref{fig::asmd}. In all settings, the BAO consistently reduces weight variability compared to the others, while remaining approximately unbiased when the tolerance $\delta$ is selected with the tuning algorithm. Furthermore, BAO exhibits robustness to misspecification in the outcome models.

\begin{table}[htbp]
\centering
\footnotesize
\caption{Results of the first simulation study for three sample sizes, $n = 500, 1000, 2000$, and different methods. In the table, ``Cvge." and ``Lgth." represent the coverage and length of 95\% confidence interval; ``r.s.'' stands for robust sandwich estimator-based version. 
}\label{tab:first}
\setlength{\tabcolsep}{5pt}
\def\arraystretch{0.85}
\begin{tabular}{lrrrrrrrrrrrr}
  \hline
   \multicolumn{1}{l}{Method} & \multicolumn{4}{c}{$\tau_0$} & \multicolumn{4}{c}{$\tau_1$} & \multicolumn{4}{c}{$\tau_2$} \\   \cmidrule(lr){2-5} \cmidrule(lr){6-9} \cmidrule(lr){10-13}
 & Bias & RMSE & Cvge. & Lgth. & Bias & RMSE & Cvge. & Lgth. & Bias & RMSE & Cvge. & Lgth. \\ 
  \hline
$ \hspace{-.2cm} n = 500$ &   &   &   &   &   &   &   &   &   &   &   &   \\ 
  Unadj. & -37.66 &  38.03 &   0.00 &  20.44 &  22.84 &  23.73 &   5.50 &  24.88 &  50.43 &  50.82 &   0.00 &  24.88 \\
    LR &  -2.90 &  14.30 &  84.30 &  41.94 &   0.41 &  19.13 &  89.50 &  51.82 &   4.68 &  16.88 &  84.50 &  51.66 \\ 
  LR stab. &  -3.07 &  14.51 &  84.40 &  43.07 &   0.56 &  19.18 &  89.50 &  51.77 &   4.50 &  16.29 &  85.10 &  50.88 \\ 
  LR trun. & -16.77 &  17.78 &  28.80 &  26.56 &   7.80 &  10.97 &  83.00 &  31.66 &  22.98 &  23.81 &  12.20 &  31.04 \\ 
  SL &  -8.12 &  13.70 &  72.90 &  36.79 &   3.53 &  14.41 &  89.30 &  44.41 &  11.62 &  16.76 &  70.60 &  44.08 \\ 
  SL stab. &  -8.49 &  13.92 &  72.30 &  37.54 &   3.92 &  14.55 &  88.60 &  44.36 &  11.32 &  16.39 &  70.60 &  43.47 \\ 
  SL trun. & -18.51 &  19.41 &  16.90 &  25.77 &   9.25 &  11.93 &  78.10 &  30.61 &  24.92 &  25.68 &   6.10 &  30.00 \\ 
  EL &  -9.53 &  14.47 &  67.60 &  35.70 &   4.56 &  14.79 &  87.50 &  43.08 &  13.30 &  18.25 &  64.90 &  42.82 \\ 
  EL stab. &  -9.94 &  14.77 &  66.80 &  36.42 &   5.00 &  14.98 &  87.00 &  43.06 &  13.01 &  17.73 &  65.30 &  42.24 \\ 
  EL trun. & -19.18 &  20.07 &  14.80 &  25.53 &   9.84 &  12.38 &  74.90 &  30.30 &  25.66 &  26.42 &   5.00 &  29.71 \\ 
  CBPSe & -10.81 &  13.61 &  74.60 &  36.51 &  10.45 &  15.14 &  80.70 &  43.44 &  12.38 &  14.32 &  82.40 &  42.54 \\ 
  CBPSo &  -9.17 &  13.01 &  81.80 &  41.03 &   8.48 &  14.78 &  86.80 &  48.17 &   9.69 &  12.64 &  91.80 &  47.82 \\ 
  RBW &  -5.36 &   8.06 &  82.10 &  21.90 &   7.65 &  12.17 &  84.40 &  35.15 &   5.09 &   5.99 &  68.80 &  11.67 \\ 
  g pool. &   0.07 &   3.92 &  94.70 &  15.37 &  -0.08 &   6.66 &  94.90 &  26.57 &   0.01 &   0.59 &  95.40 &   2.28 \\ 
  g stra. &   0.01 &   4.01 &  94.60 &  15.70 &  -0.11 &   6.77 &  94.80 &  27.10 &   0.02 &   0.66 &  94.40 &   2.61 \\ 
  LTMLE &  -0.01 &   4.37 &  94.70 &  16.74 &  -0.17 &   7.50 &  94.30 &  28.91 &   0.23 &   1.43 &  93.90 &   5.38 \\ 
  BAO &  -1.73 &   4.46 &  92.00 &  15.77 &   1.11 &   6.74 &  94.00 &  26.34 &   2.05 &   2.46 &  68.70 &   5.44 \\ 
  BAO(r.s.) & -- & -- & 98.90 & 25.06 & -- & -- & 97.30  & 30.00 &-- & -- & 100.00 & 29.55 \\
   \hline
  \hspace{-.2cm} $n = 1000$ &   &   &   &   &   &   &   &   &   &   &   &   \\ 
  Unadj. & -37.38 &  37.58 &   0.00 &  14.41 &  22.62 &  23.08 &   0.10 &  17.59 &  50.10 &  50.33 &   0.00 &  17.59 \\ 
  LR &  -1.88 &  10.81 &  88.60 &  34.47 &   0.43 &  14.19 &  93.30 &  42.62 &   3.07 &  13.22 &  88.70 &  42.63 \\ 
  LR stab. &  -1.98 &  11.06 &  88.50 &  35.33 &   0.58 &  14.30 &  92.80 &  42.47 &   2.88 &  12.78 &  88.70 &  41.87 \\ 
  LR trun. & -16.23 &  16.76 &   6.60 &  18.74 &   7.49 &   9.08 &  76.40 &  22.42 &  22.48 &  22.91 &   0.60 &  21.94 \\ 
  SL &  -5.40 &  10.14 &  77.80 &  30.69 &   2.13 &  11.34 &  92.20 &  37.12 &   7.92 &  12.76 &  75.80 &  36.99 \\ 
  SL stab. &  -5.65 &  10.36 &  78.20 &  31.28 &   2.46 &  11.48 &  91.60 &  37.01 &   7.61 &  12.35 &  77.20 &  36.41 \\ 
  SL trun. & -17.26 &  17.76 &   3.40 &  18.40 &   8.33 &   9.74 &  70.60 &  21.96 &  23.63 &  24.05 &   0.30 &  21.49 \\ 
  EL &  -6.18 &  11.09 &  73.00 &  30.18 &   3.03 &  11.89 &  90.80 &  36.36 &   9.08 &  14.14 &  68.60 &  36.19 \\ 
  EL stab. &  -6.46 &  11.36 &  72.30 &  30.75 &   3.39 &  11.94 &  90.20 &  36.17 &   8.76 &  13.73 &  69.00 &  35.63 \\ 
  EL trun. & -17.72 &  18.21 &   2.40 &  18.27 &   8.71 &  10.07 &  66.70 &  21.80 &  24.13 &  24.56 &   0.10 &  21.34 \\ 
  CBPSe &  -9.19 &  11.23 &  72.90 &  29.11 &  10.05 &  13.06 &  76.30 &  34.58 &  10.51 &  12.05 &  78.70 &  33.83 \\ 
  CBPSo &  -8.04 &  10.81 &  80.50 &  32.57 &   8.68 &  12.60 &  83.40 &  38.14 &   7.98 &  10.26 &  89.90 &  38.03 \\ 
  RBW &  -4.93 &   6.29 &  75.40 &  14.92 &   6.98 &   9.43 &  77.40 &  24.08 &   4.66 &   5.06 &  30.90 &   7.59 \\ 
  g pool. &  -0.07 &   2.80 &  94.30 &  10.80 &  -0.03 &   4.75 &  94.50 &  18.75 &   0.00 &   0.41 &  94.90 &   1.60 \\ 
  g stra. &  -0.03 &   2.88 &  94.20 &  11.03 &  -0.01 &   4.83 &  94.80 &  19.09 &   0.00 &   0.45 &  95.50 &   1.79 \\ 
  LTMLE &  -0.14 &   3.07 &  93.70 &  11.78 &   0.15 &   5.22 &  94.30 &  20.54 &   0.15 &   0.96 &  93.50 &   3.70 \\ 
  BAO &  -0.67 &   2.95 &  93.90 &  11.11 &   0.40 &   4.79 &  94.30 &  18.86 &   0.80 &   1.11 &  87.90 &   3.38 \\  
  BAO(r.s.) & -- & -- & 99.60 & 17.80 & -- & -- & 97.10  & 21.44 & -- & -- & 100.00 & 21.06 \\
   \hline
  \hspace{-.2cm} $n = 2000$ &   &   &   &   &   &   &   &   &   &   &   &   \\ 
  Unadj. & -37.34 &  37.44 &   0.00 &  10.21 &  22.71 &  22.92 &   0.00 &  12.44 &  50.13 &  50.22 &   0.00 &  12.44 \\ 
    LR &  -1.15 &   8.50 &  87.50 &  26.54 &  -0.17 &  11.08 &  93.70 &  34.00 &   2.12 &  10.62 &  87.40 &  34.01 \\ 
  LR stab. &  -1.21 &   8.72 &  87.10 &  27.36 &  -0.09 &  11.19 &  93.20 &  34.05 &   2.00 &  10.29 &  87.80 &  33.25 \\ 
  LR trun. & -16.03 &  16.31 &   0.00 &  13.31 &   7.43 &   8.28 &  56.20 &  15.88 &  22.39 &  22.59 &   0.00 &  15.56 \\ 
  SL &  -3.36 &   8.20 &  81.00 &  24.25 &   1.13 &   9.16 &  93.40 &  30.18 &   5.28 &  10.17 &  78.60 &  30.12 \\ 
  SL stab. &  -3.54 &   8.37 &  80.90 &  24.83 &   1.37 &   9.24 &  92.20 &  30.18 &   5.08 &   9.89 &  79.10 &  29.50 \\ 
  SL trun. & -16.60 &  16.86 &   0.00 &  13.17 &   7.93 &   8.72 &  49.10 &  15.69 &  23.00 &  23.20 &   0.00 &  15.37 \\ 
  EL &  -4.12 &   8.77 &  76.40 &  23.76 &   1.69 &   9.25 &  92.00 &  29.50 &   6.25 &  11.17 &  71.70 &  29.43 \\ 
  EL stab. &  -4.32 &   8.99 &  76.70 &  24.33 &   1.95 &   9.40 &  90.90 &  29.48 &   6.03 &  10.87 &  72.40 &  28.84 \\ 
  EL trun. & -16.90 &  17.17 &   0.00 &  13.10 &   8.17 &   8.94 &  45.90 &  15.61 &  23.35 &  23.55 &   0.00 &  15.29 \\ 
  CBPSe &  -8.56 &  10.02 &  61.80 &  22.31 &   9.79 &  11.82 &  63.60 &  26.89 &   9.77 &  10.85 &  67.90 &  26.14 \\ 
  CBPSo &  -7.48 &   9.39 &  72.80 &  25.03 &   8.49 &  11.05 &  73.60 &  29.71 &   7.33 &   8.88 &  84.40 &  29.56 \\ 
  RBW &  -4.87 &   5.63 &  57.00 &  10.66 &   7.03 &   8.33 &  64.30 &  17.02 &   4.70 &   4.91 &   2.60 &   5.33 \\ 
  g pool. &   0.13 &   1.96 &  94.80 &   7.64 &  -0.13 &   3.36 &  95.40 &  13.22 &   0.01 &   0.29 &  94.90 &   1.13 \\ 
  g stra. &   0.16 &   1.98 &  95.10 &   7.79 &  -0.12 &   3.41 &  95.60 &  13.45 &   0.01 &   0.33 &  94.20 &   1.26 \\ 
  LTMLE &   0.07 &   2.15 &  94.90 &   8.35 &  -0.04 &   3.72 &  95.00 &  14.50 &   0.19 &   0.69 &  92.40 &   2.54 \\ 
  BAO &  -0.25 &   1.99 &  95.20 &   7.80 &   0.15 &   3.41 &  95.70 &  13.44 &   0.55 &   0.68 &  74.60 &   1.65 \\ 
  BAO(r.s.) & -- & -- & 99.90 & 12.64 & -- & -- & 96.80  & 15.14 & -- & -- & 100.00 & 14.88 \\
   \hline
  \end{tabular}
\end{table}

\begin{table}[!htbp]
\centering
\footnotesize
\caption{Results of the second simulation study for three sample sizes, $n = 500, 1000, 2000$, and different methods. In the table, ``Cvge." and ``Lgth." stands for the coverage and length of 95\% confidence interval; ``r.s.'' stands for robust sandwich estimator-based version.}\label{tab:secondcum}
\def\arraystretch{0.85}
\begin{tabular}{lrrrrrrrr}
  \hline
   \multicolumn{1}{l}{Methods} & \multicolumn{4}{c}{$\tau_0$} & \multicolumn{4}{c}{$\widetilde{\tau}_1$} \\   \cmidrule(lr){2-5} \cmidrule(lr){6-9}
    & Bias & RMSE & Cvge. & Lgth. & Bias & RMSE & Cvge. & Lgth. \\ 
  \hline
\hspace{-.2cm}$n = 500$ &   &   &   &   &   &   &   &   \\ 
  Unadj. & -10.28 & 10.56 & 1.10 & 9.07 & 6.25 & 6.40 & 0.20 & 5.17 \\ 
  LR & -0.82 &  5.21 & 88.00 & 13.78 &  0.43 &  2.72 & 90.90 &  7.61 \\ 
  LR stab. & -0.75 &  5.00 & 88.30 & 13.05 &  0.42 &  2.59 & 90.50 &  7.21 \\ 
  LR trun. & -3.38 &  3.98 & 66.60 &  8.88 &  1.97 &  2.26 & 66.10 &  4.99 \\ 
  SL & -2.38 &  4.25 & 78.90 & 11.37 &  1.29 &  2.27 & 81.90 &  6.32 \\ 
  SL stab. & -2.18 &  4.15 & 79.70 & 10.93 &  1.27 &  2.23 & 81.20 &  6.12 \\ 
  SL trun. & -4.00 &  4.50 & 55.10 &  8.60 &  2.37 &  2.61 & 49.50 &  4.85 \\ 
  EL & -2.73 &  4.31 & 75.50 & 11.23 &  1.51 &  2.41 & 76.40 &  6.27 \\ 
  EL stab. & -2.51 &  4.16 & 75.60 & 10.81 &  1.48 &  2.34 & 75.80 &  6.04 \\ 
  EL trun. & -4.18 &  4.67 & 52.90 &  8.58 &  2.50 &  2.73 & 48.10 &  4.84 \\ 
  CBPSe & -4.41 &  4.98 & 54.40 &  9.54 &  2.94 &  3.23 & 42.60 &  5.51 \\ 
  CBPSo & -2.47 &  4.21 & 83.30 & 13.08 &  1.79 &  2.63 & 79.50 &  7.53 \\
  RBW & -2.92 &  4.37 & 89.30 & 14.51 &  2.12 &  2.83 & 87.50 &  8.52 \\ 
  g pool. & -0.07 &  1.75 & 93.80 &  6.70 &  0.03 &  0.82 & 95.10 &  3.26 \\ 
  g stra. &  0.13 &  1.91 & 95.90 & 15.48 &  0.02 &  0.89 & 97.90 &  6.61 \\ 
  LTMLE & -1.97 &  2.99 & 83.20 &  8.27 &  1.27 &  1.72 & 76.40 &  4.30 \\ 
  BAO & -1.04 &  2.04 & 90.70 &  6.74 &  0.60 &  1.02 & 89.90 &  3.34 \\ 
  BAO(r.s.) & -- & -- &  95.40 &   8.68 & --   &  --  &  97.90 &  4.91 \\
   \hline
\hspace{-.2cm}$n = 1000$ &   &   &   &   &   &   &   &   \\ 
  Unadj. & -10.18 & 10.31 & 0.00 & 6.39 & 6.21 & 6.28 & 0.00 & 3.65 \\ 
  LR & -0.31 &  4.37 & 90.80 & 10.91 &  0.12 &  2.34 & 93.30 &  6.12 \\ 
  LR stab. & -0.31 &  4.18 & 90.20 & 10.35 &  0.12 &  2.25 & 92.90 &  5.86 \\ 
  LR trun. & -3.25 &  3.57 & 46.50 &  6.26 &  1.91 &  2.07 & 42.50 &  3.52 \\ 
  SL & -1.49 &  2.92 & 82.90 &  9.08 &  0.75 &  1.60 & 86.80 &  5.10 \\ 
  SL stab. & -1.40 &  2.82 & 83.70 &  8.72 &  0.75 &  1.58 & 85.40 &  4.97 \\ 
  SL trun. & -3.58 &  3.86 & 36.30 &  6.13 &  2.12 &  2.26 & 32.10 &  3.46 \\ 
  EL & -1.73 &  3.66 & 78.00 &  8.93 &  0.93 &  1.91 & 80.30 &  4.99 \\ 
  EL stab. & -1.61 &  3.55 & 80.10 &  8.56 &  0.92 &  1.87 & 80.20 &  4.84 \\ 
  EL trun. & -3.72 &  3.99 & 32.50 &  6.10 &  2.21 &  2.35 & 26.60 &  3.45 \\ 
  CBPSe & -3.98 &  4.33 & 39.00 &  7.16 &  2.69 &  2.86 & 28.20 &  4.16 \\ 
  CBPSo & -2.13 &  3.27 & 81.20 & 10.02 &  1.54 &  2.11 & 77.70 &  5.86 \\ 
  RBW & -2.87 &  3.71 & 68.30 &  8.42 &  2.11 &  2.55 & 58.30 &  5.02 \\ 
  g pool. & -0.01 &  1.22 & 94.80 &  4.74 &  0.00 &  0.60 & 95.00 &  2.32 \\ 
  g stra. &  0.13 &  1.31 & 94.80 &  5.08 &  0.00 &  0.62 & 95.40 &  2.49 \\ 
  LTMLE & -1.95 &  2.51 & 73.90 &  5.99 &  1.26 &  1.51 & 63.70 &  3.15 \\ 
  BAO & -0.47 &  1.32 & 93.30 &  4.83 &  0.27 &  0.66 & 92.80 &  2.38 \\ 
  BAO(r.s.) & -- & -- &  97.60 &   6.33 & --   &  --  &  99.60 &   3.61 \\ 
   \hline
   \hspace{-.2cm}$n = 2000$ &   &   &   &   &   &   &   &   \\ 
  Unadj. & -10.17 & 10.24 & 0.00 & 4.54 & 6.21 & 6.24 & 0.00 & 2.59 \\ 
  LR & -0.12 &  4.93 & 91.00 &  8.49 &  0.07 &  2.44 & 92.50 &  4.59 \\ 
  LR stab. & -0.16 &  4.75 & 89.40 &  7.94 &  0.10 &  2.31 & 91.20 &  4.39 \\ 
  LR trun. & -3.23 &  3.40 & 17.80 &  4.45 &  1.93 &  2.01 & 12.00 &  2.50 \\ 
   SL & -0.91 &  3.54 & 84.00 &  7.37 &  0.48 &  1.81 & 87.80 &  4.02 \\ 
  SL stab. & -0.86 &  3.54 & 84.20 &  7.00 &  0.49 &  1.77 & 85.70 &  3.89 \\ 
  SL trun. & -3.41 &  3.57 & 12.60 &  4.39 &  2.04 &  2.12 &  8.80 &  2.47 \\ 
  EL & -1.23 &  2.88 & 79.30 &  7.04 &  0.66 &  1.48 & 82.90 &  3.84 \\ 
  EL stab. & -1.15 &  2.85 & 80.10 &  6.70 &  0.67 &  1.47 & 80.90 &  3.73 \\ 
  EL trun. & -3.51 &  3.67 & 11.50 &  4.37 &  2.10 &  2.18 &  7.30 &  2.46 \\ 
  CBPSe & -3.66 &  3.89 & 25.60 &  5.40 &  2.50 &  2.62 & 12.80 &  3.14 \\ 
  CBPSo & -1.91 &  2.76 & 79.20 &  7.86 &  1.41 &  1.84 & 73.10 &  4.63 \\ 
  RBW & -2.78 &  3.30 & 50.80 &  5.98 &  2.09 &  2.35 & 36.00 &  3.59 \\  
  g pool. &  0.02 &  0.88 & 94.10 &  3.36 &  0.00 &  0.43 & 94.60 &  1.64 \\ 
  g stra. &  0.16 &  0.95 & 93.90 &  3.57 & -0.01 &  0.46 & 94.80 &  1.74 \\ 
  LTMLE & -1.96 &  2.27 & 57.00 &  4.34 &  1.27 &  1.42 & 39.60 &  2.29 \\ 
  BAO & -0.15 &  0.91 & 94.10 &  3.36 &  0.15 &  0.46 & 92.80 &  1.65 \\ 
  BAO(r.s.) & -- & -- &  98.50 &  4.63 &  --  &  --  &  99.50 &   2.60 \\
  \hline
  \end{tabular}
\end{table}

\begin{table}[!htbp]
\centering
\footnotesize
\caption{Results of the third simulation study for three sample sizes, $n = 500, 1000, 2000$, and different methods. In the table, ``Cvge." and ``Lgth." stands for the coverage and length of 95\% confidence interval; ``r.s.'' stands for robust sandwich estimator-based version.}\label{tab:third}
\setlength{\tabcolsep}{5pt}
\def\arraystretch{0.85}
\begin{tabular}{lrrrrrrrrrrrr}
  \hline
   \multicolumn{1}{l}{Method} & \multicolumn{4}{c}{$\tau_0$} & \multicolumn{4}{c}{$\tau_1$} & \multicolumn{4}{c}{$\tau_2$} \\
\cmidrule(lr){2-5} \cmidrule(lr){6-9} \cmidrule(lr){10-13}
    & Bias & RMSE & Cvge. & Lgth. & Bias & RMSE & Cvge. & Lgth. & Bias & RMSE & Cvge. & Lgth. \\ 
  \hline
\hspace{-.2cm}$n = 500$ &   &   &   &   &   &   &   &   &   &   &   &   \\ 
  Unadj. & -4.21 & 6.51 & 72.90 & 18.95 & 15.11 & 17.34 & 47.00 & 29.43 & -6.89 & 10.79 & 92.60 & 29.43 \\
  LR &  -5.96 &  12.33 &  53.30 &  23.37 &   2.62 &  15.20 &  87.90 &  30.13 &   0.43 &  14.58 &  89.90 &  30.41 \\ 
  LR stab. &  -5.42 &  11.03 &  50.60 &  21.29 &   3.12 &  15.05 &  84.50 &  30.59 &  -0.03 &  13.91 &  91.10 &  31.52 \\ 
  LR trun. &  -6.27 &   7.45 &  53.70 &  15.67 &   6.92 &   8.67 &  69.80 &  21.12 &  -1.29 &   5.53 &  97.60 &  22.24 \\ 
  SL &  -7.00 &   9.80 &  49.30 &  19.54 &   3.02 &   9.20 &  88.20 &  24.98 &   0.31 &   8.44 &  94.50 &  25.34 \\ 
  SL stab. &  -6.11 &   9.11 &  51.40 &  18.72 &   3.87 &   9.59 &  83.10 &  24.81 &  -0.50 &   8.94 &  94.10 &  26.02 \\ 
  SL trun. &  -6.01 &   7.23 &  55.10 &  15.62 &   7.35 &   8.77 &  65.40 &  20.10 &  -1.66 &   5.23 &  97.30 &  21.34 \\ 
  EL &  -6.79 &  10.03 &  50.50 &  19.31 &   2.93 &   8.95 &  87.90 &  24.79 &   0.15 &   8.10 &  94.20 &  25.13 \\ 
  EL stab. &  -5.92 &   9.52 &  51.80 &  18.58 &   3.79 &   9.09 &  82.10 &  24.48 &  -0.61 &   8.55 &  94.20 &  25.71 \\ 
  EL trun. &  -5.85 &   7.11 &  56.90 &  15.76 &   7.42 &   8.86 &  64.80 &  20.22 &  -1.78 &   5.33 &  97.00 &  21.51 \\ 
  CBPSe &  -6.33 &   9.04 &  56.30 &  22.30 &   2.67 &   8.93 &  92.20 &  30.56 &  -1.01 &   8.39 &  96.40 &  31.90 \\ 
  CBPSo &  -6.70 &   9.82 &  52.10 &  23.44 &   2.37 &  10.24 &  93.10 &  33.97 &  -0.28 &   9.51 &  97.60 &  34.79 \\ 
  RBW &  -7.24 &   8.71 &  60.90 &  20.98 &   2.08 &   7.54 &  94.20 &  30.14 &  -1.22 &   5.81 &  97.80 &  25.25 \\ 
  g pool. &   2.85 &   4.92 &  92.00 &  15.56 &   0.65 &   5.61 &  94.90 &  20.64 &  -5.56 &   7.26 &  65.20 &  16.11 \\ 
  g stra. & -12.47 &  13.75 &  35.30 &  22.85 &  -0.28 &   6.92 &  96.10 &  35.17 &   2.04 &   6.70 &  95.00 &  33.23 \\ 
  LTMLE &  -1.82 &   5.94 &  79.30 &  17.33 &   0.51 &   7.08 &  91.30 &  21.56 &  -0.89 &   3.20 &  88.90 &  10.00 \\ 
  BAO &  -6.17 &   8.04 &  57.40 &  17.86 &   0.71 &   6.16 &  93.90 &  21.35 &  -2.22 &   5.63 &  91.90 &  17.97 \\ 
  BAO(r.s.) &  -- & --  &  57.70 &  18.72 & --  & -- &  94.20 &  21.43 & -- & -- &  98.70 &  23.15 \\ 
   \hline
\hspace{-.2cm}$n = 1000$ &   &   &   &   &   &   &   &   &   &   &   &   \\ 
  Unadj. & -4.13 & 5.46 & 69.00 & 13.87 & 15.01 & 16.08 & 14.10 & 21.80 & -6.97 & 9.04 & 85.00 & 21.80 \\ 
  LR &  -4.72 &  14.00 &  52.90 &  21.06 &   1.29 &  14.48 &  88.90 &  27.56 &   0.42 &  13.64 &  92.30 &  27.66 \\ 
  LR stab. &  -4.49 &  14.11 &  48.40 &  18.77 &   1.63 &  14.30 &  85.80 &  27.91 &   0.22 &  14.92 &  93.30 &  28.70 \\ 
  LR trun. &  -5.91 &   6.63 &  44.10 &  11.80 &   7.07 &   8.08 &  53.30 &  16.11 &  -1.73 &   4.29 &  97.50 &  16.96 \\ 
  SL &  -6.16 &  10.27 &  45.50 &  17.43 &   1.80 &   9.80 &  88.80 &  22.51 &   0.56 &   8.60 &  94.60 &  22.54 \\ 
  SL stab. &  -5.48 &  10.11 &  45.40 &  16.32 &   2.46 &  10.16 &  82.90 &  22.59 &   0.01 &   9.79 &  95.20 &  23.45 \\ 
  SL trun. &  -5.84 &   6.54 &  44.50 &  11.63 &   7.25 &   8.12 &  48.70 &  15.30 &  -1.79 &   4.07 &  97.20 &  16.15 \\ 
  EL &  -6.30 &   9.31 &  45.30 &  16.81 &   2.24 &   9.72 &  87.40 &  22.38 &   0.57 &   8.85 &  95.30 &  22.56 \\ 
  EL stab. &  -5.67 &   8.77 &  44.70 &  15.68 &   2.87 &  10.28 &  81.00 &  22.60 &   0.01 &   9.86 &  95.60 &  23.49 \\ 
  EL trun. &  -5.76 &   6.48 &  45.80 &  11.66 &   7.35 &   8.19 &  46.40 &  15.20 &  -1.85 &   4.07 &  96.40 &  16.06 \\ 
  CBPSe &  -6.02 &   8.61 &  52.90 &  19.67 &   2.47 &   7.83 &  91.70 &  26.44 &  -0.51 &   7.29 &  96.70 &  27.55 \\ 
  CBPSo &  -6.07 &   9.34 &  51.10 &  20.95 &   1.96 &   8.89 &  92.80 &  29.48 &  -0.19 &   8.47 &  98.00 &  30.25 \\ 
  RBW &  -7.09 &   7.82 &  36.30 &  12.42 &   2.45 &   6.03 &  87.80 &  19.10 &  -1.26 &   4.44 &  93.30 &  15.11 \\ 
  g pool. &   3.00 &   4.22 &  85.80 &  11.32 &   0.46 &   4.00 &  94.20 &  14.97 &  -5.61 &   6.55 &  48.20 &  12.10 \\ 
  g stra. & -12.11 &  12.88 &  19.30 &  15.54 &  -0.66 &   5.13 &  94.40 &  18.35 &   2.09 &   5.17 &  91.50 &  16.69 \\ 
  LTMLE &  -1.62 &   4.95 &  78.20 &  13.52 &   0.34 &   6.11 &  93.00 &  17.37 &  -0.83 &   2.37 &  88.50 &   7.61 \\ 
  BAO &  -6.08 &   7.22 &  49.40 &  13.77 &   1.14 &   4.53 &  94.30 &  16.28 &  -2.07 &   4.28 &  91.00 &  13.68 \\  
  BAO(r.s.) & --  & -- &  53.00 &  15.17 & -- &  --  &  95.60 &  17.26 & --  & -- &  99.30 &  18.52 \\  
   \hline
\hspace{-.2cm}$n = 2000$ &   &   &   &   &   &   &   &   &   &   &   &   \\ 
  Unadj. & -4.20 & 4.96 & 56.70 & 9.84 & 14.72 & 15.30 & 0.60 & 15.39 & -6.64 & 7.89 & 62.90 & 15.39 \\ 
  LR &  -4.10 &  10.36 &  54.40 &  18.80 &   0.66 &  10.77 &  91.80 &  25.00 &   0.93 &  10.31 &  92.60 &  24.95 \\ 
  LR stab. &  -3.91 &  10.22 &  52.10 &  16.46 &   1.02 &  10.98 &  86.70 &  25.76 &   0.75 &  11.25 &  94.80 &  26.28 \\ 
  LR trun. &  -5.86 &   6.28 &  28.10 &   8.69 &   6.92 &   7.47 &  33.70 &  11.80 &  -1.56 &   3.35 &  95.20 &  12.43 \\ 
  SL &  -5.41 &   8.59 &  46.70 &  15.78 &   1.34 &   7.97 &  91.10 &  20.48 &   0.81 &   7.73 &  93.70 &  20.52 \\ 
  SL stab. &  -4.83 &   8.26 &  46.30 &  14.48 &   1.99 &   8.37 &  84.60 &  20.77 &   0.29 &   8.38 &  95.30 &  21.51 \\ 
  SL trun. &  -5.83 &   6.25 &  27.30 &   8.59 &   7.02 &   7.51 &  29.40 &  11.38 &  -1.57 &   3.25 &  94.90 &  12.00 \\ 
  EL &  -5.63 &   8.22 &  45.70 &  15.07 &   1.58 &   7.02 &  90.20 &  19.72 &   0.78 &   6.78 &  94.30 &  19.71 \\ 
  EL stab. &  -5.00 &   7.80 &  45.50 &  13.88 &   2.28 &   7.25 &  82.50 &  20.06 &   0.20 &   7.05 &  95.60 &  20.65 \\ 
  EL trun. &  -5.79 &   6.21 &  28.10 &   8.59 &   7.10 &   7.60 &  28.20 &  11.36 &  -1.62 &   3.27 &  94.40 &  11.98 \\ 
  CBPSe &  -5.39 &   7.48 &  53.60 &  17.13 &   2.27 &   7.06 &  88.30 &  23.66 &  -0.27 &   6.50 &  94.10 &  24.44 \\ 
  CBPSo &  -5.33 &   7.86 &  53.00 &  18.51 &   1.64 &   7.84 &  90.70 &  26.22 &   0.11 &   7.23 &  97.10 &  26.79 \\ 
  RBW &  -6.95 &   7.41 &  20.50 &   8.85 &   2.11 &   4.63 &  86.80 &  14.02 &  -0.84 &   3.29 &  90.70 &  10.69 \\ 
  g pool. &   2.97 &   3.62 &  73.40 &   8.14 &   0.36 &   2.92 &  93.90 &  10.88 &  -5.39 &   5.91 &  31.70 &   8.94 \\ 
  g stra. & -12.06 &  12.43 &   5.50 &  11.45 &  -0.80 &   3.62 &  94.10 &  13.05 &   2.36 &   4.05 &  85.20 &  11.82 \\ 
  LTMLE &  -1.30 &   3.72 &  79.40 &  10.77 &   0.05 &   4.28 &  93.50 &  13.89 &  -0.71 &   1.74 &  89.20 &   5.81 \\ 
  BAO &  -6.91 &   7.39 &  24.40 &   9.73 &   1.00 &   3.36 &  93.10 &  11.82 &  -0.77 &   2.78 &  93.60 &   9.65 \\ 
  BAO(r.s.) &--  & -- &  30.70 &   11.05 & --   &  --  &  95.70 &  12.99 &  -- & -- &  99.10 &  13.80 \\
   \hline
  \end{tabular}
\end{table}

\section{Case study} \label{CaseStudy}

%%%%%%%%%%%%%%%%%%%%
%%%%%%%%%%%%%%%%%%%%
\subsection{Voucher schools in Chile} 

In 1981, Chile introduced a universal educational voucher system for elementary and secondary schools, which, in a modified form, remains in place today. 
The program provided government-funded vouchers for students to attend public or private schools, including for-profit and non-profit institutions. Since the introduction of the voucher system, enrollment in subsidized private schools has steadily increased. 
The proportion of students enrolled in subsidized private schools increased from 45\% in 2006 to 55\% in 2016, encompassing the period examined in our study \citep{mineduc2018estudiantes}.

The voucher system has generated debates, primarily due to concerns about its role in exacerbating socioeconomic segregation among schools \citep {valenzuela2014socioeconomic}. However, it remains uncertain whether the voucher system has effectively improved educational quality \citep{zubizarreta2017optimal}. As a result, ongoing discussions persist regarding whether subsidized private schools, also known as voucher schools, provide a higher quality of education compared to public schools. This issue has been explored from various perspectives in numerous studies \citep{sapelli2002performance, anand2009using, zubizarreta2017optimal}, but findings have been inconclusive.

We address this question using a comprehensive longitudinal dataset from educational administrative records, which track individual students throughout their secondary education from 2012 to 2015. The outcomes of interest are language and math scores from the University Selection Test (or PSU for \emph{Prueba de Selecci\'on Universitaria} in Spanish), a standardized exam administered nationwide in 2015 and required for admission to higher education in Chile. These test scores are standardized on a scale from 150 to 850. 
We define a sequence of four annual, time-varying treatments corresponding to school choice (subsidized private vs. public) and covering the period from eighth to twelfth grade. Our analysis proceeds in two stages: first, we estimate the mean potential PSU scores for each of the 16 possible treatment histories; second, we fit an MSM to summarize and interpret the effects across different exposure paths.

%%%%%%%%%%%%%%%%%%%%
\subsection{Educational administrative records}

We link several Chilean administrative sources for our analysis. Standardized tests provide both covariates and outcomes: the Education Quality Measurement System (or SIMCE for \emph{Sistema de Medici\'on de la Calidad de la Educaci\'on} in Spanish) is administered in eighth and tenth grades; the University Selection Test (PSU), a requirement for university admission, is taken in twelfth grade. The SIMCE assessments also collect detailed background information, including parental education, household income, and the socioeconomic status of the school, providing comprehensive data for educational research and policy evaluation.

To construct our analytic sample, we merge data from the 2011 and 2013 SIMCE assessments with 2015 PSU results using unique student identifiers. We further enhance our dataset with annual records from the Chilean Ministry of Education, which includes students' grade point averages (GPA) and school attendance rates. By combining these data sources, we create an educational census panel that tracks the same cohort of students, capturing covariates from 2011 to 2014, exposures from 2012 to 2015, and outcomes in 2015.

Our study focuses on students in Santiago, the capital and largest city of Chile. We exclude students who transferred to private non-subsidized schools, those lacking information on school type, and those who did not take the PSU, resulting in a final sample of 27,743 observations.
Baseline covariates include parental highest education level (classified as Primary, Secondary, Technical, College, or Missing), household income (grouped into six categories: 0--200, 200--400, 400--600, 600--1400, over 1400, and Missing), and gender. Additionally, we incorporate time-varying covariates such as language and math scores, school attendance, and GPA. 
See \citet{vandecandelaere2016time} for a related analysis that uses IPW to study the effects of grade retention on mathematics development throughout primary school.

%%%%%%%%%%%%%%%%%%%%
\subsection{Effects of attaining education in voucher schools on test scores}

% We begin by estimating the mean potential outcomes for the 16 possible exposure pathways corresponding to the four binary exposures (private school VS. public school) on 2015 PSU language and math scores. 
Our analysis proceeds by first estimating the mean potential outcomes for 2015 PSU language and math scores under each of the $2^4=16$ treatment paths corresponding to four annual binary school-choice decisions (private school $z_t = 1$ VS. public school $z_t = 0$). The analysis includes 12 baseline binary covariates, as well as students' GPA and school attendance measured annually at $t = 1, 2, 3, 4$. The language and math scores in eighth and tenth grades are observed at $t = 1, 3$. Missing values are present in these assessments, with 508 and 500 missing cases at $t = 1$, and 2,629 and 2,617 at $t = 3$ for language and math scores, respectively. To handle the missing values, we impute them using the sample medians and add binary covariates for missingness indicators.

\begin{figure}[!htbp]
\centering
\includegraphics[width=0.75\textwidth]{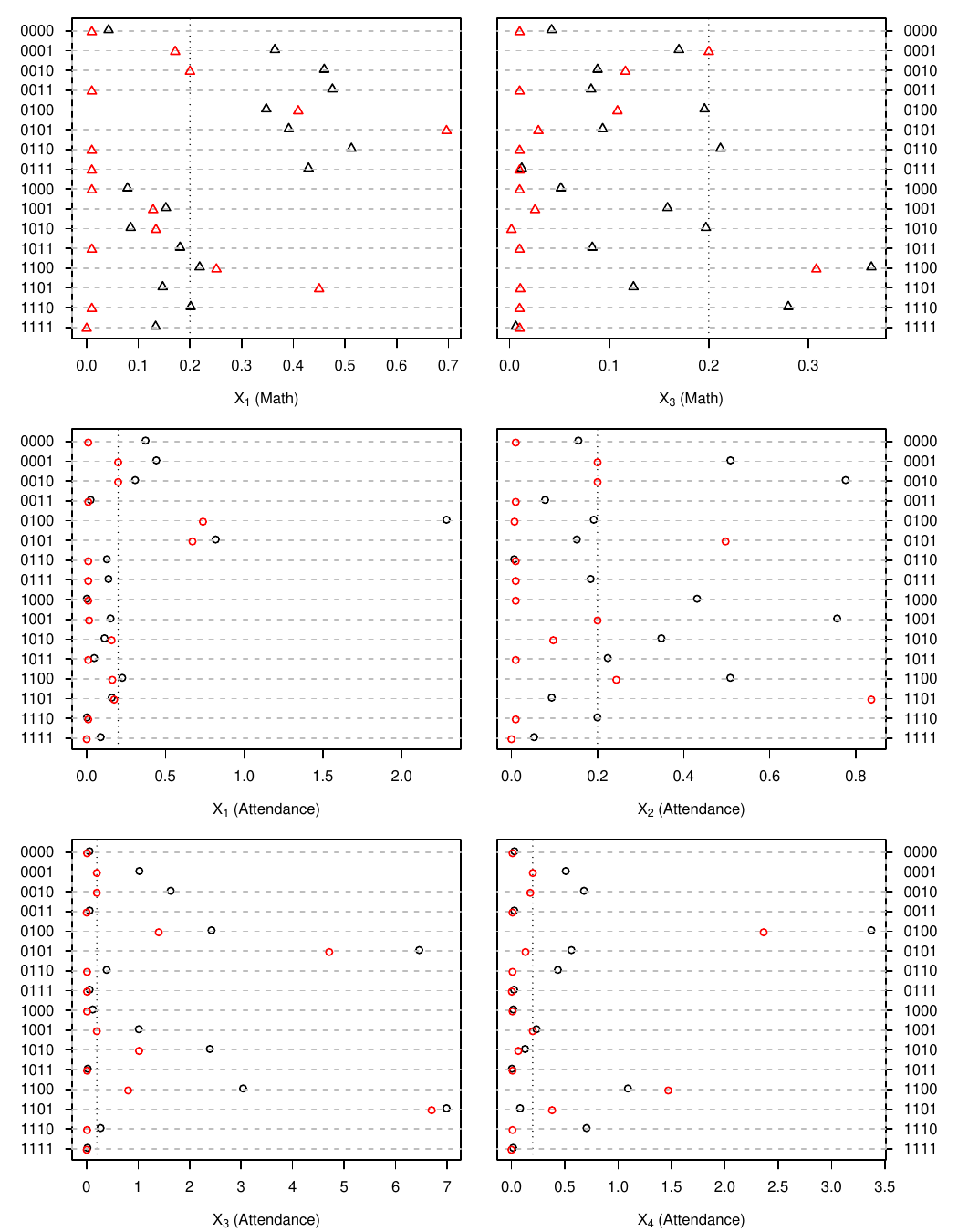} 
\caption{Covariate imbalance of $X_t$ (math scores and school attendance, measured by ASMD, $t = 1,2,3,4$), before (black) and after (red) weighting in the case study. The vertical dotted lines are the 0.2 thresholds.}
\label{fig:asmd_case}
\end{figure}

We solve weighting optimization problems for each of the 16 exposure paths, initially constraining the ASMDs to be below 0.01. For paths with infeasible problems, we incrementally relax the tolerances and reassess the ASMDs before and after weighting. Figure \ref{fig:asmd_case} illustrates the covariate imbalances after weighting, and most paths show manageable imbalances. The exceptions are paths with very few observations ($n_{0100} + n_{0101} + n_{1010} + n_{1100} + n_{1101} < \sqrt{n}$), where some residual imbalance remains.

Using the obtained weights, we estimate $\operatorname{E}_{\mathcal P}[Y(\bar{z}_4)]$ with $\operatorname{E}_{\mathcal S_{\omega}}[Y|\bar{z}_4]$, as reported in Table \ref{tab::effect_est}. 
To address certain policy-relevant questions, we fit the MSM $\text{msm}(\bar{z}_4;\boldsymbol{\tau})=\tau_0+\tau_1z_1+\tau_2(1\{z_2\neq z_1\}+1\{z_3\neq z_2\}+1\{z_4\neq z_3\}$ for language score and math score, providing the estimates $\operatorname{E}_{\text{msm}}[Y(\bar{z}_4)]$. 
Specifically, $\tau_0 = \operatorname{E}_{\mathcal P}[Y(\bar{0}_4)]$ represents the mean potential outcome if students remain in public schools for four years, $\tau_1 + \tau_0 = \operatorname{E}_{\mathcal P}[Y(\bar{1}_4)]$ is that of private schools, and $\tau_2$ is the impact of changing school type from the previous year. 
We also report other model specifications and their adjusted 
$R^2$ values. 
Our findings indicate that remaining in public schools for all four years corresponds to average language and math scores of 446.77 (95\% CI: [440.95, 452.58]) and 447.54 (95\% CI: [441.70, 453.38]), respectively. Attending private schools for the same period increases the language and math scores by 30.68 (95\% CI: [24.62, 36.73]) and 31.89 (95\% CI: [25.81, 37.97]). Changing school type once decreases language and math scores by 27.20 (95\% CI: [22.49, 31.90]) and 29.51 (95\% CI: [24.72, 34.29]), with these effects accumulating with each switch. 
Overall, these results suggest that while entering private schools improves academic performance compared to public schools, switching school types during secondary education has significant academic disadvantages.

\begin{table}[!ht]
\caption{Estimates of mean potential outcomes (language score $L$ and math score $M$). ${\mathrm{E}}_{\mathcal S}[L|\bar{z}_4]$ and ${\mathrm{E}}_{\mathcal S}[M|\bar{z}_4]$ are sample averages. ${\mathrm{E}}_{\mathcal{S}_\omega}[L|\bar{z}_4]$ and ${\mathrm{E}}_{\mathcal{S}_\omega}[M|\bar{z}_4]$ are BAO estimates by fitting saturated MSMs.
${\operatorname{E}}_{\text{msm}}[L(\bar{z}_4)]$ and ${\operatorname{E}}_{\text{msm}}[M(\bar{z}_4)]$ are BAO estimates by fitting non-saturated MSMs, $\text{msm}(\bar{z}_4;\boldsymbol{\tau}) = \tau_0+\tau_1z_1+\tau_2(1\{z_2\neq z_1\}+1\{z_3\neq z_2\}+1\{z_4\neq z_3\})$.}\label{tab::effect_est}
\centering
\footnotesize
\setlength{\tabcolsep}{5pt}
\bgroup
\def\arraystretch{1.15}
\begin{tabular}{lr@{\hspace{0.75\tabcolsep}}r@{\hspace{1.5\tabcolsep}}r@{\hspace{0.75\tabcolsep}}r@{\hspace{1.5\tabcolsep}}r@{\hspace{0.75\tabcolsep}}r@{\hspace{1.5\tabcolsep}}r@{\hspace{0.75\tabcolsep}}r@{\hspace{1.5\tabcolsep}}r@{\hspace{0.75\tabcolsep}}r@{\hspace{1.5\tabcolsep}}r@{\hspace{0.75\tabcolsep}}r@{\hspace{1.5\tabcolsep}}r@{\hspace{0.75\tabcolsep}}}
  \hline
   \multicolumn{1}{l}{$\bar{z}_4$ ($n_{\bar{z}_4}$)} & \multicolumn{2}{c}{${\mathrm{E}}_{\mathcal S}[L|\bar{z}_4]$} & \multicolumn{2}{c}{${\mathrm{E}}_{\mathcal S}[M|\bar{z}_4]$} & \multicolumn{2}{c}{${\mathrm{E}}_{\mathcal{S}_\omega}[L|\bar{z}_4]$} & \multicolumn{2}{c}{${\mathrm{E}}_{\mathcal{S}_\omega}[M|\bar{z}_4]$} & \multicolumn{2}{c}{${\operatorname{E}}_{\text{msm}}[L(\bar{z}_4)]$} & \multicolumn{2}{c}{${\operatorname{E}}_{\text{msm}}[M(\bar{z}_4)]$}\\
\cmidrule(lr){2-3} \cmidrule(lr){4-5}  \cmidrule(lr){6-7} \cmidrule(lr){8-9} \cmidrule(lr){10-11} \cmidrule(lr){12-13} 
    & Est. & 95\% CI.  & Est. & 95\% CI. & Est. & 95\% CI. & Est. & 95\% CI. & Est. & 95\% CI. & Est. & 95\% CI.\\ 
\hline
  0000 (2669) & 450 & [443, 457] & 453 & [446, 460] & 452 & [445, 459] & 454 & [446, 461] & 447 & [441, 453] & 448 & [442, 453] \\ 
  0001 (147) & 386 & [356, 417] & 384 & [354, 415] & 405 & [368, 442] & 397 & [362, 432] & 420 & [414, 425] & 418 & [412, 424] \\ 
  0010 (40) & 395 & [336, 455] & 375 & [314, 435] & 455 & [390, 521] & 421 & [352, 490] & 392 & [384, 401] & 389 & [380, 397] \\ 
  0011 (1025) & 367 & [356, 378] & 367 & [356, 378] & 410 & [393, 427] & 410 & [394, 427] & 420 & [414, 425] & 418 & [412, 424] \\
  0100 (3) & 288 & [19, 557] & 335 & [64, 606] & 362 & [147, 576] & 423 & [263, 584] & 392 & [384, 401] & 389 & [380, 397] \\ 
  0101 (18) & 436 & [376, 497] & 395 & [318, 471] & 427 & [379, 476] & 454 & [367, 541] & 365 & [353, 378] & 359 & [346, 372] \\ 
  0110 (183) & 340 & [314, 366] & 347 & [320, 375] & 404 & [366, 441] & 402 & [364, 440] & 392 & [384, 401] & 389 & [380, 397] \\ 
  0111 (3291) & 388 & [382, 394] & 390 & [384, 396] & 417 & [409, 425] & 415 & [407, 423] & 420 & [414, 425] & 418 & [412, 424] \\ 
  1000 (1288) & 467 & [457, 476] & 463 & [453, 473] & 461 & [450, 472] & 463 & [452, 474] & 450 & [446, 455] & 450 & [445, 455] \\ 
  1001 (96) & 395 & [357, 433] & 389 & [351, 427] & 432 & [392, 472] & 427 & [391, 463] & 423 & [414, 432] & 420 & [411, 430] \\ 
  1010 (33) & 430 & [366, 495] & 403 & [344, 463] & 395 & [282, 507] & 383 & [280, 485] & 396 & [382, 410] & 391 & [377, 405] \\ 
  1011 (903) & 387 & [376, 398] & 386 & [374, 398] & 417 & [404, 430] & 417 & [404, 430] & 423 & [414, 432] & 420 & [411, 430] \\ 
  1100 (2)& 246 & [-95, 588] & 281 & [196, 366] &  91 & [-109, 292] & 243 & [193, 292] & 450 & [446, 455] & 450 & [445, 455] \\ 
  1101 (27) & 473 & [421, 525] & 468 & [416, 520] & 481 & [450, 512] & 416 & [380, 452] & 423 & [414, 432] & 420 & [411, 430] \\ 
  1110 (401) & 407 & [389, 425] & 403 & [386, 421] & 449 & [425, 472] & 445 & [421, 470] & 450 & [446, 455] & 450 & [445, 455] \\ 
  1111 (17617) & 488 & [486, 490] & 489 & [487, 492] & 477 & [475, 479] & 479 & [476, 481] & 477 & [475, 480] & 479 & [477, 482] \\ 
\hline
\multicolumn{12}{l}{Models (Regressing ${\operatorname{E}}_{\mathcal{S}_\omega}[Y|\bar{z}_4]$ on $\bar{z}_4$. Significance codes:  0 `***' 0.001 `**' 0.01 `*' 0.05 `.' 0.1 ` ' 1)} & \multicolumn{1}{r}{Adj. $R^2$}\\
\hline
\multicolumn{12}{l}{msm 1 (L): $\operatorname{E}_{\mathcal{S}_\omega}[L|\bar{z}_4]=431^{***}+9^{.}(z_1+z_2+z_3+z_4)$} & \multicolumn{1}{r}{0.182} \\
\multicolumn{12}{l}{msm 1 (M): $\operatorname{E}_{\mathcal{S}_\omega}[M|\bar{z}_4] =431^{***}+10^{.}(z_1+z_2+z_3+z_4)$} & \multicolumn{1}{r}{0.173} \\
\multicolumn{12}{l}{msm 2 (L): $\operatorname{E}_{\mathcal{S}_\omega}[L|\bar{z}_4]=438^{***}+43^{***}z_1+35^{*}z_2-45^{.}z_3+3z_4$} & \multicolumn{1}{r}{0.691} \\
\multicolumn{12}{l}{msm 2 (M): $\operatorname{E}_{\mathcal{S}_\omega}[M|\bar{z}_4]=438^{***}+45^{***}z_1+36^{*}z_2-47^{.}z_3+3z_4$} & \multicolumn{1}{r}{0.676} \\
\multicolumn{12}{l}{msm 3 (L): $\operatorname{E}_{\mathcal{S}_\omega}[L|\bar{z}_4]=425^{***}+27^{*}1\{\bar{z}_4 = \bar{0}_4\}+51^{***}1\{\bar{z}_4 = \bar{1}_4\}$} & \multicolumn{1}{r}{0.801} \\
\multicolumn{12}{l}{msm 3 (M): $\operatorname{E}_{\mathcal{S}_\omega}[M|\bar{z}_4] =424^{***}+30^{*}1\{\bar{z}_4 = \bar{0}_4\}+55^{***}1\{\bar{z}_4 = \bar{1}_4\}$} & \multicolumn{1}{r}{0.817} \\
\multicolumn{12}{l}{msm 4 (L): 
 $\operatorname{E}_{\mathcal{S}_\omega}[L|\bar{z}_4]=447^{***}+31^{***}z_1-25^{***}1\{z_2\neq z_1\}-33^{***}1\{z_3\neq z_2\}-21^{*}1\{z_4\neq z_3\}$} & \multicolumn{1}{r}{0.927} \\
\multicolumn{12}{l}{msm 4 (M): $\operatorname{E}_{\mathcal{S}_\omega}[M|\bar{z}_4]=447^{***}+32^{***}z_1-27^{***}1\{z_2\neq z_1\}-35^{***}1\{z_3\neq z_2\}-29^{**}1\{z_4\neq z_3\}$} & \multicolumn{1}{r}{0.944} \\
\multicolumn{12}{l}{msm 5 (L): $\operatorname{E}_{\mathcal{S}_\omega}[L|\bar{z}_4]=447^{***}+31^{***}z_1-27^{***}(1\{z_2\neq z_1\}+1\{z_3\neq z_2\}+1\{z_4\neq z_3\})$} & \multicolumn{1}{r}{0.931} \\
\multicolumn{12}{l}{msm 5 (M): $\operatorname{E}_{\mathcal{S}_\omega}[M|\bar{z}_4]=448^{***}+32^{***}z_1-30^{***}(1\{z_2\neq z_1\}+1\{z_3\neq z_2\}+1\{z_4\neq z_3\})$} & \multicolumn{1}{r}{0.949} \\
\hline
  \end{tabular}
  \egroup
\end{table}

\section{Summary and remarks}\label{Discuss}

Estimating the effects of time-varying treatments is a central challenge in longitudinal studies. Weighting approaches are widely used to adjust for observed time-varying confounders, but their performance is often compromised by weight instability when there is poor covariate overlap.
This limitation has motivated the development of methods that improve finite-sample performance by explicitly regularizing weights and enforcing balance.

In this vein, this paper introduces balancing by adaptive orthogonalization (BAO), a novel weighting approach for estimating mean potential outcomes and treatment effects. BAO works by balancing the components of covariates that are orthogonal to their histories, thereby isolating new information available at each time point. This approach provides an implementation of sufficient covariate balance conditions under a semi-parametric outcome model. Specifically, BAO constructs stable, nonnegative weights by solving constrained quadratic programming problems, thereby directly balancing covariates across time without requiring extrapolation. This approach achieves efficiency comparable to the g-computation formula.
Importantly, rather than relying on a correct outcome model to specify its balance conditions, BAO attains robustness by balancing specified functions of the covariates while ensuring the remainder is balanced in probability as the weights converge to true inverse propensities. 

Our simulation studies confirm these theoretical advantages. Under correct model specification, BAO matches the low RMSE of the g-computation formula. Under outcome model misspecification, it exhibits robustness comparable to that of TMLE. Beyond simulations, we applied BAO to a central policy question, estimating the effects of school choice on student achievement in Chile. Our findings revealed a significant negative impact of switching school types on academic performance, providing actionable evidence for educational policy.

While BAO offers a robust and efficient framework, our work highlights exciting directions for future research. First, the current implementation relies on projection models for covariates, a requirement that could be relaxed by imputing and balancing mean potential covariates, as noted in Remark \ref{rem::condition}. Second, we have focused on continuous outcomes, yet extending BAO to handle other outcomes is a key priority. Future work will therefore focus on extending the approach to survival outcomes and continuous treatments, as well as developing scalable solutions for studies with longer time horizons or multilevel treatment structures.

%\section{Summary and remarks}
%\vspace{1cm}
%\pagebreak
% \onehalfspacing
\singlespacing
\bibliographystyle{asa}
\bibliography{msm24/ref}
%\nobibliography{append}

%%%%%%%%%%%%%%%%%%%%
%%%%%%%%%%%%%%%%%%%%
%%%%%%%%%%%%%%%%%%%%
% \clearpage
% \appendix
% \section*{Supplementary Materials} 

% \input{sec_supp}

%\bibliographystyle{asa}

\end{document}

\pagestyle{plain}

\newtheoremstyle{mystyle}% name
{\topsep}% Space above
{\topsep}% Space below
{\it}% Body font
{}% Indent amount
{\bf}% Theorem head font
{.}%Punctuation after theorem head
{.5em}%Space after theorem head
{}% theorem head spec

\renewcommand{\arraystretch}{1.3}

\newcommand{\argmin}{\mathop{\mathrm{argmin}}}
\makeatletter
\newcommand{\grande}{\bBigg@{2.25}}
\newcommand{\enorme}{\bBigg@{5}}

\newcommand{\blind}{0}

\newcommand{\tit}{Supplementary Materials for\\Adaptive Orthogonalization for Stable Estimation of the Effects of Time-Varying Treatments}

\if0\blind

{\title{\tit
\thanks{The authors thank Zach Branson, Ambarish Chattopadhyay, Eric Cohn, Larry Han, Jie Hu, Kwangho Kim, Kosuke Imai, Marshall Joffe, Bijan Niknam, Marc Ratkovic, Jamie Robins, Zhu Shen, and Yi Zhang for helpful comments and suggestions. This work was partially supported by the Alfred P. Sloan Foundation (G-2020-13946) and the Patient-Centered Outcomes Research
Initiative (PCORI, ME-2022C1-25648).
}\vspace*{.3in}}
\author{
Yige Li 
\thanks{Department of Health Care Policy, Harvard University.}	   
\and
Mar\'ia de los Angeles Resa 
\thanks{Department of Statistics, Columbia University.}
\and 
Jos\'e R. Zubizarreta$^{*}$  
\thanks{Department of Health Care Policy, Biostatistics, and Statistics, and CAUSALab, Harvard University.
Email: \url{zubizarreta@hcp.med.harvard.edu}.} 
}

\date{} 

\maketitle
}\fi

\if1\blind
\title{\tit}
\date{} 
\maketitle
\fi

\begin{abstract}
The Supplementary Materials provide additional arguments and results omitted from the main text. Section \ref{sec::methods} includes further details on the proposed method, and Section \ref{sec::extensions} contains an extension for the censoring setting. Additional figures and tables for the simulation studies and the case study are provided in Section \ref{sec::simulation} and Section \ref{sec::casestudy}, respectively.
\end{abstract}

\clearpage

\singlespacing
\pagebreak
\thispagestyle{empty} 
\tableofcontents
\pagebreak
\doublespacing

\setcounter{page}{2}

%%%%%%%%%%%%%%%%%%%%%%%%%%%%%%%%%%%%%%%%%%%
%%%%%%%%%%%%%%%%%%%%%%%%%%%%%%%%%%%%%%%%%%%
%%%%%%%%%%%%%%%%%%%%%%%%%%%%%%%%%%%%%%%%%%%

\section{Methods}\label{sec::methods}

\subsection{Identification and potential covariate distribution representation}\label{sec::identification}

As shown in Section 3.1 of the main text, Assumptions 1–3 allow for the nonparametric identification of the mean potential outcome $\mathrm{E}_{\mathcal P}[Y(\bar{z}_T)]$. This can be expressed in three equivalent forms: as the g-formula iterated integral \eqref{eqn::IdByG}, as an inverse propensity score weighted integral \eqref{eqn::IdByPS}, or as an integral over potential covariates \eqref{eqn::IdByBC}. The derivation is as follows.
\begin{eqnarray}
&& \mathrm{E}_{\mathcal P}[Y(\bar{z}_T)]  \nonumber \\
&=& \int \mathrm{E}_{\mathcal P}[Y(\bar{z}_T) | \bar{X}_T, \bar{z}_T] f_{\mathcal P}(X_T| \bar{X}_{T-1}, \bar{Z}_{T-1} = \bar{z}_{T-1})\cdots f_{\mathcal P}(X_2|X_1,Z_1 = z_1)f_{\mathcal P}(X_1)d\bar{X}_T \nonumber \\
&&\text{\small (By Assumption 1a)} \nonumber \\
&=& \int \mathrm{E}_{\mathcal P}[Y | \bar{X}_T, \bar{z}_T] f_{\mathcal P}(X_T| \bar{X}_{T-1}, \bar{Z}_{T-1} = \bar{z}_{T-1})\cdots f_{\mathcal P}(X_2|X_1,Z_1 = z_1)f_{\mathcal P}(X_1)d\bar{X}_T \label{eqn::IdByG}\\
&& \text{\small (By Assumption 3a; g-formula identification)} \nonumber \\
&=& \int Y \frac{1\{\bar{Z}_T = \bar{z}_T\}}{\prod_{t = 1}^T \mathrm{P}(Z_t= z_t|\bar{X}_t, \bar{Z}_{t-1})}f_{\mathcal P}(Y| \bar{X}_T, \bar{Z}_T) f_{\mathcal P}(\bar{X}_T, \bar{Z}_{T})dYd\bar{Z}_Td\bar{X}_T \label{eqn::IdByPS}\\
&&\text{\small (By Assumption 2; IPW identification)} \nonumber \\
&=& \int\mathrm{E}_{\mathcal P}[Y(\bar{z}_T)| \bar{X}_T=\bar{x}_T,  \bar{z}_T]f_{\mathcal P}(X_T(\bar{z}_{T-1}) = x_T | \bar{X}_{T-1}(\bar{z}_{T-2}) = \bar{x}_{T-1}, \bar{Z}_{T-1} = \bar{z}_{T-1})\cdot \nonumber \\
&& f_{\mathcal P}(X_{T-1}(\bar{z}_{T-2}) = x_{T-1} | \bar{X}_{T-2}(\bar{z}_{T-3}) = \bar{x}_{T-2},\bar{Z}_{T-2} = \bar{z}_{T-2})\cdots f_{\mathcal P}(X_1 = x_1)d\bar{x}_T \nonumber \\
&& \text{\small (By g-formula identification and Assumption 3b)} \nonumber \\
&=& \int \mathrm{E}_{\mathcal P}[Y(\bar{z}_T)| \bar{X}_T=\bar{x}_T,  \bar{z}_T]f_{\mathcal P}(X_T(\bar{z}_{T-1}) = x_T | \bar{X}_{T-1}(\bar{z}_{T-2}) = \bar{x}_{T-1},\bar{Z}_{T-2} = \bar{z}_{T-2})\cdot  \nonumber\\
&& f_{\mathcal P}(X_{T-1}(\bar{z}_{T-2}) = x_{T-1} | \bar{X}_{T-2}(\bar{z}_{T-3}) = \bar{x}_{T-2},\bar{Z}_{T-2} = \bar{z}_{T-2})\cdots f_{\mathcal P}(X_1 = x_1)d\bar{x}_T \nonumber \\
&=& \int \mathrm{E}_{\mathcal P}[Y(\bar{z}_T)| \bar{X}_T(\bar{z}_{T-1}),  \bar{z}_T]f_{\mathcal P}(\underaccent{\bar}{X}_{T-1}(\bar{z}_{T-2})| \bar{X}_{T-2}(\bar{z}_{T-3}),\bar{z}_{T-3})\cdots f_{\mathcal P}(X_1)d\bar{X}_T(\bar{z}_{T-1})\nonumber \\
&& \text{\small (By Assumption 1b)} \nonumber \\
&=& \int \mathrm{E}_{\mathcal P}[Y(\bar{z}_T)| \bar{X}_T(\bar{z}_{T-1}) = \bar{x}_T, \bar{z}_T]f_{\mathcal P}(X_T(\bar{z}_{T-1}) = x_T, \cdots, X_1 = x_1)d\bar{x}_T \label{eqn::IdByBC} \\
&&\text{\small (By Assumptions 1b and 3b; Potential covariate distribution representation)} \nonumber 
\end{eqnarray}

Now we extend from the proof in \citet{imai2015robust} and show that in population, the true inverse propensity score balances any function of covariates and treatments to its target, that is $\mathrm{E}_{\mathcal{P}_{\pi^{-1}}}[\psi\{\bar{X}_T;\bar{Z}_T\}|\bar{z}_T]= \mathrm{E}_{\mathcal P}[\psi\{\bar{X}_T(\bar{z}_{T-1});\bar{z}_T\}]$.
\begin{eqnarray*}
&&\mathrm{E}_{\mathcal{P}_{\pi^{-1}}}[\psi\{\bar{X}_T;\bar{Z}_T\}|\bar{Z}_T = \bar{z}_T]\\
&=& \int \psi\{\bar{X}_T;\bar{z}_T\}\frac{1\{\bar{Z}_{T} = \bar{z}_T\}\mathrm{P}(\bar{Z}_T = \bar{z}_T)}{\prod_{t = 1}^T \mathrm{P}(Z_{t} = z_t|\bar{Z}_{t-1} = \bar{z}_{t-1}, \bar{X}_{t})}f_{\mathcal P}(\bar{X}_T|\bar{Z}_T = \bar{z}_T) d\bar{X}_T \\
&=& \int  \psi\{\bar{X}_T;\bar{z}_T\}\mathrm{P}( \bar{z}_T)\frac{1\{{Z}_{1} = {z}_1\}f_{\mathcal P}(X_1|Z_1 = z_1)}{\frac{f_{\mathcal P}(X_1|Z_1 = z_1)\mathrm P(Z_{1} = z_1)}{f_{\mathcal P}(X_1)}}\cdots\frac{1\{{Z}_{T} = {z}_T\}\frac{f_{\mathcal P}(\bar{X}_T|\bar{Z}_T = \bar{z}_T)}{f_{\mathcal P}(\bar{X}_{T-1}|\bar{Z}_{T-1} = \bar{z}_{T-1})}}{\frac{f_{\mathcal P}(X_T|\bar{z}_T, \bar{X}_{T-1})\mathrm P(Z_{T} = z_T|\bar{z}_{T-1}, \bar{X}_{T-1})}{f_{\mathcal P}(X_T|\bar{Z}_{T-1} = \bar{z}_{T-1}, \bar{X}_{T-1})}}d\bar{X}_T \\
&=& \int \psi\{\bar{X}(\bar{z}_{T-1});\bar{z}_T\} \frac{1\{{Z}_{1} = {z}_1\}\cdots 1\{{Z}_{T} = {z}_T\}\mathrm{P}(\bar{z}_T)}{\mathrm P(Z_{1} = z_1)\cdots\mathrm P(Z_{T} = z_T|\bar{z}_{T-1})}f_{\mathcal P}(X_T(\bar{z}_{T-1}),...,X_1)d\bar{X}_T(\bar{z}_{T-1}) \\
&& \text{\small (By Assumptions 1b and 3b)} \\
&=& \int  \psi\{\bar{X}_T(\bar{z}_{T-1});\bar{z}_T\} f_{\mathcal P}(X_1, X_2(z_1),...,X_T(\bar{z}_{T-1}))d\bar{X}_T(\bar{z}_{T-1})\\
&=& \mathrm{E}_{\mathcal P}[\psi\{\bar{X}_T(\bar{z}_{T-1});\bar{z}_T\}]
\end{eqnarray*}

\subsection{Proposition 1.A: implementation for sufficient balance conditions}

\setcounter{subtheorem}{0}
\begin{subtheorem}\label{prop1} ~
\begin{enumerate}[label=(\Roman*)]
    \item Population balance (identification). 
    Suppose the following mean independence condition holds,
    \begin{equation}\label{eqn::v}
\mathrm{E}_{\mathcal P}[\mathrm{E}_{\mathcal P}[\cdots\mathrm{E}_{\mathcal P}[\mathrm{E}_{\mathcal P}[R_t|\bar{X}_{t-1}, \bar{z}_{t-1}]|\bar{X}_{t-2}, \bar{z}_{t-2}]\cdots|X_1,z_1]] = \mathrm{E}_{\mathcal P}[R_t|\bar{z}_{t-1}], ~t = 2,..., T,
\end{equation}
where $R_t(\bar{z}_{t-1}) := g_t\{X_t(\bar{z}_{t-1})\} - {\beta}_{X_t\sim \bar{X}_{t-1}|z_{t-1}}\bar{g}_{t-1}\{X_{t-1}(\bar{z}_{t-2})\}$ are the residuals at time $t$. If the weights $\omega$ satisfy the following population balance condition
\begin{eqnarray}
\Big\{\mathrm{E}_{\mathcal{P}_{\omega}}[g_1\{X_1\}|\bar{z}_T] = \mathrm{E}_{\mathcal P}[g_1\{X_1\}]; ~\mathrm{E}_{\mathcal{P}_{\omega}}[R_t|\bar{z}_T] =  \mathrm{E}_{\mathcal P}[R_t|\bar{Z}_{t-1} = \bar{z}_{t-1}], ~ t = 2,...,T\Big\}, \label{eqn::balconimp}
\end{eqnarray}
then the balance condition $\big\{\mathrm{E}_{\mathcal{P}_{\omega}}[g_t\{X_t\}|\bar{z}_T] = \mathrm{E}_{\mathcal P}[g_t\{X_t(\bar{z}_{t-1})\}], ~t = 1,..., T\big\}$ holds.

\item Sample balance (estimation). Suppose the variance-covariance matrix of each $g_t\{X_t\}$, $t = 1,...,T$, is finite and non-singular, and the estimator $\widehat{\beta}_{X_t\sim \bar{X}_{t-1}|z_{t-1}}$ is $\sqrt{n}$-consistent for ${\beta}_{X_t\sim \bar{X}_{t-1}|z_{t-1}}$, $t = 2,...,T$, then the following sample balance condition 
\begin{eqnarray}
&& \hspace{-1.2cm}\Big\{\mathrm{E}_{\mathcal{S}_{\omega}}[g_1\{X_1\}|\bar{z}_T] -  \mathrm{E}_{\mathcal S}[g_1\{X_1\}]=o(n^{-\frac 1 2}); \mathrm{E}_{\mathcal{S}_{\omega}}[\widehat{R}_t| \bar{z}_T] -  \mathrm{E}_{\mathcal S}[\widehat{R}_t|\bar{z}_{t-1}]=o(n^{-\frac 1 2}), ~ t = 2,...,T\Big\}, ~\quad
\label{eqn::balconsam}
\end{eqnarray}
where $\widehat{R}_t(\bar{z}_{t-1}) := g_t\{X_t(\bar{z}_{t-1})\} - \widehat{\beta}_{X_t\sim \bar{X}_{t-1}|z_{t-1}}\bar{g}_{t-1}\{X_{t-1}(\bar{z}_{t-2})\}$,
implies Claim \ref{prop1}. 
That is, for $t = 1,...,T$,  $\mathrm {E}_{\mathcal{S}_{\omega}}[g_t\{X_t\} | \bar{z}_T]$ is asymptotically linear and converges to $\mathrm{E}_{\mathcal P}[g_t\{X_t(\bar{z}_{t-1})\}]$.
\end{enumerate}

\begin{proof}
We prove (I) and (II) by mathematical induction. 
\begin{enumerate}[label=(\Roman*)]
    \item For $g_1\{X_1\}$ and $g_2\{X_2\}$,
\begin{eqnarray}
\mathrm{E}_{\mathcal{P}_{\omega}}[g_1\{X_1\} |\bar{z}_T] &=& \mathrm{E}_{\mathcal P}[g_1\{X_1\}]; \quad \text{\small (By the balance condition for $X_1$ in \eqref{eqn::balconimp})}%=: T(X_1)
\nonumber \\\mathrm{E}_{\mathcal{P}_{\omega}}[g_2\{X_2\} |\bar{z}_T]
&=&\mathrm{E}_{\mathcal{P}_{\omega}}[g_2\{X_2\} - \beta_{X_2\sim X_1|z_1} g_1\{X_1\}|\bar{z}_T] + \beta_{X_2\sim X_1|z_1}\mathrm{E}_{\mathcal{P}_{\omega}}[g_1\{X_1\} |\bar{z}_T] \nonumber \\
&=&\mathrm{E}_{\mathcal P}[g_2\{X_2\} -  \beta_{X_2\sim X_1|z_1} g_1\{X_1\} |Z_1 = z_1] + \beta_{X_2\sim X_1|z_1}\mathrm{E}_{\mathcal P}[g_1\{X_1\}]
\nonumber \\
&& \text{\small (By the balance condition for $X_1, X_2$ in \eqref{eqn::balconimp})} \nonumber \\
&=&\mathrm{E}_{\mathcal P}[g_2\{X_2(z_1)\} -  \beta_{X_2\sim X_1|z_1}g_1\{X_1\} |Z_1 =  z_1] + \beta_{X_2\sim X_1|z_1}  \mathrm{E}_{\mathcal P}[g_1\{X_1\}]\nonumber \\
&& \text{\small (By SUTVA)} \nonumber \\
&=&\mathrm{E}_{\mathcal P}[R_2(z_1)| Z_1 = z_1] + \beta_{X_2\sim X_1|z_1}\mathrm{E}_{\mathcal P}[g_1\{X_1\}]  \nonumber \\
&& \text{\small (By the definition $R_2(z_1) := g_2\{X_2(z_1)\} -  \beta_{X_2\sim X_1|z_1}g_1\{X_1\}$)} \nonumber \\
&=&\mathrm{E}_{\mathcal P}[R_2(z_1)] + \beta_{X_2\sim X_1|z_1}\mathrm{E}_{\mathcal P}[g_1\{X_1\}]
\nonumber \\
&& \text{\small (By the condition $\mathrm{E}_{\mathcal P}[R_2(z_1)| z_1] = \mathrm{E}_{\mathcal P}[R_2(z_1)]$, which is proved next)} \nonumber \\
&=&\mathrm{E}_{\mathcal P}[g_2\{X_2(z_1)\} -  \beta_{X_2\sim X_1|z_1}g_1\{X_1\}] + \beta_{X_2\sim X_1|z_1}\mathrm{E}_{\mathcal P}[g_1\{X_1\}]
\nonumber \\
&& \text{\small (By the definition $R_2(z_1) := g_2\{X_2(z_1)\} -  \beta_{X_2\sim X_1|z_1}g_1\{X_1\}$)} \nonumber \\
&=&\mathrm{E}_{\mathcal P}[g_2\{X_2(z_1)\}].
\nonumber 
\end{eqnarray}
The condition $\mathrm{E}_{\mathcal P}[R_2(z_1)| z_1] = \mathrm{E}_{\mathcal P}[R_2(z_1)]$ in the above proof holds under the mean independence condition \eqref{eqn::v} and Assumption 1b, because
\begin{eqnarray*}
\mathrm{E}_{\mathcal P}[R_2(z_1)] &=& \mathrm{E}_{X_1, X_2}[g_2\{X_2(z_1)\}] -\mathrm{E}_{X_1,X_2}[\beta_{X_2\sim X_1|z_1}g_1\{X_1\}] \\
&=& \mathrm{E}_{X_1}[\mathrm{E}_{X_2}[g_2\{X_2\}|X_1, z_1]]- \mathrm{E}_{X_1}[\beta_{X_2\sim X_1|z_1}g_1\{X_1\}] \\
&& \text{\small (By Assumption 1b $X_2(z_1) \ci Z_1|X_1$)}\\
&=& \mathrm{E}_{X_1}[\mathrm{E}_{X_2}[g_2\{X_2\} - \beta_{X_2\sim X_1|z_1}g_1\{X_1\}|X_1, z_1]] \\
&=& \mathrm{E}_{X_1}[\mathrm{E}_{X_2}[R_2|X_1, z_1]] \\
&=& \mathrm{E}_{\mathcal P}[R_2|z_1].\\
&& \text{\small (By mean independence conditions $\mathrm{E}_{X_1}[\mathrm{E}_{X_2}[R_2|X_1, z_1]] = \mathrm{E}_{\mathcal P}[R_2 | z_1]$ in \eqref{eqn::v})}
\end{eqnarray*}
Hence, we have verified for $g_1\{X_1\}$ and $g_2\{X_2\}$. If assuming the statement holds for $g_1\{X_1\}$, ..., $g_{t-1}\{X_{t-1}\}$, by showing it also true for $g_t\{X_t\}$, we can complete the proof.

For $g_t\{X_t\}$,
\begin{eqnarray}
&& \mathrm{E}_{\mathcal{P}_{\omega}}[g_t\{X_t\}|\bar{Z}_T = \bar{z}_T] \nonumber \\
&=& \mathrm{E}_{\mathcal{P}_{\omega}}[g_t\{X_t\} - \beta_{X_t\sim \bar{X}_{t-1}|z_{t-1}}\bar{g}_{t-1}\{X_{t-1}\}+\beta_{X_t\sim \bar{X}_{t-1}|z_{t-1}}\bar{g}_{t-1}\{X_{t-1}\}|\bar{Z}_T = \bar{z}_T]\nonumber \\
&=&\mathrm{E}_{\mathcal P}[g_t\{X_t\} - \beta_{X_t\sim \bar{X}_{t-1}|z_{t-1}}\bar{g}_{t-1}\{X_{t-1}\}| \bar{Z}_{t-1} = \bar{z}_{t-1}] + \mathrm{E}_{\mathcal P}[\beta_{X_t\sim \bar{X}_{t-1}|z_{t-1}}\bar{g}_{t-1}\{X_{t-1}(\bar{z}_{t-2})\}] \nonumber \\
&& \text{\small (By the balance condition for $X_t$ in \eqref{eqn::balconimp} and  assuming the same statement true for $1,...,t-1$)} \nonumber \\
&=&\mathrm{E}_{\mathcal P}[g_t\{X_t(\bar{z}_{t-1})\} - \beta_{X_t\sim \bar{X}_{t-1}|z_{t-1}}\bar{g}_{t-1}\{{X}_{t-1}(\bar{z}_{t-2})\}|\bar{z}_{t-1}] + \mathrm{E}_{\mathcal P}[\beta_{X_t\sim \bar{X}_{t-1}|z_{t-1}}\bar{g}_{t-1}\{X_{t-1}(\bar{z}_{t-2})\}] \nonumber \\
&=&\mathrm{E}_{\mathcal P}[g_t\{X_t(\bar{z}_{t-1})\} - \beta_{X_t\sim \bar{X}_{t-1}|z_{t-1}}\bar{g}_{t-1}\{{X}_{t-1}(\bar{z}_{t-2})\}] + \mathrm{E}_{\mathcal P}[\beta_{X_t\sim \bar{X}_{t-1}|z_{t-1}}\bar{g}_{t-1}\{X_{t-1}(\bar{z}_{t-2})\}] \nonumber \\
&& \text{\small (By SUTVA, $R_t(\bar{z}_{t-1}) := g_t\{X_t(\bar{z}_{t-1})\} -  \beta_{X_t\sim \bar{X}_{t-1}|z_{t-1}}\bar{g}_{t-1}\{X_{t-1}(\bar{z}_{t-2})\}$, } \nonumber\\
&& \text{\small and $\mathrm{E}_{\mathcal P}[R_t(\bar{z}_{t-1})| \bar{z}_{t-1}] = \mathrm{E}_{\mathcal P}[R_t(\bar{z}_{t-1})]$)} \nonumber \\
&=&\mathrm{E}_{\mathcal P}[g_t\{X_t(\bar{z}_{t-1})\}].  \nonumber 
\end{eqnarray}
The condition $\mathrm{E}_{\mathcal P}[R_t(\bar{z}_{t-1})] = \mathrm{E}_{\mathcal P}[R_t(\bar{z}_{t-1}) |\bar{z}_{t-1}]$ holds because 
\begin{eqnarray*}
&&\mathrm{E}_{\mathcal P}[R_t(\bar{z}_{t-1})] \\
&=& \mathrm{E}_{\bar{X}_t}[g_t\{X_t(\bar{z}_{t-1})\} -  \beta_{X_t\sim \bar{X}_{t-1}|z_{t-1}}\bar{g}_{t-1}\{X_{t-1}(\bar{z}_{t-2})\}] \\
&=& \mathrm{E}_{\bar{X}_t}[g_t\{X_t(\bar{z}_{t-1})\}] -  \mathrm{E}_{\bar{X}_t}[\beta_{X_t\sim \bar{X}_{t-1}|z_{t-1}}\bar{g}_{t-1}\{X_{t-1}(\bar{z}_{t-2})\}] \\
&=& \mathrm{E}_{X_1}[\mathrm{E}_{X_2}[\cdots\mathrm{E}_{X_{t-1}}[\mathrm{E}_{X_t}[g_t\{X_t\}|\bar{X}_{t-1}, \bar{z}_{t-1}]|\bar{X}_{t-2}, \bar{z}_{t-2}]\cdots|X_1,z_1]] - \\
&& \mathrm{E}_{X_1}[\mathrm{E}_{X_2}[\cdots\mathrm{E}_{X_{t-1}}[\beta_{X_t\sim \bar{X}_{t-1}|z_{t-1}}\bar{g}_{t-1}\{X_{t-1}\}|\bar{X}_{t-2}, \bar{z}_{t-2}]\cdots|X_1,z_1]] \\
&& \text{\small (By Assumption 1b $X_k(\bar{z}_{k-1}) \ci Z_{k-1}|(\bar{X}_{k-1},\bar{Z}_{k-2})$, $k = 2, ..., t$)}\\
&=& \mathrm{E}_{X_1}[\mathrm{E}_{X_2}[\cdots\mathrm{E}_{X_{t-1}}[\mathrm{E}_{X_t}[R_t|\bar{X}_{t-1}, \bar{z}_{t-1}]|\bar{X}_{t-2}, \bar{z}_{t-2}]\cdots|X_1,z_1]] \\
&=& \mathrm{E}_{\mathcal P}[R_t|\bar{z}_{t-1}]. \\
&& \text{\small  (By $\mathrm{E}_{X_1}[\mathrm{E}_{X_2}[\cdots\mathrm{E}_{X_{t-1}}[\mathrm{E}_{X_t}[R_t|\bar{X}_{t-1}, \bar{z}_{t-1}]|\bar{X}_{t-2}, \bar{z}_{t-2}]\cdots|X_1,z_1]] = \mathrm{E}_{\mathcal P}[R_t|\bar{z}_{t-1}]$ in \eqref{eqn::v})}
 \end{eqnarray*}
So, the statement for $g_t\{X_t\}$ is true. Therefore, we can complete the proof.

\item Now we prove convergence in probability with rate $1/\sqrt{n}$ for the weighted means of covariates, that is $\sqrt{n}(\mathrm{E}_{\mathcal{S}_{\omega}}[(g_1\{X_1\},..., g_T\{X_T\}) | \bar{z}_T] - \mathrm{E}_{\mathcal P}[(g_1\{X_1\},..., g_T\{X_T(\bar{z}_{T-1})\})]) = O_p(1)$, and meanwhile show $\mathrm{E}_{\mathcal{S}_{\omega}}[(g_1\{X_1\},..., g_T\{X_T\}) | \bar{Z}_T = \bar{z}_T]$ is a regular and asymptotically
linear estimator by mathematical induction.

For $g_1\{X_1\}$, by the corresponding balance condition in \eqref{eqn::balconsam}, 
\begin{eqnarray*}
&& \sqrt{n}(\mathrm{E}_{\mathcal{S}_{\omega}}[g_1\{X_1\}| \bar{Z}_T = \bar{z}_T] - \mathrm{E}_{\mathcal P}[g_1\{X_1\}]) \\
&=&  \sqrt{n}(\mathrm{E}_{\mathcal S}[g_1\{X_1\}] + o(n^{-\frac 1 2})- \mathrm{E}_{\mathcal P}[g_1\{X_1\}]) \\
&=& \frac{\sqrt n}{n}\sum_{i= 1}^n (g_1\{X_{i1}\}-\mathrm{E}_{\mathcal P}[g_1\{X_1\}])+o(1). 
\end{eqnarray*}
Since WLOG we can let $\mathrm{E}_{\mathcal P}[(g_1\{X_1\}-\mathrm{E}_{\mathcal P}[g_1\{X_1\}])(g_1\{X_1\}-\mathrm{E}_{\mathcal P}[g_1\{X_1\}])^\top]$ be finite and nonsingular, $\mathrm{E}_{\mathcal{S}_{\omega}}[g_1\{X_1\}| \bar{Z}_T = \bar{z}_T]$ is a regular and asymptotically linear (RAL) estimator. 
Therefore, by the Central Limit Theorem, $\mathrm{E}_{\mathcal{S}_{\omega}}[g_1\{X_1\}| \bar{Z}_T = \bar{z}_T]$ is $\sqrt n$-consistent and asymptotically Gaussian. The variance $V_1 := \text{Var}(\mathrm{E}_{\mathcal{S}_{\omega}}[g_1\{X_1\}| \bar{Z}_T = \bar{z}_T]) = \frac{\text{Var}(g_1\{X_1\})}{n}$.

For $g_2\{X_2\}$,
\begin{eqnarray*}
&&\sqrt{n}(\mathrm{E}_{\mathcal{S}_{\omega}}[g_2\{X_2\}| \bar{Z}_T = \bar{z}_T] - \mathrm{E}_{\mathcal P}[g_2\{X_2(z_1)\}]) \\
&=& \sqrt{n}(\mathrm{E}_{\mathcal{S}_{\omega}}[\widehat{R}_2(z_1)|\bar{z}_T] - \mathrm{E}_{\mathcal P}[{R}_2(z_1)] +  \mathrm{E}_{\mathcal{S}_{\omega}}[\widehat{\beta}_{X_2}{g}_1\{X_1\}| \bar{z}_T]  - \mathrm{E}_{\mathcal P}[{\beta}_{X_2}{g}_1\{X_1\}]) \\
&& \text{\small (Abbreviate ${\beta}_{X_2\sim X_1|z_1}$ as ${\beta}_{X_2}$; $g_2\{X_2(z_1)\} = {\beta}_{X_2}{g}_1\{X_1\}+{R}_2(z_1)= \widehat{\beta}_{X_2}{g}_1\{X_1\}+\widehat{R}_2(z_1)$)}\\
&=& \sqrt{n}(\mathrm{E}_{\mathcal S}[\widehat{R}_2(z_1)|z_1]+ o(n^{-\frac 1 2}) - \mathrm{E}_{\mathcal P}[{R}_2(z_1)]) +  \sqrt{n}{\beta}_{X_2}(\mathrm{E}_{\mathcal{S}_{\omega}}[{g}_1\{X_1\}|\bar{z}_T]  - \mathrm{E}_{\mathcal P}[{g}_1\{X_1\}]) +\\
&&\sqrt{n}(\widehat{\beta}_{X_2}-\beta_{X_2})\mathrm{E}_{\mathcal{S}_{\omega}}[{g}_1\{X_1\}|\bar{z}_T] \\
&& \text{\small (By the implemented balance condition for $\widehat{R}_2$ in \eqref{eqn::balconsam})}\\
&=& \sqrt{n}(\mathrm{E}_{\mathcal S}[\widehat{R}_2(z_1)|z_1] + o(n^{-\frac 1 2})- \mathrm{E}_{\mathcal P}[{R}_2|z_1]) + \sqrt{n} {\beta}_{X_2}(\mathrm{E}_{\mathcal{S}_{\omega}}[{g}_1\{X_1\}|\bar{z}_T]  - \mathrm{E}_{\mathcal P}[{g}_1\{X_1\}]) +\\
&& \sqrt{n}(\widehat{\beta}_{X_2}-\beta_{X_2})\mathrm{E}_{\mathcal{S}_{\omega}}[{g}_1\{X_1\}|\bar{z}_T] \\
&& \text{\small (By the mean independence condition $\mathrm{E}_{\mathcal P}[{R}_2(z_1)] 
 = \mathrm{E}_{\mathcal P}[{R}_2(z_1)|z_1] $ proved in (I))}\\
&=& \sqrt{n}(\mathrm{E}_{\mathcal S}[{R}_2(z_1)|z_1] + o(n^{-\frac 1 2})- \mathrm{E}_{\mathcal P}[{R}_2(z_1)|z_1]) +  \sqrt{n}{\beta}_{X_2}(\mathrm{E}_{\mathcal{S}_{\omega}}[{g}_1\{X_1\}|\bar{z}_T]  - \mathrm{E}_{\mathcal P}[{g}_1\{X_1\}]) + \\
&& \sqrt{n}(\widehat{\beta}_{X_2}-\beta_{X_2})(\mathrm{E}_{\mathcal{S}_{\omega}}[{g}_1\{X_1\}|\bar{z}_T] - \mathrm{E}_{\mathcal S}[g_1\{X_1\}|z_1]) \\
&& \text{\small (By $g_2\{X_2(z_1)\} = \widehat{R}_2(z_1) + \widehat{\beta}_{X_2}{g}_1\{X_1\} = {R}_2(z_1) + {\beta}_{X_2}{g}_1\{X_1\}$)}\\
&=& \frac{\sqrt n}{|\mathcal{I}_{z_1}|}\sum_{i\in \mathcal{I}_{z_1}} (R_{i2}-\mathrm{E}_{\mathcal P}[R_2|z_1])+ o(1) + \frac{\sqrt n}{n}\beta_{X_2}\sum_{i\in \mathcal{I}} (g_1\{X_{i1}\}-\mathrm{E}_{\mathcal P}[g_1\{X_1\}]) + \\
&& \sqrt{n}(\widehat{\beta}_{X_2}-\beta_{X_2})(\mathrm{E}_{\mathcal{P}}[{g}_1\{X_1\}] - \mathrm{E}_{\mathcal P}[g_1\{X_1\}|z_1]) +{\beta}_{X_2}o(1) +(\widehat{\beta}_{X_2}-{\beta}_{X_2})o(1) + \\
&& \sqrt{n}(\widehat{\beta}_{X_2}-\beta_{X_2})(\mathrm{E}_{\mathcal{S}}[{g}_1\{X_1\}] - \mathrm{E}_{\mathcal S}[g_1\{X_1\}|z_1]-\mathrm{E}_{\mathcal{P}}[{g}_1\{X_1\}] + \mathrm{E}_{\mathcal P}[g_1\{X_1\}|z_1]) \\
&& \text{\small (By the RAL estimator $\mathrm{E}_{\mathcal{S}_{\omega}}[g_1\{X_1\}| \bar{Z}_T = \bar{z}_T]$)}\\
&=& \frac{\sqrt n}{|\mathcal{I}_{z_1}|}\sum_{i\in \mathcal{I}_{z_1}} (R_{i2}-\mathrm{E}_{\mathcal P}[R_2|z_1]) + \frac{\sqrt n}{n}\beta_{X_2}\sum_{i\in \mathcal{I}} (g_1\{X_{i1}\}-\mathrm{E}_{\mathcal P}[g_1\{X_1\}])+\\
&& \sqrt{n}(\widehat{\beta}_{X_2}-\beta_{X_2})(\mathrm{E}_{\mathcal{P}}[{g}_1\{X_1\}] - \mathrm{E}_{\mathcal P}[g_1\{X_1\}|z_1]) +  (\widehat{\beta}_{X_2}-\beta_{X_2})O_p(1) +o(1)\\
&=& \frac{\sqrt n}{|\mathcal{I}_{z_1}|}\sum_{i\in \mathcal{I}_{z_1}} (R_{i2}-\mathrm{E}_{\mathcal P}[R_2|z_1])+O(\sqrt{n}(\widehat{\beta}_{X_2}-\beta_{X_2}))+ (\widehat{\beta}_{X_2}-\beta_{X_2})O_p(1)+o(1). 
\end{eqnarray*}
$\mathrm{E}_{\mathcal P}[g_1\{X_1\}] - \mathrm{E}_{\mathcal P}[g_1\{X_1\}|Z_1 = z_1] = 0$ is not likely to happen if $Z_1$ depends on $X_1$, therefore, we consider it a finite constant. To get the convergence rate of $\widehat{\beta}_{X_2\sim X_1|z_1}-\beta_{X_2\sim X_1|z_1}$, we first check when we can find $\beta_{X_2\sim X_1|z_1}$ such that
$\mathrm{E}_{X_1}[\mathrm{E}_{X_2}[g_2\{X_2\} -  \beta_{X_2\sim X_1|z_1}g_1\{X_1\}|X_1, z_1]] = \mathrm{E}_{X_2}[g_2\{X_2\} -  \beta_{X_2\sim X_1|z_1}g_1\{X_1\}| z_1]$ in \eqref{eqn::v}. 
It is straightforward to see the condition holds when $g_2\{X_2(z_1)\} = \beta_{X_2\sim X_1|z_1}g_1\{X_1\}+\varepsilon_2$, $\varepsilon_2\ci X_1|Z_1$. In this case, $\widehat{\beta}_{X_2\sim X_1|z_1}$ estimated by regressing $g_2\{X_2\}$ on $g_1\{X_1\}$ within the strata $Z_1 = z_1$ is a consistent estimator.

Let's have $\widehat{\beta}_{X_2\sim X_1|z_1}$ estimated by OLS, then $\widehat{\beta}_{X_2\sim X_1|z_1}-\beta_{X_2\sim X_1|z_1} =  ({g}_1\{\boldsymbol{X}_1\}^\top \text{diag}({1}\{\boldsymbol{Z}_1 = z_1\}){g}_1\{\boldsymbol{X}_1\})^{-1}{g}_1\{\boldsymbol{X}_1\}^\top \text{diag}({1}\{\boldsymbol{Z}_1 = z_1\})(g_2\{\boldsymbol{X}_2(z_1)\}-g_1\{\boldsymbol{X}_1\}\beta_{X_2\sim X_1|z_1}) = O_p(\sqrt{\frac{P_{g_1}}{|\mathcal{I}_{z_1}|}})$ is asymptotically Gaussian. With OLS estimator for $\beta_{X_2\sim X_1|z_1}$, $\mathrm{E}_{\mathcal{S}_{\omega}}[g_2\{X_2\}| \bar{Z}_T = \bar{z}_T]$ is a regular and asymptotically linear estimator, thus it is $\sqrt{n}$-consistent and asymptotically Gaussian with variance $V_2 = \frac{\text{Var}(R_2)}{|\mathcal{I}_{z_1}|/n}+\beta_{X_2}^\top V_1\beta_{X_2}+(\mathrm{E}_{\mathcal{P}}[{g}_1\{X_1\}] - \mathrm{E}_{\mathcal P}[g_1\{X_1\}|z_1])^\top\frac{\text{Var}(R_2)}{(|\mathcal{I}_{z_1}|-P_{g_1})/n}(\mathrm{E}_{\mathcal{P}}[{g}_1\{X_1\}] - \mathrm{E}_{\mathcal P}[g_1\{X_1\}|z_1]) $.

We assume the statement holds for $g_1\{X_1\},...,g_{t-1}\{X_{t-1}\}$ and now prove it for $g_t\{X_t\}$.
\begin{eqnarray*}
&&\sqrt{n}(\mathrm{E}_{\mathcal{S}_{\omega}}[g_t\{X_t\}| \bar{Z}_T = \bar{z}_T] - \mathrm{E}_{\mathcal P}[g_t\{X_t(\bar{z}_{t-1})\}])\\
&=& \sqrt{n}(\mathrm{E}_{\mathcal{S}_{\omega}}[\widehat{R}_t(\bar{z}_{t-1})|\bar{z}_T] - \mathrm{E}_{\mathcal P}[{R}_t(\bar{z}_{t-1})] + \\
&& \mathrm{E}_{\mathcal{S}_{\omega}}[\widehat{\beta}_{X_t}\bar{g}_{t-1}\{X_{t-1}\}| \bar{z}_T]  - \mathrm{E}_{\mathcal P}[{\beta}_{X_t}\bar{g}_{t-1}\{X_{t-1}(\bar{z}_{t-2})\}]) \\
&& \text{\small (Abbreviate ${\beta}_{X_t\sim \bar{X}_{t-1}|z_{t-1}}$ as ${\beta}_{X_t}$; }\\
&& \text{\small $g_{t}\{X_{t}(\bar{z}_{t-1})\} = \widehat{R}_t(\bar{z}_{t-1}) + \widehat{\beta}_{X_t}\bar{g}_{t-1}\{X_{t-1}(\bar{z}_{t-2})\} = {R}_t(\bar{z}_{t-1}) + {\beta}_{X_t}\bar{g}_{t-1}\{X_{t-1}(\bar{z}_{t-2})\}$)}\\
&=& \sqrt{n}(\mathrm{E}_{\mathcal S}[\widehat{R}_t(\bar{z}_{t-1})|\bar{z}_{t-1}] - \mathrm{E}_{\mathcal P}[{R}_t(\bar{z}_{t-1})]) + \\
&& \sqrt{n}{\beta}_{X_t}(\mathrm{E}_{\mathcal{S}_{\omega}}[\bar{g}_{t-1}\{X_{t-1}\}|\bar{z}_T]  - \mathrm{E}_{\mathcal P}[\bar{g}_{t-1}\{X_{t-1}(\bar{z}_{t-2})\}]) + \\
&&\sqrt{n}(\widehat{\beta}_{X_t}-\beta_{X_t})\mathrm{E}_{\mathcal{S}_{\omega}}[\bar{g}_{t-1}\{X_{t-1}\}|\bar{z}_T] + o(1)\\
&& \text{\small (By the implemented balance condition for $\widehat{R}_t$ in \eqref{eqn::balconsam})}\\
&=&\sqrt{n}(\mathrm{E}_{\mathcal S}[{R}_t(\bar{z}_{t-1})|\bar{z}_{t-1}] - \mathrm{E}_{\mathcal P}[{R}_t(\bar{z}_{t-1})|\bar{z}_{t-1}]) + \\
&& \sqrt{n}{\beta}_{X_t}(\mathrm{E}_{\mathcal{S}_{\omega}}[\bar{g}_{t-1}\{X_{t-1}\}|\bar{z}_T]  - \mathrm{E}_{\mathcal P}[\bar{g}_{t-1}\{X_{t-1}(\bar{z}_{t-2})\}]) +\\
&&\sqrt{n}(\widehat{\beta}_{X_t}-\beta_{X_t})(\mathrm{E}_{\mathcal{S}_{\omega}}[\bar{g}_{t-1}\{X_{t-1}\}|\bar{z}_T] - \mathrm{E}_{\mathcal S}[\bar{g}_{t-1}\{X_{t-1}\}|\bar{z}_{t-1}])+o(1)\\
&& \text{\small (By $\widehat{R}_t(\bar{z}_{t-1}) + \widehat{\beta}_{X_t}\bar{g}_{t-1}\{X_{t-1}(\bar{z}_{t-2})\} = {R}_t(\bar{z}_{t-1}) + {\beta}_{X_t}\bar{g}_{t-1}\{X_{t-1}(\bar{z}_{t-2})\}$ }\\
&& \text{\small and $\mathrm{E}_{\mathcal P}[{R}_t(\bar{z}_{t-1})] 
 = \mathrm{E}_{\mathcal P}[{R}_t(\bar{z}_{t-1})|\bar{z}_{t-1}] $)}\\
&=&\frac{\sqrt n}{|\mathcal{I}_{\bar{z}_{t-1}}|}\sum_{i\in \mathcal{I}_{\bar{z}_{t-1}}} (R_{it}-\mathrm{E}_{\mathcal P}[R_t|\bar{z}_{t-1}])  + \\
&& \sqrt{n}{\beta}_{X_t}(\mathrm{E}_{\mathcal{S}_{\omega}}[\bar{g}_{t-1}\{X_{t-1}\}|\bar{z}_T]  - \mathrm{E}_{\mathcal P}[\bar{g}_{t-1}\{X_{t-1}(\bar{z}_{t-2})\}]) + \\
&& \sqrt{n} (\widehat{\beta}_{X_t}-\beta_{X_t})(\mathrm{E}_{\mathcal P}[\bar{g}_{t-1}\{X_{t-1}(\bar{z}_{t-2})\}] - \mathrm{E}_{\mathcal P}[\bar{g}_{t-1}\{X_{t-1}\}|\bar{z}_{t-1}])+ \\
&& (\widehat{\beta}_{X_t}-\beta_{X_t}) O_p(1) +o(1) \\
&=& \frac{\sqrt n}{|\mathcal{I}_{\bar{z}_{t-1}}|}\sum_{i\in \mathcal{I}_{\bar{z}_{t-1}}} (R_{it}-\mathrm{E}_{\mathcal P}[R_t|\bar{z}_{t-1}]) +O(\sqrt{n}(\widehat{\beta}_{X_t}-\beta_{X_t})) +(\widehat{\beta}_{X_t}-\beta_{X_t})O_p(1)+o(1). 
\end{eqnarray*}
Again, we consider $\mathrm{E}_{\mathcal P}[\bar{g}_{t-1}\{X_{t-1}(\bar{z}_{t-2})\}] - \mathrm{E}_{\mathcal P}[\bar{g}_{t-1}\{X_{t-1}\}|\bar{z}_{t-1}]$ a finite constant if $Z_{t-1}$ depends on $\bar{X}_{t-1}$. Now we see when we can find coefficient $\beta_{X_t\sim \bar{X}_{t-1}|z_{t-1}}$ such that \\
$\mathrm{E}_{X_1}[\mathrm{E}_{X_2}[\cdots\mathrm{E}_{X_{t-1}}[\mathrm{E}_{X_t}[R_t|\bar{X}_{t-1}, \bar{z}_{t-1}]|\bar{X}_{t-2}, \bar{z}_{t-2}]\cdots|X_1,z_1]] = \mathrm{E}_{\mathcal P}[R_t|\bar{z}_{t-1}]$ in \eqref{eqn::v}. It is straightforward to see that holds when $g_t\{X_t(\bar{z}_{t-1})\} = \beta_{X_t\sim \bar{X}_{t-1}|z_{t-1}}\bar{g}_{t-1}\{X_{t-1}\}+\varepsilon_t$, $\varepsilon_t\ci \bar{X}_{t-1}|Z_{t-1}$. In this case, $\beta_{X_t\sim \bar{X}_{t-1}|z_{t-1}}$ estimated by regressing $g_t\{X_t\}$ on $\bar{g}_{t-1}\{X_{t-1}\}$ within the strata $Z_{t-1} = z_{t-1}$ is a consistent estimator. 
Let's have $\widehat{\beta}_{X_t\sim \bar{X}_{t-1}|z_{t-1}}$ estimated by OLS, then $\widehat{\beta}_{X_t\sim \bar{X}_{t-1}|z_{t-1}}-{\beta}_{X_t\sim \bar{X}_{t-1}|z_{t-1}}  = O_p(\sqrt{\sum_{s = 1}^{t-1} P_{g_s}/|\mathcal{I}_{z_{t-1}}|})$ is asymptotically Gaussian. With OLS estimator for ${\beta}_{X_t\sim \bar{X}_{t-1}|z_{t-1}}$, $\mathrm{E}_{\mathcal{S}_{\omega}}[g_t\{X_t\}| \bar{Z}_T = \bar{z}_T]$ is a regular and asymptotically linear estimator, thus it is $\sqrt{n}$-consistent and asymptotically Gaussian with variance $\frac{\text{Var}(R_t)}{|\mathcal{I}_{\bar{z}_{t-1}}|/n}+\beta_{X_t}^{\top}\bar{V}_{t-1}\beta_{X_t}+\\
\frac{(\mathrm{E}_{\mathcal P}[\bar{g}_{t-1}\{X_{t-1}(\bar{z}_{t-2})\}] - \mathrm{E}_{\mathcal P}[\bar{g}_{t-1}\{X_{t-1}\}|\bar{z}_{t-1}])^\top\text{Var}(R_t)(\mathrm{E}_{\mathcal P}[\bar{g}_{t-1}\{X_{t-1}(\bar{z}_{t-2})\}] - \mathrm{E}_{\mathcal P}[\bar{g}_{t-1}\{X_{t-1}\}|\bar{z}_{t-1}])}{(|\mathcal{I}_{z_{t-1}}|-\sum_{s = 1}^{t-1}P_{g_s})/n}$.
\end{enumerate}
\end{proof}
\end{subtheorem}

\subsection{Proposition 1.B: robustness to misspecification of balance conditions}\label{section::dual}

\begin{subtheorem}\label{prop2}
Under Assumptions 4, 5, and 6, as well as the regularity conditions in
Assumption \ref{regularity}, the weights derived from \eqref{eqn::opt0} satisfy Claim 1.B.
\begin{align}
& \Big\{\underset{\omega}{\text{minimize}}\sum_{i: Z_{iT} =\bar{z}_{T}} \omega_i^2 : \big|\mathrm{E}_{\mathcal{S}_{\omega}}[g_1\{X_1\}|\bar{z}_T] -  \mathrm{E}_{\mathcal{S}}[g_1\{X_1\}]\big| \preceq \delta_{1}, ~\big|\mathrm{E}_{\mathcal{S}_{\omega}}[\widehat{R}_t|\bar{z}_T] -  \mathrm{E}_{\mathcal{S}}[\widehat{R}_t|\bar{z}_{t-1}]\big| \preceq \delta_{t},  \nonumber \\
& t=2,...,T, \text{ where the tolerances } \delta_t = o(n^{-\frac{1}{2}}) \text{ are vectors of non-negative values}\Big\}. \label{eqn::opt0} 
\end{align}
\begin{proof}
The proof consists of the following parts.
\begin{enumerate}[label=(\Roman*)]
    \item Under Assumptions 4 and 5, $\mathrm{E}_{\mathcal{S}_{\pi^{-1}}}[h_T\{\bar{X}_T;\bar{Z}_T\}|\bar{z}_T]$ is $\sqrt{n}$-consistent for $\mathrm{E}_{\mathcal P}[h_T\{\bar{X}_T(\bar{z}_{T-1});\bar{z}_T\}]$ and is asymptotically Gaussian.

    \item Under the regularity conditions in Assumption \ref{regularity} which are extended from Assumption 1 in \citet{wang2020minimal} and Assumptions 2 and 4 in \citet{chattopadhyay2023onestep}, the
weights converge to the vector of true inverse propensity scores, $n^{-\frac 1 2}\|{1}\{\bar{\boldsymbol Z}_{T} = \bar{z}_T\} n\omega^{\text{bao}} - {1}\{\bar{\boldsymbol Z}_{T} = \bar{z}_T\}{\pi}^{-1}(\bar{\boldsymbol X}_{T};\bar{z}_T)\|_2 = o_p(1)$, where $\bar{\boldsymbol Z}_T$ and $\bar{\boldsymbol X}_T$ are matrices with $n$ rows, ${1}\{\bar{\boldsymbol Z}_{T} = \bar{z}_T\}\omega^{\text{bao}}$ is a vector of length $n$ solved from \eqref{eqn::opt0}.

    \item With (I) and (II), $\mathrm{E}_{\mathcal{S}_{\omega^{\text{bao}}}}[h_T\{\bar{X}_T;\bar{z}_T\}|\bar{z}_T]$ is consistent for $\mathrm{E}_{\mathcal P}[h_T\{\bar{X}_T(\bar{z}_{T-1});\bar{z}_T\}]$. 
    Moreover, under Assumption 6, $\mathrm{E}_{\mathcal{S}_{\omega^{\text{bao}}}}[h_T\{\bar{X}_T;\bar{z}_T\}|\bar{z}_T]$ is asymptotically Gaussian.  
\end{enumerate}
Here are the detailed arguments.
\begin{enumerate}[label=(\Roman*)]
    \item According to Section 1.1, when identifying $\mathrm{E}_{\mathcal P}[Y(\bar{z}_T)]$ by weighting the population $\mathcal P$ with the true stabilized inverse propensity scores $1\{\bar{Z}_{T} = \bar{z}_T\}\mathrm{P}(\bar{z}_T)\pi^{-1}(\bar{X}_T;\bar{z}_T) := \frac{1\{\bar{Z}_{T} = \bar{z}_T\}\mathrm{P}(\bar{z}_T)}{\prod_{t = 1}^T\mathrm{P}(z_t|\bar{z}_{t-1},\bar{X}_{t})}$, any functional form of the covariates $\psi\{\bar{X}_T;\bar{Z}_T\}$ including $h_T\{\bar{X}_T;\bar{Z}_T\}$ is balanced to their population potential means, such that
\begin{eqnarray*}
\mathrm{E}_{\mathcal{P}_{\pi^{-1}}}[h_T\{\bar{X}_T;\bar{Z}_T\}|\bar{Z}_T = \bar{z}_T]= \mathrm{E}_{\mathcal P}[h_T\{\bar{X}_T(\bar{z}_{T-1});\bar{z}_T\}].
\end{eqnarray*}

Then we consider the sample $\mathcal S$ where each unit $i$ is weighted by the true inverse propensity score $1\{\bar{Z}_{iT} = \bar{z}_T\}n^{-1}\pi^{-1}(\bar{X}_{iT};\bar{z}_T)$. Under Assumptions 4 and 5, by Theorem 1 in \citet{ma2020robust}, $\mathrm{E}_{\mathcal{S}_{\pi^{-1}}}[h_T\{\bar{X}_T;\bar{Z}_T\}|\bar{z}_T]$ converges to $\mathrm{E}_{\mathcal{P}_{\pi^{-1}}}[h_T\{\bar{X}_T;\bar{Z}_T\}|\bar{z}_T]$ in probability with rate $1/\sqrt{n}$ and is asymptotically Gaussian. Since $\mathrm{E}_{\mathcal{P}_{\pi^{-1}}}[h_T\{\bar{X}_T;\bar{Z}_T\}|\bar{Z}_T = \bar{z}_T]= \mathrm{E}_{\mathcal P}[h_T\{\bar{X}_T(\bar{z}_{T-1});\bar{z}_T\}]$, we complete the proof for (I).

\item The proof is in Lemma \ref{lemma::2}.

\item From (I),  $\mathrm{E}_{\mathcal{S}_{\pi^{-1}}}[h_T\{\bar{X}_T;\bar{Z}_T\}|\bar{z}_T]$ converges to $\mathrm{E}_{\mathcal P}[h_T\{\bar{X}_T(\bar{z}_{T-1});\bar{z}_T\}]$ and is asymptotically Gaussian. Now we will show $\mathrm{E}_{\mathcal{S}_{\omega^{\text{bao}}}}[h_T\{\bar{X}_T;\bar{z}_T\}|\bar{z}_T]-\mathrm{E}_{\mathcal{S}_{\pi^{-1}}}[h_T\{\bar{X}_T;\bar{z}_T\}|\bar{z}_T]$ converges to 0. 
\begin{align}
& \mathrm{E}_{\mathcal{S}_{\omega^{\text{bao}}}}[h_T\{\bar{X}_T;\bar{z}_T\}|\bar{z}_T]-\mathrm{E}_{\mathcal{S}_{\pi^{-1}}}[h_T\{\bar{X}_T;\bar{z}_T\}|\bar{z}_T] \nonumber\\
= & \beta_{h\sim g}(\mathrm{E}_{\mathcal{S}_{\omega^{\text{bao}}}}[\bar{g}_T\{X_T\}|\bar{z}_T]-\mathrm{E}_{\mathcal{P}}[\bar{g}_T\{X_T(\bar{z}_{T-1})\}]+\nonumber \\
& \mathrm{E}_{\mathcal{P}}[\bar{g}_T\{X_T(\bar{z}_{T-1})\}]-\mathrm{E}_{\mathcal{S}_{\pi^{-1}}}[\bar{g}_T\{X_T\}|\bar{z}_T])+ \nonumber\\
&(\mathrm{E}_{\mathcal{S}_{\omega^{\text{bao}}}}[h_T\{\bar{X}_T;\bar{z}_T\}-\beta_{h\sim g}\bar{g}_T\{X_T\}|\bar{z}_T]-\mathrm{E}_{\mathcal{S}_{\pi^{-1}}}[h_T\{\bar{X}_T;\bar{z}_T\}-\beta_{h\sim g}\bar{g}_T\{X_T\}|\bar{z}_T]).\nonumber
\end{align} 
By Theorem 1.A and (I), the first line of the above quantity is asymptotically Gaussian.
By Cauchy–Schwarz inequality, the second line of the above quantity is bounded by $n^{-\frac 1 2}\|{1}\{\bar{\boldsymbol Z}_{T} = \bar{z}_T\} n\omega^{\text{bao}} - \frac{{1}\{\bar{\boldsymbol Z}_{T} = \bar{z}_T\}}{{\pi}(\bar{\boldsymbol X}_{T};\bar{z}_T)}\|_2\sqrt{\frac{\|{h}_T\{\bar{\boldsymbol X}_T;\bar{z}_T\}-\bar{g}_T\{\boldsymbol{X}_T(\bar{z}_{T-1})\}\beta_{h\sim g}\|_2^2}{n}}$. Therefore, $\mathrm{E}_{\mathcal{S}_{\omega^{\text{bao}}}}[h_T\{\bar{X}_T;\bar{z}_T\}|\bar{z}_T]$ is consistent for $\mathrm{E}_{\mathcal P}[h_T\{\bar{X}_T(\bar{z}_{T-1});\bar{z}_T\}]$ if $n^{-\frac 1 2}\|{1}\{\bar{\boldsymbol Z}_{T} = \bar{z}_T\} n\omega^{\text{bao}} - {1}\{\bar{\boldsymbol Z}_{T} = \bar{z}_T\}{\pi}^{-1}(\bar{\boldsymbol X}_{T};\bar{z}_T)\|_2 = o_p(1)$, which is shown in (II). 
$\mathrm{E}_{\mathcal{S}_{\omega^{\text{bao}}}}[h_T\{\bar{X}_T;\bar{z}_T\}|\bar{z}_T]$ is asymptotically normal if by Assumption 6, $\|{h}_T\{\bar{\boldsymbol X}_T;\bar{z}_T\}-\bar{g}_T\{\boldsymbol{X}_T\}\beta_{h\sim g}\|_2 = O_p(1)$. 
\end{enumerate}
\end{proof}
\end{subtheorem}

We will prove Theorem 1.B (II) in Lemma \ref{lemma::2}. Assumption \ref{regularity} includes the regularity conditions for showing $n^{-\frac 1 2}\|{1}\{\bar{\boldsymbol Z}_{T} = \bar{z}_T\} n\omega^{\text{bao}} - {1}\{\bar{\boldsymbol Z}_{T} = \bar{z}_T\}{\pi}^{-1}(\bar{\boldsymbol X}_{T};\bar{z}_T)\|_2 = o_p(1)$.

\begin{assumption}\label{regularity} ~

a. There exists a constant $C > 0$, $\Lambda_{\text{min}}\{\mathrm{E}_{\mathcal P}[\bar{g}_{T}\{X_{T}\}^\top \text{diag}(1\{\bar{Z}_T = \bar{z}_T\})\bar{g}_{T}\{X_{T}\}]\} > C$, where $\Lambda_{\text{min}}\{\cdot\}$ denotes the smallest eigenvalue.

b. There exists a smooth function of covariates and treatments, $m^*\{\bar{x}_T;\bar{z}_T\}$, such that the true inverse probability weights for group $\bar{Z}_T = \bar{z}_T$ satisfy $1/\pi(\bar{x}_T;\bar{z}_T) = n\rho'(m^*\{\bar{x}_T;\bar{z}_T\})$ for all $\bar{x}_T \in\bar{\mathcal X}_T$ and a linear combination of $\bar{g}_T\{x_T\}$ satisfies $\sup_{\bar{x}_T\in\bar{\mathcal X}_T} |m^*\{\bar{x}_T;\bar{z}_T\} -
\bar{g}_T\{x_T\}^\top \lambda^*| = O(P_{\bar{g}_T}^{-r_{\pi}})$ for some $\lambda^* \in \mathbb{R}^{P_{\bar{g}_T}}$ and $r_{\pi} > 1$, where $P_{\bar{g}_T}$ denotes the dimension of $\bar{g}_T\{X_T\}$.

c.  There exists a constant $0<c_0 <1/2$, such that $c_0 \leq n\rho'(v)\leq 1-c_0 $ for any $v=\bar{g}_T\{x_{T}\}^\top\lambda$ with $\lambda\in
int(\Theta)$, where $\Theta$ is a compact set and $int(\cdot)$ is the interior of a set. Also, there exist constants $c_1 < c_2 < 0$, such that $c_1 \leq n\rho''(v) \leq c_2 < 0$ in some small neighborhood $\mathcal{V}$ of $v^*=\bar{g}_T\{x_{T}(\bar{z}_{T-1})\}^\top\lambda^\dagger$, where $\lambda^\dagger$ is the solution to the dual optimization problem $\underset{\lambda}{\text{minimize}} ~ \frac{1}{n}\big[\sum_{i = 1}^n -1\{\bar{Z}_{iT} = \bar{z}_T\}n\rho(\bar{g}_T\{X_{iT}\}^\top\lambda)\big] + {\mathrm E}_{\mathcal S}[\bar{g}_T\{X_T(\bar{z}_{T-1})\}]^\top\lambda$.

d. There exists a constant $C>0$ such that $\sup_{\bar{x}_T\in \bar{\mathcal X}_T} \|\bar{g}_T\{x_T\}\|_2 \leq CP_{\bar{g}_T}^{1/2}$ and \\
$\|\mathrm{E}_{\mathcal P}[\bar{g}_{T}\{X_{T}\}\bar{g}_{T}\{X_{T}\}^\top]\|_F \leq C$, where $\|\cdot\|_F$ denotes the Frobenius norm.

e. $\|\bar{\delta}_{T}\|_2$ = $O_p[P_{\bar{g}_T}^{1/4}\{(\log P_{\bar{g}_T})/n\}^{1/2} + P_{\bar{g}_T}^{1/2-r_{\pi}}]$.

f. $P_{\bar{g}_T} = O(n^r)$ for some $0<r<2/3$.
\end{assumption}

\begin{lemma}\label{lemma::2}
Under the regularity conditions in Assumption \ref{regularity}, which are extended from Assumption 1 in \citet{wang2020minimal} and Assumptions 2 and 4 in \citet{chattopadhyay2023onestep}, the
weights converge to the vector of true inverse propensity scores in $\ell_2$-norm, that is, $n^{-1/2}\|{1}\{\bar{\boldsymbol Z}_{T} = \bar{z}_T\} n\omega^{\text{bao}} - {1}\{\bar{\boldsymbol Z}_{T} = \bar{z}_T\}{\pi}^{-1}(\bar{\boldsymbol X}_{T};\bar{z}_T)\|_2 = o_p(1)$, where $\omega^{\text{BAO}}$ is solved from the optimization problem \eqref{eqn::opt0}. 

This statement will be proved by four steps, (I)--(IV).
\begin{enumerate}[label=(\Roman*)]
    \item The dual problem of \eqref{eqn::opt0}
    is an empirical loss minimization problem with $\ell_1$-regularization:
    \begin{eqnarray}
\underset{\lambda}{\text{minimize}} ~&& \big[\sum_{i = 1}^n -1\{\bar{Z}_{iT} = \bar{z}_T\}\rho(\bar{\widehat R}_{iT}^\top\lambda)\big] + \nonumber\\
&& ({\mathrm E}_{\mathcal S}[g_{1}\{X_1\}]^\top,{\mathrm E}_{\mathcal S}[{\widehat R}_{2}|z_{1}]^\top,...,{\mathrm E}_{\mathcal S}[{\widehat R}_{T}|\bar{z}_{T-1}]^\top)^\top \lambda +\bar{\delta}_{T}^\top |\lambda|, \label{eqn::dual0}
\end{eqnarray}
where 
$\rho(v) = -\frac{v^2}{4}$, $\bar{\widehat R}_{iT} = (g_1\{X_{i1}\}^\top,\widehat{R}_{i2}^\top,...,\widehat{R}_{iT}^\top)^\top$ is a $P_{\bar{g}_T}\times 1$ vector of residuals, $\lambda$ is a $P_{\bar{g}_T}\times 1$ vector of dual variables corresponding to the $P_{\bar{g}_T}$ balancing constraints. 

The dual \eqref{eqn::dual0} is equivalent to the following problem involving mean potential covariates:
    \begin{eqnarray}
\underset{\lambda}{\text{minimize}} ~\big[\sum_{i = 1}^n -1\{\bar{Z}_{iT} = \bar{z}_T\}\rho(\bar{g}_T\{X_{iT}\}^\top\lambda)\big] + ({\mathrm E}_{\mathcal S}[\bar{g}_T\{X_T(\bar{z}_{T-1})\}]+\bar{\Delta}_T)^\top \lambda, \label{eqn::dual1} 
\end{eqnarray}
where the bias $\bar{\Delta}_{T} := (\Delta_1,...,\Delta_T)$ is asymptotically Gaussian, and $\|{\Delta}_{t}\|_2 =O_p(P_{\bar{g}_t}^{\frac 1 2}n^{-\frac 1 2})$.
\item If $\omega^{\text{BAO}}$ and $\lambda_R^\dagger$ are solutions to the primal \eqref{eqn::opt0} and the dual \eqref{eqn::dual0} respectively, 
then for the units $i: \bar{Z}_{iT} = \bar{z}_T$, $\omega^{\text{bao}}_i = \rho'(\bar{\widehat R}_{iT}^\top\lambda_R^\dagger)$, where $\rho'(v)$ is the first derivative of $\rho(v)$. 

Let $\lambda^\dagger :=
\argmin_{\lambda}  \big[\sum_{i = 1}^n -1\{\bar{Z}_{iT} = \bar{z}_T\}\rho(\bar{g}_T\{X_{iT}\}^\top\lambda)\big] + {\mathrm E}_{\mathcal S}[\bar{g}_T\{X_T(\bar{z}_{T-1})\}]^\top \lambda $, then by Assumption \ref{regularity}a,
\begin{eqnarray} 
    n^{-1/2}\|{1}\{\bar{\boldsymbol Z}_{T} = \bar{z}_T\}n\rho'(\bar{\widehat {\boldsymbol R}}_{T}^\top\lambda_R^\dagger)- {1}\{\bar{\boldsymbol Z}_{T} = \bar{z}_T\}n\rho'(\bar{g}_{T}\{\boldsymbol{X}_{T}\}^\top\lambda^\dagger)\|_2 = O_p(P_{\bar{g}_T}^{\frac 1 2} n^{-\frac 1 2}).\label{eqn::diff}
\end{eqnarray}

\item Let $\tilde\lambda \in \argmin_{\lambda} \mathrm{E}_{\mathcal P}[-n 1\{\bar{Z}_{T} = \bar{z}_T\}\rho(\bar{g}_T\{X_{T}\}^\top\lambda)|\bar{x}_{T}(\bar{z}_{T-1})] + \bar{g}_T\{x_{T}(\bar{z}_{T-1})\}^\top \lambda$, then for the units $i: \bar{Z}_{iT} = \bar{z}_T$, $\tilde\lambda$ satisfies 
\begin{eqnarray*}
n\rho'(\bar{g}_T\{x_{T}\}^\top\tilde\lambda) = \frac{1}{\mathrm{P}(\bar{Z}_{iT} = \bar{z}_T|\bar{x}_{T}(\bar{z}_{T-1}))} = \pi^{-1}(\bar{x}_T;\bar{z}_T).
\end{eqnarray*}

\item Under Assumption \ref{regularity}, the weights $\omega^{\text{BAO}}$ converge to the true inverse propensity scores, $n^{-1/2}\|{1}\{\bar{\boldsymbol Z}_{T} = \bar{z}_T\} n\omega^{\text{bao}} - {1}\{\bar{\boldsymbol Z}_{T} = \bar{z}_T\}{\pi}^{-1}(\bar{\boldsymbol X}_{T};\bar{z}_T)\|_2 = o_p(1)$.
\end{enumerate}

\begin{proof} We follow the proof in \citet{wang2020minimal, chattopadhyay2023onestep}.
\begin{enumerate}[label=(\Roman*)]
\item We prove the equivalence of two duals, \eqref{eqn::dual0} and \eqref{eqn::dual1}, by showing their primals are identical.

By Theorem \ref{prop1} (II), the weights $\omega^{\text{BAO}}$ from  the primal optimization problem \eqref{eqn::opt0} imply 
\begin{eqnarray*}
\mathrm {E}_{\mathcal{S}_{\omega^{\text{bao}}}}[(g_1\{X_1\},..., g_T\{X_T\}) | \bar{Z}_T = \bar{z}_T] -\mathrm{E}_{\mathcal P}[(g_1\{X_1\},..., g_T\{X_T(\bar{z}_{T-1})\})] = O_p(P_{\bar{g}_T}^{\frac 1 2}n^{-\frac 1 2}).
\end{eqnarray*}
Let $\bar{\Delta}_T :=\mathrm {E}_{\mathcal{S}_{\omega^{\text{bao}}}}[(g_1\{X_1\},..., g_T\{X_T\}) | \bar{Z}_T = \bar{z}_T] -\mathrm{E}_{\mathcal S}[(g_1\{X_1\},..., g_T\{X_T(\bar{z}_{T-1})\})]$, which is also $O_p(P_{\bar{g}_T}^{\frac 1 2}n^{-\frac 1 2})$ and asymptotically Gaussian. Thus, if the global minimum is unique, \eqref{eqn::opt0} and the following primal problem give the same solution,
\begin{eqnarray}
&& \underset{\omega}{\text{minimize}} \sum_{i \in \mathcal{I}_{\bar{z}_{T}}} \omega_i^2, \nonumber\\
&& \text{ subject to } \sum\limits_{i\in \mathcal{I}_ {\bar{z}_{T}}} \omega_{i}g_{t}\{X_{it}\}-
({\mathrm E}_{\mathcal S}[g_t\{X_t(\bar{z}_{t-1})\}]+\Delta_t) =0, ~ t=1,...,T. ~~ \label{eqn::opt1}
\end{eqnarray} 

According to Theorem 1 in \citet{wang2020minimal} and Theorem 1 in \citet{chattopadhyay2020balancing}, the dual of the above primal optimization problem is \eqref{eqn::dual1}, that is,
\begin{eqnarray*}
\underset{\lambda}{\text{minimize}} ~\big[\sum_{i = 1}^n -1\{\bar{Z}_{iT} = \bar{z}_T\}\rho\{\bar{g}_T\{X_{iT}\}^\top\lambda\}\big] + ({\mathrm E}_{\mathcal S}[\bar{g}_T\{X_T(\bar{z}_{T-1})\}]+\bar{\Delta}_T)^\top \lambda,
\end{eqnarray*}
where 
$\rho(v) = -v^2/4$. Meanwhile, the dual of \eqref{eqn::opt0} is \eqref{eqn::dual0}. So, \eqref{eqn::dual0} and \eqref{eqn::dual1} are equivalent.

\item According to Theorem 1 in \citet{wang2020minimal} and Theorem 1 in \citet{chattopadhyay2020balancing}, $\omega^{\text{bao}}_i = \rho'(\bar{\widehat R}_{iT}^\top\lambda_R^\dagger)$. Now, we'd like to show that 
\begin{eqnarray*} 
    n^{-1/2}\|{1}\{\bar{\boldsymbol Z}_{T} = \bar{z}_T\}n\rho'(\bar{\widehat {\boldsymbol R}}_{T}^\top\lambda_R^\dagger)- {1}\{\bar{\boldsymbol Z}_{T} = \bar{z}_T\}n\rho'(\bar{g}_{T}\{\boldsymbol{X}_{T}\}^\top\lambda^\dagger)\|_2 = O_p(P_{\bar{g}_T}^{\frac 1 2} n^{-\frac 1 2}),
\end{eqnarray*}
where $\lambda^\dagger = \argmin_{\lambda}  \big[\sum_{i = 1}^n -1\{\bar{Z}_{iT} = \bar{z}_T\}\rho(\bar{g}_T\{X_{iT}\}^\top\lambda)\big] + {\mathrm E}_{\mathcal S}[\bar{g}_T\{X_T(\bar{z}_{T-1})\}]^\top \lambda$. 

Let $\lambda^\dagger_{\Delta} := \argmin_{\lambda}  \big[\sum_{i = 1}^n -1\{\bar{Z}_{iT} = \bar{z}_T\}\rho(\bar{g}_T\{X_{iT}\}^\top\lambda)\big] + ({\mathrm E}_{\mathcal S}[\bar{g}_T\{X_T(\bar{z}_{T-1})\}]+\Delta)^\top \lambda$. From (I), we know that ${1}\{\bar{\boldsymbol Z}_{T} = \bar{z}_T\}n\rho'(\bar{\widehat {\boldsymbol R}}_{T}^\top\lambda_R^\dagger) = {1}\{\bar{\boldsymbol Z}_{T} = \bar{z}_T\}n\rho'(\bar{g}_{T}\{\boldsymbol{X}_{T}\}^\top\lambda_{\bar{\Delta}_T}^\dagger)$.

The first derivative of $L_{\Delta}(\lambda) := \big[\sum_{i = 1}^n -1\{\bar{Z}_{iT} = \bar{z}_T\}\rho(\bar{g}_T\{X_{iT}\}^\top\lambda)\big] + ({\mathrm E}_{\mathcal S}[\bar{g}_T\{X_T(\bar{z}_{T-1})\}]+\Delta)^\top \lambda$ is $\frac{\partial L_{\Delta}(\lambda)}{\partial \lambda} = \big[\sum_{i = 1}^n -1\{\bar{Z}_{iT} = \bar{z}_T\}\bar{g}_T\{X_{iT}\}\rho'(\bar{g}_T\{X_{iT}\}^\top\lambda)\big] + ({\mathrm E}_{\mathcal S}[\bar{g}_T\{X_T(\bar{z}_{T-1})\}]+\Delta)$. Thus,
$\frac{1}{2}\big[\sum_{i = 1}^n 1\{\bar{Z}_{iT} = \bar{z}_T\}(\bar{g}_T\{X_{iT}\}\bar{g}_T\{X_{iT}\}^\top)\lambda^\dagger_{\Delta}\big] + ({\mathrm E}_{\mathcal S}[\bar{g}_T\{X_T(\bar{z}_{T-1})\}]+\Delta) = 0$,
$n\rho'(\bar{g}_{T}\{X_{iT}\}^\top\lambda^\dagger_{\Delta}) = \bar{g}_{T}\{X_{iT}\}^\top (n^{-1}\bar{g}_T\{\boldsymbol{X}_T\}^\top\text{diag}(1\{\bar{\boldsymbol Z}_T = \bar{z}_T\})\bar{g}_{T}\{\boldsymbol{X}_T\})^{-1} ({\mathrm E}_{\mathcal S}[\bar{g}_T\{X_T(\bar{z}_{T-1})\}]+\Delta)$.

By Assumption \ref{regularity}a, if for some constant $C > 0$, we have $\Lambda_{\text{min}}\{\mathrm{E}_{\mathcal S}[\bar{g}_{T}\{X_{T}\}\text{diag}(1\{\bar{Z}_T = \bar{z}_T\})\bar{g}_{T}\{X_{T}\}^\top]\} > C$, then \eqref{eqn::diff} holds.

\item For $\tilde\lambda \in \argmin_{\lambda} \mathrm{E}_{\mathcal P}[-n 1\{\bar{Z}_{T} = \bar{z}_T\}\rho(\bar{g}_T\{X_{T}\}^\top\lambda)|\bar{X}_{T}(\bar{z}_{T-1}) = \bar{x}_T] + \bar{g}_T\{x_{T}\}^\top \lambda$, from Theorem 8.1 in \citet{chattopadhyay2023onestep}, we can conclude that for the units $i: \bar{Z}_{iT} = \bar{z}_T$,
\begin{eqnarray*}
n\rho'(\bar{g}_T\{x_{T}\}^\top\tilde\lambda) = \frac{1}{\mathrm{P}(\bar{Z}_{iT} = \bar{z}_T|\bar{X}_{T}(\bar{z}_{T-1}) = \bar{x}_T)}.
\end{eqnarray*}

By Assumptions 1b and 3b, $\mathrm{P}(\bar{Z}_T = \bar{z}_T|\bar{X}_{T}(\bar{z}_{T-1})) = \mathrm{P}(Z_{1} = z_1|X_{1}, \underaccent{\bar}{X}_2(z_1))
[\prod_{t = 2}^{T-1}\mathrm{P}(Z_{t} = z_t|\bar{z}_{t-1},\bar{X}_t(\bar{z}_{t-1}), \underaccent{\bar}{X}_{t+1}(\bar{z}_t))]\mathrm{P}(Z_{T} = z_T|\bar{z}_{T-1},\bar{X}_T(\bar{z}_{T-1}))  = \mathrm{P}(Z_{1} = z_1|X_{1})\prod_{t = 2}^T\mathrm{P}(Z_{t} = z_t|\bar{z}_{t-1},\bar{X}_{t})= \pi(\bar{X}_T;\bar{z}_T)$. This connection means the product of treatment assignment probabilities $\pi(\bar{X}_T;\bar{z}_T)$ can be alternatively considered as by fitting a binary indicator $1\{\bar{Z}_T = \bar{z}_T\}$ on the sequence of potential covariates $\bar{X}_{T}(\bar{z}_{T-1})$. 

\item Now, let's see how $1\{\bar{\boldsymbol Z}_T = \bar{z}_T\}n\omega^{\text{bao}}$ converges to $1\{\bar{\boldsymbol Z}_T = \bar{z}_T\}\pi^{-1}(\bar{\boldsymbol X}_T;\bar{z}_T)$. 
\begin{eqnarray*}
    && n^{-1/2}\|{1}\{\bar{\boldsymbol Z}_{T} = \bar{z}_T\}n\omega^{\text{bao}} - {1}\{\bar{\boldsymbol Z}_{T} = \bar{z}_T\}{\pi}^{-1}(\bar{\boldsymbol X}_{T};\bar{z}_T)\|_2 \\
    &=& n^{-1/2} \|{1}\{\bar{\boldsymbol Z}_{T} = \bar{z}_T\}n\rho'(\bar{\widehat{\boldsymbol R}}_{T}^\top\lambda_R^\dagger) - {1}\{\bar{\boldsymbol Z}_{T} = \bar{z}_T\}n\rho'(m^*\{\bar{\boldsymbol X}_T(\bar{z}_{T-1});\bar{z}_T\})\|_2 \\
    &=& n^{-1/2} \|{1}\{\bar{\boldsymbol Z}_{T} = \bar{z}_T\}n[\rho'(\bar{\widehat {\boldsymbol R}}_T^\top\lambda_R^\dagger)-\rho'(\bar{g}_T\{\boldsymbol{X}_{T}(\bar{z}_{T-1})\}^\top\lambda^\dagger)]\|_2 + \\
    &&  n^{-1/2}\|{1}\{\bar{\boldsymbol Z}_{T} = \bar{z}_T\}n[\rho'(\bar{g}_T\{\boldsymbol{X}_{T}(\bar{z}_{T-1})\}^\top\lambda^\dagger)-\rho'(\bar{g}_T\{\boldsymbol{X}_{T}(\bar{z}_{T-1})\}^\top\lambda^*)]\|_2 + \\
    && n^{-1/2}\|{1}\{\bar{\boldsymbol Z}_{T} = \bar{z}_T\}n[\rho'(\bar{g}_T\{\boldsymbol{X}_{T}(\bar{z}_{T-1})\}^\top\lambda^*)-\rho'(m^*\{\bar{\boldsymbol X}_T(\bar{z}_{T-1});\bar{z}_T\})]\|_2 \\
    && \text{\small (By the triangle inequality)} \\
    &\leq&  O_p(P_{\bar{g}_T}^{1/2} n^{-1/2})+ C\sup_{\bar{x}_T\in\bar{\mathcal X}_T}\|\bar{g}_T\{x_{T}\}^\top(\lambda^\dagger-\lambda^*)\|_2 +  C\sup_{\bar{x}_T\in\bar{\mathcal X}_T} |m^*\{\bar{x}_T;\bar{z}_T\} -
\bar{g}_T\{x_T\}^\top \lambda^*| \\
  && \text{\small (By (II) and the mean value theorem)} \\
  &=& O_p(P_{\bar{g}_T}^{1/2}n^{-1/2})+O_p\{P_{\bar{g}_T}^{1/2}P_{\bar{g}_T}^{1/4}(\frac{\log P_{\bar{g}_T}}{n})^{1/2} + P_{\bar{g}_T}^{1/2}P_{\bar{g}_T}^{-r_{\pi}+1/2}\} + O(P_{\bar{g}_T}^{-r_{\pi}}) \\
   && \text{\small (By Theorem 8.2 in \citet{chattopadhyay2023onestep})} \\
    &=& O_p\{\frac{P_{\bar{g}_T}^{1/2}}{n^{1/2}}+\frac{P_{\bar{g}_T}^{3/4}\log P_{\bar{g}_T}}{n^{1/2}} + P_{\bar{g}_T}^{1-r_{\pi}}\} \\
    &=& o_p(1) \quad \text{\small (By Assumption \ref{regularity}f)} 
\end{eqnarray*}
\end{enumerate}
\end{proof}
\end{lemma}

% %%%%%%%%%%%%%%%%%%%%%
% %%%%%%%%%%%%%%%%%%%%%
\section{Extension for censored data}\label{sec::extensions}

% %%%%%%%%%%%%%%%%%%%%
% %%%%%%%%%%%%%%%%%%%%

The identification is similar to the setting without censoring in Section \ref{sec::identification}.
\begin{eqnarray}
&&\mathrm{E}_{\mathcal P}[Y(\bar{z}_T)] \nonumber \\
&=& \int  \mathrm{E}_{\mathcal P}[Y(\bar{z}_T) | \bar{x}_T, \bar{z}_T, \bar{C}_T = \bar{0}_T] f_{\mathcal P}(X_T = x_T | \bar{X}_{T-1} = \bar{x}_{T-1}, \bar{Z}_{T-1} = \bar{z}_{T-1}, \bar{C}_{T-1} = \bar{0}_{T-1})\nonumber \\
&& f_{\mathcal P}(X_{T-1} = x_{T-1} | \bar{X}_{T-2} = \bar{x}_{T-2}, \bar{Z}_{T-2} = \bar{z}_{T-2}, \bar{C}_{T-2} = \bar{0}_{T-2})\cdots f_{\mathcal P}(X_1 = x_1)d\bar{x}_T \nonumber \\
&& \text{\small (By sequentially ignorability assumption $Y(\bar{z}_T) \ci (Z_t, C_t)| (\bar{Z}_{t-1}, \bar{C}_{t-1}, \bar{X}_t$))} \nonumber \\
&=& \int  \mathrm{E}_{\mathcal P}[Y(\bar{z}_T)| \bar{X}_T, \bar{Z}_T = \bar{z}_T, \bar{C}_T = \bar{0}_T]\nonumber \\
&& f_{\mathcal P}(X_T(\bar{z}_{T-1}) = x_T | \bar{X}_{T-1}(\bar{z}_{T-2}) = \bar{x}_{T-1}, \bar{z}_{T-1}, \bar{C}_{T-1} = \bar{0}_{T-1}) \cdots f_{\mathcal P}(X_1 = x_1)d\bar{x}_T \nonumber \\
&& \text{\small (By the stable unit treatment value assumption,} \nonumber \\
&& \text{\small $Y = \sum_{\bar{z}_T \in \bar{\mathcal{Z}}_T}1\{\bar{Z}_{T} = \bar{z}_T\} Y(\bar{z}_T)$, when $\bar{C}_{T} = \bar{0}_T$; }  \nonumber \\
&& \text{\small $X_{t} = \sum_{\bar{z}_{t-1} \in \bar{\mathcal{Z}}_{t-1}}1\{\bar{Z}_{t-1} = \bar{z}_{t-1}\} X_{t}(\bar{z}_{t-1})$, when $\bar{C}_{t-1} = \bar{0}_{t-1}$)} \nonumber \\
&=& \int \mathrm{E}_{\mathcal P}[Y(\bar{z}_T)| \bar{X}_T, \bar{Z}_T = \bar{z}_T, \bar{C}_T = \bar{0}_T]f_{\mathcal P}(X_T(\bar{z}_{T-1}) = x_T, \cdots, X_1 = x_1)d\bar{x}_T \nonumber \\
&& \text{\small (By sequentially ignorability assumption $X_{t+1}(\bar{z}_{t}) \ci (Z_s, C_s) | (\bar{X}_s, \bar{Z}_{s-1}, \bar{C}_{s-1})$, $s\leq t$)} \nonumber \\
&=& \int \mathrm{E}_{\mathcal P}[Y(\bar{z}_T)| \bar{X}_T(\bar{z}_{T-1}), \bar{Z}_T = \bar{z}_T, \bar{C}_T = \bar{0}_T]f_{\mathcal P}(X_T(\bar{z}_{T-1}) = x_T, \cdots, X_1 = x_1)d\bar{x}_T \label{eqn::IdCen1} 
\end{eqnarray}

We still assume $Y(\bar{z}_T) = \alpha_{0,\bar{z}_T} +  \sum_{t = 1}^T \alpha_{t,\bar{z}_{T}}^\top g_t\{X_t(\bar{z}_{t-1})\} + h_T\{\bar{X}_T(\bar{z}_{T-1});\bar{z}_T\} + \varepsilon$.
To let identification happen, which means finding weights such that $\eqref{eqn::IdCen1}=\mathrm{E}_{\mathcal{P}_{\omega}}[Y| \bar{z}_T, \bar{C}_T = \bar{0}_T]$, the following set of balance conditions suffices, 
\begin{eqnarray}\label{eqn::balconcen}
\mathrm{E}_{\mathcal{P}_{\omega}}[(\bar{g}_T\{X_T\}, h_T\{\bar{X}_T;\bar{z}_T\}) | \bar{z}_T, \bar{C}_T = \bar{0}_T] = \mathrm{E}_{\mathcal P}[(\bar{g}_T\{X_T(\bar{z}_{T-1})\}, h_T\{\bar{X}_T(\bar{z}_{T-1});\bar{z}_T\})]. 
\end{eqnarray}

The following set of balance conditions implies $\mathrm{E}_{\mathcal{P}_{\omega}}[\bar{g}_T\{X_T\} | \bar{z}_T, \bar{C}_T = \bar{0}_T] = \mathrm{E}_{\mathcal P}[\bar{g}_T\{X_T(\bar{z}_{T-1})\}]$,
\begin{align}
\Big\{& \mathrm{E}_{\mathcal{P}_{\omega}}[g_1\{X_1\}|\bar{z}_T, \bar{C}_T = \bar{0}_T] = \mathrm{E}_{\mathcal{P}}[g_1\{X_1\}]; ~ \mathrm{E}_{\mathcal{P}_{\omega}}[R_t|\bar{z}_{T}, \bar{C}_T = \bar{0}_T] =  \mathrm{E}_{\mathcal{P}}[R_t|\bar{z}_{t-1}, \bar{C}_{t-1} = \bar{0}_{t-1}],\nonumber\\
& \text{where }R_t(\bar{z}_{t-1})= g_t\{X_t(\bar{z}_{t-1})\} - \beta_{X_t \sim \bar{X}_{t-1}|z_{t-1}, C_{t-1} = 0}\bar{g}_{t-1}\{{X}_{t-1}(\bar{z}_{t-2})\}\Big\}. \nonumber
\end{align}

\begin{proof}
We prove the statement for $T = 2$. The rest will be generalized from Theorem \ref{prop1}.
\begin{eqnarray}
&& \mathrm{E}_{\mathcal{P}_{\omega}}[g_1\{X_1\} |\bar{Z}_2 = \bar{z}_2, \bar{C}_2 = \bar{0}_2] = \mathrm{E}_{\mathcal P}[g_1\{X_1\}] ~ \text{\small (By the balance condition for $X_1$)}%=: T(X_1)
\nonumber \\
&& \mathrm{E}_{\mathcal{P}_{\omega}}[g_2\{X_2\} |\bar{Z}_2= \bar{z}_2, \bar{C}_2 = \bar{0}_2] \nonumber \\
&=&\mathrm{E}_{\mathcal{P}_{\omega}}[g_2\{X_2\} -  \beta_{X_2\sim X_1|z_1,C_1 = 0}g_1\{X_1\} |\bar{z}_2, \bar{C}_2 = \bar{0}_2] + \mathrm{E}_{\mathcal{P}_{\omega}}[\beta_{X_2\sim X_1|z_1,C_1 = 0}g_1\{X_1\}|\bar{z}_2, \bar{C}_2 = \bar{0}_2] \nonumber \\
&=&\mathrm{E}_{\mathcal P}[g_2\{X_2\}- \beta_{X_2\sim X_1|z_1,C_1 = 0} g_1\{X_1\}|z_1, C_1 = 0] +\beta_{X_2\sim X_1|z_1,C_1 = 0}\mathrm{E}_{\mathcal P}[g_1\{X_1\}]
\nonumber \\
&& \text{\small (By the balance conditions for $X_1, X_2$)} \nonumber \\
&=&\mathrm{E}_{\mathcal P}[g_2\{X_2(z_1)\} - \beta_{X_2\sim X_1|z_1,C_1 = 0} g_1\{X_1\}|z_1, C_1 = 0] + \beta_{X_2\sim X_1|z_1,C_1 = 0}\mathrm{E}_{\mathcal P}[g_1\{X_1\}]
\nonumber \\
&& \text{\small (By SUTVA)} \nonumber \\
&=&\mathrm{E}_{\mathcal P}[g_2\{X_2(z_1)\}- \beta_{X_2\sim X_1|z_1,C_1 = 0} g_1\{X_1\}] + \beta_{X_2\sim X_1|z_1,C_1 = 0}\mathrm{E}_{\mathcal P}[g_1\{X_1\}]
\nonumber \\
&& \text{\small (By the extra condition stated below)} \nonumber \\
&=&\mathrm{E}_{\mathcal P}[g_2\{X_2(z_1)\}] 
\nonumber 
\end{eqnarray}
The condition $\mathrm{E}_{\mathcal P}[g_2\{X_2(z_1)\} - \beta_{X_2\sim X_1|z_1,C_1 = 0}g_1\{X_1\}] = \mathrm{E}_{\mathcal P}[g_2\{X_2\} - \beta_{X_2\sim X_1|z_1,C_1 = 0}g_1\{X_1\}|z_1, C_1 = 0]$ holds if $X_2(z_1) \ci (Z_1,C_1)|X_1$ and $\mathrm{E}_{\mathcal P}[\mathrm{E}_{\mathcal P}[g_2\{X_2\} - \beta_{X_2\sim X_1|z_1,C_1 = 0}g_1\{X_1\}|X_1, z_1, C_1 = 0]] = \mathrm{E}_{\mathcal P}[g_2\{X_2\} - \beta_{X_2\sim X_1|z_1,C_1 = 0}g_1\{X_1\}|z_1,C_1 = 0]$. The proof is as follows.
\begin{eqnarray*}
&& \mathrm{E}_{\mathcal P}[g_2\{X_2(z_1)\} - \beta_{X_2\sim X_1|z_1,C_1 = 0}g_1\{X_1\}] \\
&=&  \mathrm{E}_{\mathcal P}[g_2\{X_2(z_1)\}] -\beta_{X_2\sim X_1|z_1,C_1 = 0}\mathrm{E}_{\mathcal P}[g_1\{X_1\}] \\
&=& 
\mathrm{E}_{\mathcal P}[\mathrm{E}_{\mathcal P}[g_2\{X_2\}|X_1, z_1,C_1 = 0]]- \beta_{X_2\sim X_1|z_1,C_1 = 0}\mathrm{E}_{\mathcal P}[g_1\{X_1\}] \\
&=& 
\mathrm{E}_{\mathcal P}[\mathrm{E}_{\mathcal P}[g_2\{X_2\} - \beta_{X_2\sim X_1|z_1,C_1 = 0}g_1\{X_1\}|X_1, z_1, C_1 = 0]] \\
&=& \mathrm{E}_{\mathcal P}[g_2\{X_2\} - \beta_{X_2\sim X_1|z_1,C_1 = 0}g_1\{X_1\}|z_1, C_1 = 0].
\end{eqnarray*}
\end{proof}

\clearpage
\section{DAGs for the simulation settings}\label{sec::simulation}

\begin{figure}[htbp]
\caption{Causal graphs for the simulation studies}
\begin{center}

\begin{subfigure}[t]{1.0\textwidth}
\centering
 \includegraphics[scale = 0.3]{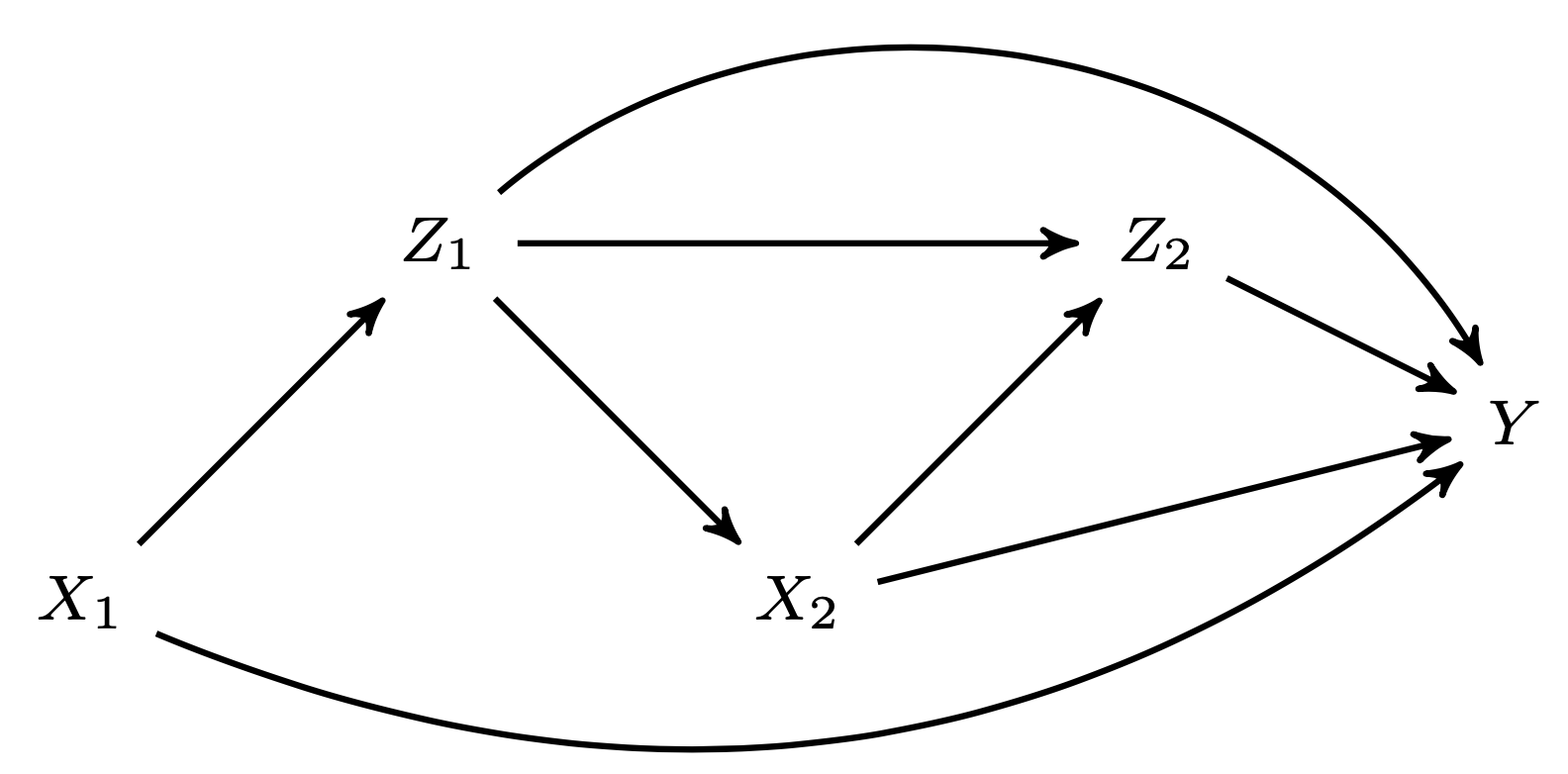}
\caption{Simulation study 1}

\label{fig:structure1}
\end{subfigure}

%%%%%%%

\begin{subfigure}[t]{1.0\textwidth}
\centering
  \includegraphics[scale = 0.3]{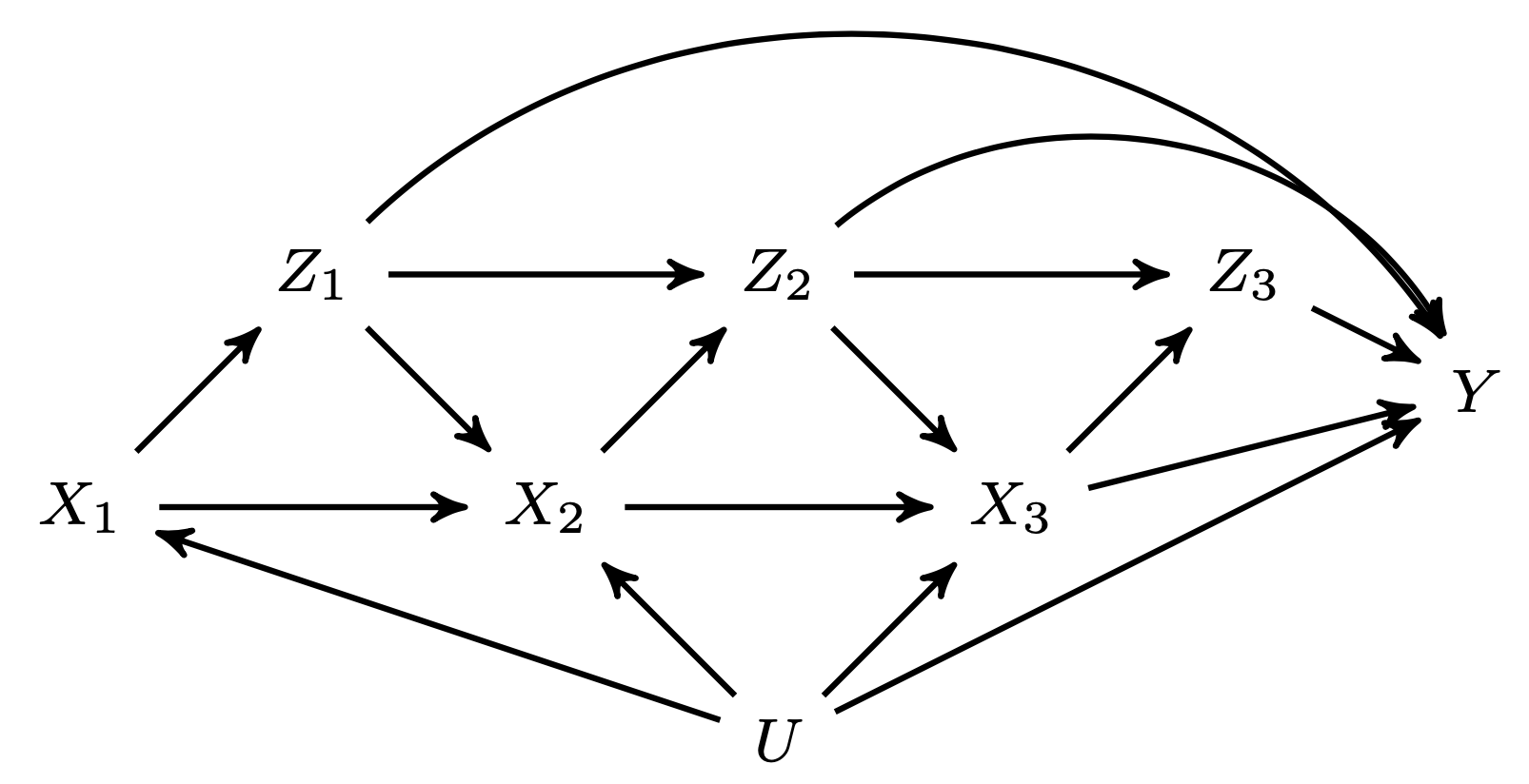}
\caption{Simulation study 2}
 
\label{fig:structure3}
\end{subfigure}

%%%%%%%

\begin{subfigure}[t]{1.0\textwidth}
\centering
  \includegraphics[scale = 0.3]{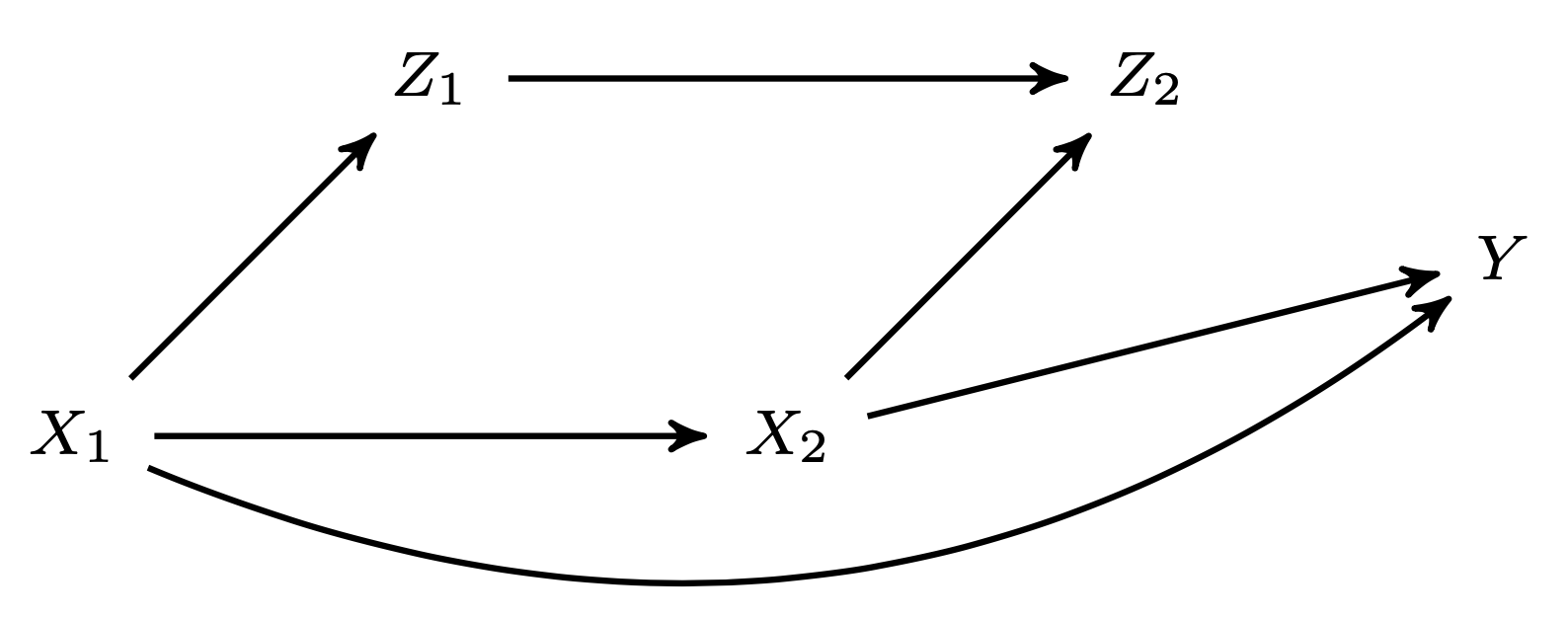}
\caption{Simulation study 3}

\label{fig:structure2}
\end{subfigure}

\end{center}
\label{fig:causal}
\end{figure}

\clearpage
\section{Additional tables for the case study}\label{sec::casestudy}

\begin{table}[!htbp] 
\begin{center}
\caption{Regression coefficients in $\text{logit}\{\mathrm{P}(Z_t = 1|\bar{X}_t, \bar{Z}_{t-1})\}$ and $\mathrm{E}_{\mathcal S}[X_t|\bar{X}_{t-1},\bar{Z}_{t-1}]$. The continuous covariates are scaled to have mean 0 and variance 1.}
\scriptsize
\setstretch{1}
\setlength{\tabcolsep}{1.pt}
\begin{tabular}{lrrrrrrrrrrrrrr}
  \hline
 & $Z_1$ & $Z_2$ & $Z_3$ & $Z_4$ & $X_{2,\text{GPA}}$ & $X_{2,\text{Att.}}$ & $X_{3,\text{Lang.}}$ & $X_{3,\text{Math}}$ & $X_{3,\text{GPA}}$ & $X_{3,\text{Att.}}$ & $X_{4,\text{GPA}}$ & $X_{4,\text{Att.}}$ & $Y_{\text{Language}}$ & $Y_{\text{Math}}$ \\ 
  \hline
Intercept & *1.04 & *-0.11 & *-1.36 & *-3.02 & *-0.26 & *-0.12 & *-0.11 & *-0.04 & *-0.05 & *-0.07 & *-0.05 & *-0.13 & *464.21 & *466.08 \\ 
\multicolumn{3}{l}{Parents' Education} \\
  \hspace{2pt}Primary & *-0.70 & -0.06 & 0.30 & 0.18 & 0.06 & *0.13 & *-0.08 & *-0.15 & 0.00 & 0.01 & 0.01 & *0.08 & *-29.87 & *-23.82 \\ 
  \hspace{2pt}Secondary & *-0.24 & -0.07 & 0.26 & 0.03 & *0.08 & *0.18 & -0.06 & *-0.11 & 0.00 & 0.03 & 0.02 & *0.08 & -4.24 & -1.11 \\ 
  \hspace{2pt}Technical & 0.10 & 0.12 & -0.08 & -0.07 & *0.08 & *0.19 & -0.01 & -0.03 & 0.02 & 0.04 & 0.01 & 0.05 & *10.71 & *10.76 \\ 
  \hspace{2pt}College & 0.19 & 0.14 & *-0.61 & -0.02 & *0.11 & *0.15 & 0.06 & 0.03 & 0.04 & 0.04 & 0.05 & 0.03 & *18.01 & *22.02 \\ 
  \multicolumn{3}{l}{Household Income} \\
  \hspace{2pt}$< 200$ & *-0.27 & -0.19 & *0.35 & -0.20 & -0.03 & *-0.12 & -0.06 & *-0.06 & -0.03 & -0.01 & -0.04 & -0.01 & *-26.52 & *-20.85 \\ 
  \hspace{2pt}200-400 & 0.14 & -0.10 & 0.30 & 0.06 & -0.05 & -0.06 & -0.02 & 0.01 & -0.02 & 0.04 & -0.01 & 0.00 & *-12.49 & -6.93 \\ 
  \hspace{2pt}400-600 & *0.43 & 0.14 & 0.04 & 0.09 & -0.02 & -0.02 & 0.01 & *0.06 & -0.02 & 0.02 & 0.00 & 0.01 & -4.58 & -2.25 \\ 
  \hspace{2pt}600-1400 & *0.79 & *0.38 & -0.09 & 0.28 & 0.02 & 0.02 & 0.03 & *0.11 & 0.01 & 0.04 & 0.00 & 0.03 & -0.06 & 5.33 \\ 
  \hspace{2pt}$>1400$ & *1.02 & *0.64 & -0.14 & 0.60 & *0.12 & 0.04 & 0.07 & *0.15 & 0.07 & 0.08 & 0.00 & 0.00 & 10.56 & *17.65 \\ 
  Female & *0.24 & *0.22 & -0.03 & *0.25 & *0.14 & *-0.06 & *0.09 & *-0.15 & *0.05 & *-0.09 & *0.20 & *-0.12 & 2.30 & *-6.92 \\ 
  \hline
  $X_{1,\text{Language}}$ & *0.25 & 0.00 & *-0.10 & 0.00 & *0.13 & *0.03 & *0.42 & *0.10 & *0.03 & *0.02 & *0.02 & -0.01 & *24.25 & 1.49 \\ 
  $X_{1,\text{Math}}$ & *0.38 & *-0.08 & *-0.12 & -0.04 & *0.20 & *0.09 & *0.13 & *0.48 & *0.03 & 0.01 & *0.01 & -0.01 & *11.84 & *23.09 \\ 
  $X_{1,\text{GPA}}$ & *-0.80 & -0.03 & 0.02 & 0.03 & *0.54 & *-0.14 & *0.05 & *0.08 & *0.05 & *-0.02 & *0.06 & *-0.05 & *9.94 & *10.40 \\ 
  $X_{1,\text{Attendance}}$ & *0.82 & 0.03 & *0.11 & -0.05 & *-0.31 & *0.34 & *-0.04 & *-0.05 & *-0.03 & *0.05 & *-0.04 & *0.08 & *-8.76 & *-8.98 \\ 
  $X_{1,\text{NA in Lang.}}$ & 0.12 & 0.03 & 0.00 & 0.04 & *-0.09 & *-0.22 & *-0.14 & -0.04 & -0.04 & -0.07 & -0.01 & 0.04 & -6.54 & -0.34 \\ 
  $X_{1,\text{NA in Math}}$ & -0.06 & 0.19 & *0.43 & -0.06 & *-0.10 & *-0.35 & 0.00 & *-0.10 & *-0.08 & *-0.09 & -0.04 & -0.02 & -8.69 & *-12.73 \\ 
  $Z_1$ &  & *2.05 & *0.70 & *0.54 & *0.18 & *0.09 & *0.09 & *0.15 & 0.01 & -0.01 & -0.01 & *-0.04 & *26.85 & *25.16 \\ 
  \hline
  $X_{2,\text{GPA}}$ &  & -0.04 & *-0.19 & 0.01 &  &  & *0.11 & *0.15 & *0.43 & *0.04 & *0.39 & *0.25 & *5.68 & *7.32 \\ 
  $X_{2,\text{Attendance}}$ &  & *0.39 & *0.11 & -0.03 &  &  & 0.01 & *0.05 & *0.05 & *0.30 & *0.08 & *0.25 & 1.23 & *3.96 \\  
  &  &  & *6.89 & *0.35 &  &  & *0.05 & *0.07 & 0.03 & *0.11 & 0.01 & *0.08 & *24.79 & *27.48 \\ 
  \hline
  $X_{3,\text{Language}}$ &  &  & 0.04 & -0.03 &  &  &  &  &  &  & *0.07 & *0.04 & *25.30 & *3.07 \\ 
  $X_{3,\text{Math}}$ &  &  & -0.07 & *0.23 &  &  &  &  &  &  & *0.05 & *0.05 & *18.77 & *41.51 \\
  $X_{3,\text{GPA}}$ &  &  & -0.01 & 0.05 &  &  &  &  &  &  & *0.45 & *-0.42 & -3.42 & -1.00 \\ 
  $X_{3,\text{Attendance}}$ &  &  & *0.19 & *-0.29 &  &  &  &  &  &  & *-0.33 & *0.51 & *7.10 & *5.58 \\ 
  $Z_3$ &  &  &  & *5.79 &  &  &  &  &  &  & *-0.07 & *0.12 & *-46.35 & *-47.26 \\ 
  \hline
  $X_{4,\text{GPA}}$ &  &  &  & *-0.18 &  &  &  &  &  &  &  &  & *26.53 & *26.41 \\ 
  $X_{4,\text{Attendance}}$ &  &  &  & *0.50 &  &  &  &  &  &  &  &  & *10.14 & *14.07 \\ 
  $Z_4$ &  &  &  &  &  &  &  &  &  &  &  &  & *8.76 & 5.33 \\ 
   \hline
\end{tabular}
\end{center}
\end{table}

\pagebreak
\onehalfspacing
\bibliographystyle{asa}
\bibliography{msm24/ref}